\documentclass{article}
\usepackage{algorithm}
\usepackage[algo2e]{algorithm2e}
\usepackage{algpseudocode}
\usepackage{amsfonts}
\usepackage{eurosym}
\usepackage{amsmath}
\usepackage{rotating}
\usepackage{graphicx}
\usepackage{caption}
\usepackage{lscape}
\usepackage{xcolor}
\usepackage[colorlinks = true,
            linkcolor = blue,
            urlcolor  = blue,
            citecolor = blue,
            anchorcolor = blue]{hyperref}
\usepackage[nottoc]{tocbibind}
\usepackage{subcaption}
\usepackage{bm}
\usepackage{cancel}
\usepackage{soul}
\usepackage{booktabs}
\usepackage{tablefootnote} 
\usepackage{threeparttable}
\usepackage{tabularx}
\usepackage{float}
\usepackage{authblk}
\usepackage{tabularx}
\usepackage{booktabs}

\begin{document}

\title{A Weibull Mixture Cure Frailty Model\\ for High-dimensional Covariates}

\date{}
\author[1]{Fatih K\i z\i laslan}
\author[2]{David Michael Swanson}
\author[1]{Valeria Vitelli}
\affil[1]{Oslo Centre for Biostatistics and Epidemiology, Department of Biostatistics, University of Oslo, Oslo, Norway}
\affil[2]{Department of Biostatistics, The University of Texas MD Anderson Cancer Center, Houston, TX, USA}
\maketitle

\begin{abstract}

A novel mixture cure frailty model is introduced for handling censored survival data. Mixture cure models are preferable when the existence of a cured fraction among patients can be assumed. However, such models are heavily underexplored: frailty structures within cure models remain largely undeveloped, and furthermore, most existing methods do not work for high-dimensional datasets, when the number of predictors is significantly larger than the number of observations. In this study, we introduce a novel extension of the Weibull mixture cure model that incorporates a frailty component, employed to model an underlying latent population heterogeneity with respect to the outcome risk.
Additionally, high-dimensional covariates are integrated into both the cure rate and survival part of the model, providing a comprehensive approach to employ the model in the context of high-dimensional omics data. We also perform variable selection via an adaptive elastic-net penalization, and propose a novel approach to inference using the expectation–maximization (EM) algorithm. Extensive simulation studies are conducted across various scenarios to demonstrate the performance of the model, and results indicate that our proposed method outperforms competitor models. We apply the novel approach to analyze RNAseq gene expression data from bulk breast cancer patients included in The Cancer Genome Atlas (TCGA) database. A set of prognostic biomarkers is then derived from selected genes, and subsequently validated via both functional enrichment analysis and comparison to the existing biological literature. Finally, a prognostic risk score index based on the identified biomarkers is proposed and validated by exploring the patients' survival.


\end{abstract}

\section{Introduction}

When analyzing time-to-event data, the presence of substantial censoring following a prolonged follow-up period often indicates the presence of ``long-term survivors'' or ``cured individuals'', i.e., individuals who may never experience the event of interest. This phenomenon can be observed in certain clinical investigations such as cancer studies, where successful treatment can effectively prevent disease recurrence. This scenario becomes more noticeable among patients who are diagnosed in the early stages of cancer development.  	
The presence of a cured fraction violates a key assumption in traditional survival models, like the Cox Proportional Hazards (PH) model, which assumes that all subjects will inevitably experience the event of interest. When a cured fraction exists, this assumption no longer holds true. Hence, the use of a Cox PH model in such cases can lead to an underestimation of hazard rates and an overestimation of survival probabilities for individuals susceptible to the event \cite{price2001modelling}. In practice, the cured subjects cannot be observed directly, as they are censored together with the individuals who will eventually experience the outcome. However, the Kaplan-Meier (KM) estimated survival curve can be used for determining the existence of a cured fraction when the follow-up period is sufficiently long. A long and consistent plateau in this curve could imply the existence of a cured subset of subjects.	
Cure models expand the scope of survival analysis by incorporating a fraction of cured individuals. The mixture cure (MC) model is the most commonly employed cure rate model, which was first introduced by Boag \cite{boag1949maximum} and improved by Berkson and Gage \cite{berkson1952survival}. The MC model incorporates two components: one component represents the cured fraction, i.e., the fraction of individuals who will never experience the event of interest, having a survival probability of one. The other component instead captures the susceptible fraction, i.e., the fraction of individuals for whom the event occurrence is governed by a proper survival distribution, which is often referred to as the ``latency distribution''. 

In medical and epidemiological studies, it is common to assume the existence of unobserved factors generating heterogeneity among individuals that cannot be explained via the observed covariates. Frailty models can be employed to incorporate and account for unobserved heterogeneity among individuals, to allow for a more accurate modeling of the survival outcome. 
These models introduce a random frailty term, often assumed to follow a specific distribution, which captures individual-specific characteristics that affect the hazard or risk of the event. Price and Manatunga \cite{price2001modelling} introduced a gamma frailty term into the latency distribution to address the presence of unobserved risks within the MC model. Peng and Zhang \cite{peng2008estimation} extended this model by incorporating covariates into both the cure rate and the latency distribution, and they named this model the mixture cure gamma frailty model. Due to the presence of missing variables in cure models, Sy and Taylor \cite{sy2000estimation} as well as Peng and Dear \cite{peng2000nonparametric} employed the EM algorithm \cite{dempster1977maximum} to derive maximum likelihood estimates of the model parameters. Moreover, Peng and Zhang \cite{peng2008estimation} studied semi-parametric estimation methods for the mixture cure gamma frailty model using the EM algorithm and a multiple imputation method with low-dimensional covariates. They also examined the identifiability of the general mixture cure frailty model (MCFM) in \cite{peng2008identifiability}. Cai et al. \cite{cai2012smcure} developed an R package, called \texttt{smcure} \cite{smcure}, to facilitate the application of semi-parametric estimation techniques to both the proportional hazards MC model and the accelerated failure time MC model. Finally, frailty models have also been used in this context when dealing with correlated data, and particularly to be able to take into account heterogeneity in the presence of recurrent events. For instance, Rondeau et al. \cite{rondeau2013cure} studied the analysis of recurrent time-to-event data using frailty models with a cured fraction.

The focus of the present paper is on situations when the available covariates are way more than the subjects included in the sample. Thanks to advancements in biomedicine, we can now generate and gather molecular data from diverse modalities, such as genomics, epigenomics, transcriptomics, proteomics, and metabolomics, often resulting in high-dimensional data sets. Due to the volume and complexity of such data, and related challenges and opportunities, more sophisticated methods than the traditional ones are required, especially to deal with variable selection. To this purpose, the use of advanced regularization methods such as least absolute shrinkage and selection operator (lasso) \cite{tibshirani1996regression}, elastic net \cite{zou2005regularization}, the adaptive elastic net \cite{zou2009adaptive}, and others has become a standard practice for enhancing accuracy and interpretation. The three methods mentioned above are established popular choices for performing simultaneous prediction and variable selection in high-dimensional linear regression problems. These techniques allow discovering meaningful patterns, such as identifying relevant prognostic biomarkers for a certain disease, from high-dimensional molecular data. When it comes to generalized linear models, Friedman et al. \cite{friedman2010regularization} introduced a novel approach utilizing coordinate descent to compute regularization paths, and also implemented their method in the now very popular R package \texttt{glmnet} \cite{glmnet}. Simon et al. \cite{simon2011regularization} extended the previous work to Cox models for right-censored data, incorporating elastic net regularization, with a very efficient algorithm that was then integrated into the \texttt{glmnet} R package, enabling the solution of significantly larger problems than previously possible. Tay et al. \cite{tay2023elastic} expanded the scope of the elastic net-regularized regression to encompass all generalized linear model families, including Cox models with (start, stop] data and strata, also implemented within the \texttt{glmnet} R package.
Adaptive versions of both the lasso \cite{zou2005regularization} and elastic net \cite{zou2009adaptive} have also been developed, and their properties explored under specific conditions. In a discussion article, B\"{u}hlmann and Meier \cite{buhlmann2008discussion} emphasized the relevance of a multi-step procedure for the adaptive lasso, showing that it allows for sparser models at each step.  Furthermore, Xiao and Xu \cite{xiao2015multi} introduced the multi-step adaptive elastic-net, showing via extensive numerical exploration that this method effectively reduces the false positives in the variable selection process while preserving estimation accuracy, and implemented it into the \texttt{msaenet} R package \cite{msaenet}. For further insights into regularization methods and their applications, interested readers may refer to B\"{u}hlmann and van de Geer \cite{buhlmann2011statistics}, Hastie et al. \cite{hastie2015statistical} and James et al. \cite{james2021statistical}.

In the context of survival data within high-dimensional settings, the traditional Cox PH model has been commonly used. Although there has recently been a growing interest in the application of cure models to high-dimensional data, the existing literature on this subject has remained somewhat limited.
Liu et al. \cite{liu2012variable} introduced a variable selection procedure for the semi-parametric MC model using penalties based on the smoothly clipped absolute deviation (SCAD) and lasso. Fan et al. \cite{Fan2017promoting} explored the minimax concave penalty (MCP), ridge, and lasso penalties, to estimate the MC model in high-dimensional data sets. Masud et al. \cite{masud2018variable} studied variable selection problems for both the mixture and promotion time cure models, employing penalty terms such as lasso and adaptive lasso. Baretta and Heuchenne \cite{beretta2019variable} conducted a study on variable selection using SCAD penalties for the semi-parametric PH cure model, considering time-varying covariates, and implemented the method in the \texttt{penPHcure} \cite{penPHcure} R package. Sun et al. \cite{sun2019variable} developed a variable selection methodology that incorporates lasso, adaptive lasso, and SCAD-type penalties, tailored for the semi-parametric promotion time cure model, and particularly suitable for interval-censored data. 
Bussy et al. \cite{bussy2019c} proposed a novel statistical model known as C-mix, specifically designed as a mixture-of-experts model for handling censored survival outcomes, to model the patients' heterogeneity by detecting distinct subgroups, and allowed high-dimensional covariates by employing an elastic net penalization. They also conducted a comparison of the C-mix model with both the MC and Cox PH models through simulation experiments. Shi et al. \cite{shi2020promoting} developed a novel penalty method for the MC model that includes both the MCP and a new penalty term that promotes sign consistency using the covariate effects. Xie and Yu \cite{xie2021mixture} proposed the use of neural networks to address inference in the MC model, demonstrating their favorable predictive ability in high-dimensional settings. Xu et al. \cite{xu2021variable} considered the variable selection problem by adopting penalties such as lasso, adaptive lasso, and SCAD for the generalized odds rate MC model, particularly in the context of interval-censored data. Lastly, Fu et al. \cite{fu2022controlled} investigated the Weibull MC model, incorporating generalized monotone incremental forward stagewise (GMIFS) \cite{hastie2007stagewise} and EM algorithm techniques for inference. They applied the EM algorithm to a penalized Weibull MC model, employing a lasso type penalty. Their findings indicated a superior performance of their penalized MC model as compared to alternative methods in various simulation scenarios. It is noteworthy to emphasize that all aforementioned studies on cure models exhibit significant limitations concerning the size of covariates relative to the sample size that can be handled by the model, in both simulation experiments and real-data applications, with the only exception being the case studies conducted by Bussy et al. \cite{bussy2019c} and Fu et al. \cite{fu2022controlled}. Moreover, neither of the two latter studies include a frailty term in the model.

This study is therefore aimed at developing a first novel version of the MCFM useful in instances where high-dimensional covariates are potentially associated with both the cured rate and uncured subjects in the sample. To this purpose, a new penalized EM algorithm is introduced, incorporating adaptive elastic net penalties specifically tailored for the MCFM. In particular, the model assumes that the distribution of uncured subjects follows a Weibull distribution, and that their survival probability at each time depends on a subset of the high-dimensional set of covariates, which is also a target of inference. To the best of our knowledge, this study represents the first exploration of MCFM, i.e. the first attempt at incorporating frailty into the MCM, within high-dimensional settings. Nonetheless, surpassing the predominant focus of the existing literature on MCFM on low-dimensional settings, our study explores scenarios in which the sizes of covariates significantly exceed the sample sizes in both simulated and real-data applications.
The rest of this article is organized as follows: in Section \ref{sec:PenMCFM}, we present the classical MCFM and its extension to high-dimensional settings. In Section \ref{sec:EM}, we introduce the novel penalized MCFM (penMCFM), with the associated adaptive EM algorithm for inference, in details.
The performances of our proposed methods are compared with some competitors through a comprehensive Monte Carlo simulation study in Section \ref{sec:Simulation}. 
In Section \ref{sec:Application}, all considered methods are applied to publicly available breast cancer RNA-seq data, with the aim of identifying potential biomarker genes. We also conduct functional enrichment analyses for better interpretation, and determine a prognostic risk score for enhanced validation, of all the identified biomarker genes. Finally, in Section \ref{sec:Conclusion} we provide a brief summary and discussion of the study, also mentioning potential future research directions.

\section{The Penalized Mixture Cure Frailty Model}\label{sec:PenMCFM}

\subsection{Classical Mixture Cure Frailty Models}

Let the random variable $T$ represent the lifetime of interest with 
survival function denoted by $S_{pop}(t), t\in[0,+\infty)$. Let $Y$ be the indicator for a
subject eventually $(Y=1)$ or never $(Y=0)$ experiencing the event of
interest, with $\pi =P(Y=1)$ representing the probability of a subject being
susceptible (or uncured) for the event of interest. Among the subjects for whom 
$Y=0$, the survival function is $S(t|Y=0)=1, \forall t\in[0,+\infty)$, and for those who experience the event ($Y=1$)%
, the survival function and the probability density function (pdf) are $%
S(t|Y=1)$ and $f(t|Y=1)$, respectively. $Y$ is not observed for a censored subject. The population survival function is therefore defined as

\begin{equation}
S_{pop}(t)=1-\pi +\pi S(t|Y=1).  \label{S_pop0}
\end{equation}%
Note that since $S_{pop}(t)\rightarrow 1-\pi $ as $t\rightarrow +\infty $, $%
S_{pop}(t)$ is not a proper survival function. The uncured rate $\pi $ and the
survival function of the uncured subjects $S(t|Y=1)$ are also referred to as
the incidence and the latency distribution, respectively.

The basic model introduced in \eqref{S_pop0} can be extended to include the covariates associated with
the incidence and latency distributions. Let us denote via $\mathbf{x}$ and $\mathbf{z}$ the covariates that have effect on the latency
distribution and the incidence, respectively. The model (\ref{S_pop0}) can be then rewritten as

\begin{equation}
S_{pop}(t|\mathbf{x,z})=1-\pi (\mathbf{z})+\pi (\mathbf{z})S(t|Y=1,\mathbf{x}%
),  \label{S_pop1}
\end{equation}%
where $\pi (\mathbf{z})$ is the probability of a subject being uncured
conditionally on $\mathbf{z}$, and $S(t|Y=1,\mathbf{x})$ is the survival
function of the lifetime distribution of uncured subjects conditionally on $%
\mathbf{x}$. 
Concerning the modeling of the effect of the covariates $\mathbf{z}$ on the incidence, as previously proposed in \cite{farewell1982use} 
we use a logistic regression model of the form $\pi(\mathbf{z}) = e^{\mathbf{z}^\top \mathbf{b}}/(1+e^{\mathbf{z}^\top \mathbf{b}}),$ where $\mathbf{z}^\top \in\mathbb{R}^{n \times P_1+1} $ is a covariate matrix, with columns $\mathbf{z}_1, \cdots, \mathbf{z}_n \in\mathbb{R}^{P_1+1}$, and $\mathbf{b}= (b_0,b_1, \cdots, b_{P_1})^\top \in\mathbb{R}^{P_1+1} $ is a vector of unknown regression coefficients. When the mixture cure model defined in (\ref{S_pop1}) is specified via proportional hazards, we get the following PH mixture cure model

\begin{equation}
S_{pop}(t|\mathbf{x,z})=1-\pi (\mathbf{z})+\pi (\mathbf{z}) S_{0}(t)^{\exp(\mathbf{x}^\top \bm{\beta} )},  \label{S_pop2}
\end{equation}%
where $S_{0}(t)$ is the baseline survival function,  $\mathbf{x}^\top \in\mathbb{R}^{n \times P_2}$ is the covariate matrix, with columns $\mathbf{x_1}, \cdots, \mathbf{x_n} \in\mathbb{R}^{P_2}$, and $\bm{\beta} = (\beta_1, \cdots, \beta_{P_{2}})^ \top \in\mathbb{R}^{P_2}$ is the vector of unknown regression coefficients for the latency distribution.

In medical and epidemiological studies, frailty models extend the Cox PH model to account for unobservable heterogeneity among individuals. This extension provides a more flexible structure for analysis.
A frailty is defined as an unobservable, random, multiplicative factor acting on the hazard function. Let $W_i$ be a non-negative frailty
random variable associated to the $i$th subject, $i=1,\ldots,n$, with cumulative distribution function (cdf) $F_{W_i}(w)$.
The hazard function of the $i$th subject with frailty $W_i$ is 
\begin{equation*}
h(t|W_i) = W_i \; h_{0}(t)  \exp (\mathbf{x_i}^\top \bm{\beta}),
\end{equation*}%
where $h_{0}(t)$ is a baseline hazard function common for all subjects and $\mathbf{x_i}$ is the covariate vector for the $i$th subject. If
we include the frailty in the latency distribution in model (%
\ref{S_pop2}), the conditional survival function given the frailty $W$ takes the form
\begin{equation*}
S(t|Y=1,W,\mathbf{x})=\exp (-W e^{ \mathbf{x}^\top \bm{\beta}} H_{0}(t)),
\end{equation*}%
where $H_{0}(t)$ is the baseline cumulative hazard function. Then, the
marginal survival function of uncured subjects based on the frailty model is
given by 
\begin{eqnarray}
S(t|Y=1,\mathbf{x}) &=&\int_0^{+\infty}S(t|Y=1,W=w,\mathbf{x})dF_{W}(w) \notag \\
&=& L_{W}(e^{\mathbf{x}^\top \bm{\beta}}H_{0}(t)),  \notag  
\end{eqnarray}%
where $L_{W}(s)=E(e^{-ws})$ is the Laplace transformation of the frailty $W$%
. Hence, model (\ref{S_pop2}) with frailty becomes 
\begin{equation}
S_{pop}(t|\mathbf{x,z})=1-\pi (\mathbf{z})+\pi (\mathbf{z})L_{W}(e^{\mathbf{x}^\top \bm{%
\beta}}H_{0}(t)).  \label{S_pop3}
\end{equation}%
Model \eqref{S_pop3} was first introduced by Peng and Zhang \cite%
{peng2008estimation} and called the \emph{mixture cure frailty model}. Model (\ref{S_pop3}) reduces to the PH mixture cure model in (\ref{S_pop2}) when
there is no frailty effect, namely $W\equiv 1$, and it reduces to a standard
frailty model when there is no cure fraction existing in the population,
namely $\pi (\mathbf{z})\equiv 1$.

In this study, we consider a fully parametric version of model (\ref%
{S_pop3}). We assume that the baseline of the latency distribution follows
a Weibull distribution $WE(\alpha ,\gamma )$ with $\alpha $ and $\gamma $ being the scale and shape parameters, respectively.
The hazard and cumulative hazard functions for the
baseline of the latency distribution then become $h_{0}(t)=\alpha \gamma
t^{\gamma -1}$ and $H_{0}(t)=\alpha t^{\gamma }$, respectively. Because of the identifiability issues
associated with the mixture cure frailty model \cite{peng2008identifiability}, we need to fix the mean of the frailty distribution
to $1$.
Therefore, we assume that the frailty $W$ follows a gamma distribution
with mean $1$ and variance $1/\theta $. The Laplace transformation of the frailty is then $L_{w}(s)=(1+s/\theta )^{-\theta }$.
Therefore, given our parametric assumptions, the survival function of model (\ref{S_pop3}) can be
rewritten as 
\begin{eqnarray}
S_{pop}(t|\mathbf{x,z}) 
&=&\frac{1}{1+e^{\mathbf{z}^\top \mathbf{b}}}+\frac{e^{\mathbf{z}^\top \mathbf{b}}}{1+e^{\mathbf{z}^\top \mathbf{b}}}  \left( 1+\frac{\alpha t^{\gamma }e^{\mathbf{x}^\top\bm{\beta}}}{\theta }\right)
^{-\theta }. \label{S_pop_frailty}
\end{eqnarray}%
Note that, with the parametric assumptions as detailed above, the hazard function of the latency, i.e., of the uncured subjects, does
not satisfy the PH assumption \cite{peng2008estimation}.

\subsection{Mixture Cure Frailty Models with high-dimensional covariates}

In the context of medical applications, and particularly in cancer studies, molecular data of several kinds can be used to better estimate both the cured fraction, and the uncured patients survival. This is particularly relevant in cancer studies, since tissue samples originating from biopsies of the tumor are often available, and these can provide excellent information on the tumor composition and characterization. Such molecular omics data often include genomics, epigenomics, transcriptomics, proteomics, metabolomics and radiomics information, often generated using advanced high-throughput technologies to allow the understanding of complex biological systems and their underlying mechanisms. Common characteristic of all these data layers is their high-dimensionality, meaning that the number of variables $p$ included in each omics data layer is much larger than the sample size $n$. We plan to modify the mixture cure frailty model as discussed above so that high-dimensional omics data can be used as covariates in the model. 

In this set-up, we consider both covariates $\mathbf{x}$ and $\mathbf{z}$, related to the latency part and the uncured rate of the model, respectively, as potentially high-dimensional, for example including some kind of omics information. These covariates can also include lower-dimensional information associated to the patients' demographics or clinical characteristics, such as age, sex, treatment method. This means that the two vectors of variables in $\mathbf{x}$ and $\mathbf{z}$ could potentially combine very diverse exogenous information, including both clinical variables and high-dimensional omics features, which often show varying degrees of accuracy, redundancy, and noise.

We apply well-known regularization approaches such as lasso \cite{tibshirani1996regression} and the adaptive elastic net \cite{zou2009adaptive} to regression models to find the relevant variables from our complete set of omics covariates in both $\mathbf{x}$ and $\mathbf{z}$. The high-dimensional covariate matrices $\mathbf{x}$ and $\mathbf{z}$ can share features or
not, depending on the available domain knowledge, and moreover the low-dimensional covariates (e.g., demographics and clinical information) in the matrices are left unpenalized. In what follows, we will refer to the low- and high-dimensional parts of the covariate matrices via ``unpenalized'' and ``penalized'' variables, respectively. For these covariates and for their corresponding regression coefficients, we use the following notations. The covariate matrices $\mathbf{x}$ and $\mathbf{z}$ can include both types of unpenalized and penalized variables, which are represented as $\mathbf{x}=(\mathbf{x}_{u}, \mathbf{x}_{p})$ and $\mathbf{z}=(\mathbf{z}_{u}, \mathbf{z}_{p}),$ where $\mathbf{x}_u, \: \mathbf{z}_{u}$ and $\mathbf{x}_p, \: \mathbf{z}_{p}$ represent the unpenalized and penalized variables in $\mathbf{x}$ and $\mathbf{z}$, respectively. The vector of regression coefficients $\bm{\beta}$ and $\mathbf{b}$ are also split in the same way into $\bm{\beta}=(\bm{\beta}_{u},\bm{\beta}_{p})$ and $\mathbf{b}=(\mathbf{b}_{0},\mathbf{b}_{u},\mathbf{b}_{p}),$ where $b_{0}$ is
the intercept, and $\bm{\beta}_{u}, \: \mathbf{b}_{u}$ and $\bm{\beta}_{p}, \: \mathbf{b}_{p}$ represent the unpenalized and penalized regression coefficients, respectively.

\section{Inference in the Penalized Mixture Cure Frailty Model}\label{sec:EM}

Let the observed data be $\mathbf{D}_{i}=(\delta _{i},t_{i},\mathbf{x}_{u,i},\mathbf{x}_{p,i},%
\mathbf{z}_{u,i},\mathbf{z}_{p,i}),$ $i=1,...,n,$ where $t_{i}$ is the
observed survival time for the $i$th subject, $\delta _{i}$ is an indicator
function of censoring, with $\delta _{i}=1$ for the uncensored time and $%
\delta _{i}=0$ for the censored time. $\mathbf{z}_{u,i}\in\mathbb{R}^{P^u_1}$ and $\mathbf{z}_{p,i}\in\mathbb{R}^{P^p_1},$ are respectively the unpenalized and penalized observed covariates associated to the cure part of the model for the $i$th subject, $i=1,...,n$, while $
\mathbf{x}_{u,i}\in\mathbb{R}^{P^u_2}$ and $\mathbf{x}_{p,i}\in\mathbb{R}^{P^p_2}$ are the same for the survival part. When possible, we will use the compact notation $\mathbf{z}_i=( \mathbf{1}, \mathbf{z}_{u,i}, \mathbf{z}_{p,i})$ and $\mathbf{x}_i=(\mathbf{x}_{u,i}, \mathbf{x}_{p,i})$, with $\mathbf{z}_i\in\mathbb{R}^{P_1+1}$ and $\mathbf{x}_i\in\mathbb{R}^{P_2}$, $P_1 = P^u_1+P^p_1$ and $P_2 = P^u_2+P^p_2$. 

The likelihood function for the right-censored
observed survival data $\mathbf{D}=(\mathbf{D}_{1},...,\mathbf{D}_{n})$ is 
\begin{eqnarray}
L(\alpha ,\gamma ,\theta, \bm{\beta}, \mathbf{b}|\mathbf{D})   &=&  \prod_{i=1}^{n}\left\{ f_{pop}(t_{i}|\mathbf{x}_{i}\mathbf{,z}%
_{i})\right\} ^{\delta _{i}}\left\{ S_{pop}(t_{i}|\mathbf{x}_{i}\mathbf{,z}%
_{i})\right\} ^{1-\delta _{i}}  \notag \\
 &=&  \prod_{i=1}^{n}\left\{ \frac{e^{\mathbf{z}_{i}^\top \mathbf{b} }}{1+e^{\mathbf{z}_{i}^\top \mathbf{b} }}%
\alpha \gamma t_{i}^{\gamma -1}e^{\mathbf{x}_{i}^\top \bm{\beta} }\left( 1+\frac{\alpha
t_{i}^{\gamma }e^{\mathbf{x}_{i}^\top \bm{\beta} }}{\theta }\right) ^{-\theta-1 }\right\}
^{\delta _{i}}  \notag \\
 && \times \left\{ \frac{1}{1+e^{\mathbf{z}_{i}^\top \mathbf{b} } }\left( 1+e^{\mathbf{z}_{i}^\top \mathbf{b} }\left(
1+\frac{\alpha t_{i}^{\gamma}e^{ \mathbf{x}_{i}^\top \bm{\beta} }}{\theta }\right)
^{-\theta }\right) \right\} ^{1-\delta _{i}}   \label{l_observed} 
\end{eqnarray}
where $\alpha ,\gamma $ are the unknown parameters of the Weibull distribution, $%
\theta $ is the unknown parameter of the gamma distribution used for the frailty, $\mathbf{b}\in\mathbb{R}^{P_1+1}$ and $\bm{\beta}\in\mathbb{R}^{P_2}$ are vectors of
unknown regression coefficients for the covariates $\mathbf{x}$ and $\mathbf{z}$, respectively, where we indicate $\mathbf{z}_{i}^\top \mathbf{b} = b_{0} + \mathbf{z}_{u,i}^\top \mathbf{b}_{u}+ \mathbf{z}_{p,i}^\top \mathbf{b}_{p}$ and $\mathbf{x}_{i}^\top \bm{\beta}  = \mathbf{x}_{u,i}^\top \bm{\beta}_u  + \mathbf{x}_{p,i}^\top \bm{\beta}_{p}$, $i=1,...,n$. The maximum likelihood estimator of the unknown parameters can be obtained by direct
maximization of the observed likelihood function in (\ref{l_observed}) when the
number of unknown parameters is not large. Since we would like to
consider high-dimensional covariates and perform variable selection via
penalization, direct maximization of the likelihood function will not
provide parameter estimates. We therefore propose to adapt the E-M algorithm to include regularization methods to the purposes of parameter estimation and variable selection in penMCFM. Details on the proposed implementation of the EM for penMCFM are given in Section \ref{sec:EMpenMCFM}, after recalling the standard EM for MCFM in Section \ref{sec:EMforMCFM}. For the sake of comparisons in the simulation studies, we have also adapted the GMIFS method from \cite{hastie2007stagewise} to obtain parameter estimations in penMCFM using the observed likelihood function in (\ref{l_observed}), and details on the implementation of this method are given in Section \ref{sec:GMIFS}.  

\subsection{EM Algorithm for the standard MCFM}\label{sec:EMforMCFM}

Recall that the cure indicator $Y$ is a latent boolean variable such that $Y=1$ if an individual is
susceptible (uncured), and $Y=0$ if non-susceptible (cured). It follows from the censoring assumption that
if $\delta _{i}=1$ then $y_{i}=1$, and if $\delta _{i}=0$ then $y_{i}$ is
not observable; in this case, $y_{i}$ is latent and can assume either values (one or zero). Hence, $\mathbf{y}=(y_{1},...,y_{n})$ is only partially observed, which means that we need an EM algorithm to carry out inference on the latent variables. 

The conditional survival function of model (\ref{S_pop_frailty}) for the $i$-th subject, given
the covariates and the latent value of the frailty $w_{i}$, is 
\begin{equation*}
S_{pop}(t_{i}|\mathbf{x}_{i}\mathbf{,z}_{i}\mathbf{,}w_{i})=1-\pi (\mathbf{z}%
_{i})+\pi (\mathbf{z}_{i})\exp \left( -w_{i}\alpha t_{i}^{\gamma }e^{\mathbf{x}_i^\top \bm{\beta} }\right) ,
\end{equation*}%
while the conditional survival and hazard functions of the susceptible
subjects, i.e. those such that $Y_{i}=1,$ are respectively
\begin{equation*}
S(t_{i}|\mathbf{x}_{i}\mathbf{,}w_{i},Y_{i}\mathbf{=}1)=\exp \left(
-w_{i}\alpha t_{i}^{\gamma} e^{\mathbf{x}_i^\top \bm{\beta}} \right),
\end{equation*}
and
\begin{equation*}
h(t_{i}|\mathbf{x,}w_{i},Y_{i}\mathbf{=}1)=w_{i}\alpha \gamma t_{i}^{\gamma
-1}e^{\mathbf{x}_i^\top \bm{\beta}}.
\end{equation*}

Note that, for the $i$-th subject, we have that $1-\pi (%
\mathbf{z}_{i})=P(y_{i}=0|\mathbf{z}_{i})$. Hence, given the frailty $w_{i}$, the contribution of the $i$-th subject to the likelihood
function is $1-\pi (\mathbf{z}_{i})$ when $Y_{i}=0$ and 
\begin{equation*}
\pi (\mathbf{z}_{i})S(t_{i}|\mathbf{x}_{i}\mathbf{,}w_{i},Y_{i}\mathbf{=}1)%
\left[ h(t_{i}|\mathbf{x_i},w_{i},Y_{i}\mathbf{=}1)\right] ^{\delta _{i}}
\end{equation*}%
when $Y_{i}=1$. Therefore, the complete-data likelihood function corresponding to \eqref{l_observed} can be expressed as 
\begin{equation*}
L_{c}(\alpha ,\gamma ,\theta, \bm{\beta}, \mathbf{b}|\mathbf{D,y,w}%
)=\prod_{i=1}^{n}\left[ 1-\pi (\mathbf{z}_{i})\right] ^{1-y_{i}}\pi (\mathbf{%
z}_{i})^{y_{i}} 
\left[ \left\{ h(t|\mathbf{x,}y_{i}\mathbf{=}1)\right\}
^{\delta _{i}}S(t_{i}|\mathbf{x}_{i}\mathbf{,}y_{i}\mathbf{=}1)\right]
^{y_{i}}f_{W}(w_{i}),
\end{equation*}
given the values of the latent outcome $Y_{i}=y_{i}$ and frailty $W_{i}=w_{i}$, for $i=1,...n,$ and where $%
f_{W}(w_{i})$ is the pdf of the gamma distribution. Hence, the corresponding complete-data
log-likelihood function $l_c(\bm{\Lambda}),$ where $\bm{\Lambda} =\left( \alpha ,\gamma
,\theta, \bm{\beta}, \mathbf{b}\right),$ can be written as the sum of the log-likelihood functions corresponding to the different model parts, and specifically
\begin{equation} \label{eq:loglik}
l_{c}(\bm{\Lambda}) = l_{c}(\alpha ,\gamma ,\theta, \bm{\beta}, \mathbf{b} | \mathbf{D,y,w}) = l_{c_{1}}(\mathbf{b}) + l_{c_{2}}(\alpha ,\gamma ,\bm{\beta})  + l_{c_{3}}(\theta),
\end{equation}
where 
\begin{equation*}
l_{c_{1}}(\mathbf{b)} = \sum_{i=1}^{n}\left[ (1-y_{i})\log \left( 1-\pi (%
\mathbf{z}_{i})\right) +y_{i}\log \pi (\mathbf{z}_{i})\right] ,
\end{equation*}%
\begin{equation*}
l_{c_{2}}(\alpha ,\gamma ,\bm{\beta}) = \sum_{i=1}^{n}\left[ \delta
_{i}\log \left( \alpha \gamma t_{i}^{\gamma -1}e^{\mathbf{x}_{i}^\top \bm{\beta} }\right) -w_{i}y_{i}\alpha t_{i}^{\gamma }e^{\mathbf{x}_{i}^\top \bm{\beta} }\right],
\end{equation*}%
\begin{equation*}
l_{c_{3}}(\theta \mathbf{)=}\sum_{i=1}^{n}\left[ \theta \log \theta -\log
\Gamma (\theta )-w_{i}\theta +(\delta _{i}+\theta -1)\log w_{i}\right] ,
\end{equation*}%
and $\delta _{i}y_{i}=\delta _{i}.$

In the E-step, we evaluate the conditional expectation of the complete-data
log-likelihood function with respect to the latent variables $y_{i}$'s and $%
w_{i}$'s given the parameter estimates at the $m$-th iteration of the algorithm $\bm{\Lambda}
^{(m)}=(\alpha ^{(m)},\gamma ^{(m)},\theta ^{(m)},\bm{\beta}^{(m)},%
\mathbf{b}^{(m)}),$ and given the observed data $\mathbf{D}$. We then need to compute
the conditional expectations of $y_{i},$ $w_{i}$, $\log w_{i}$ and $%
w_{i}y_{i}$ given $\left( \bm{\Lambda} ^{(m)},\mathbf{D}\right) $. These
expectations for the $m$-th iteration of the E-step are obtained as follows 

\begin{equation}
p_{i}^{(m)} \equiv E( y_{i}|\bm{\Lambda}^{(m)},\mathbf{D} ) = \delta _{i}+(1-\delta _{i})
 \left. \frac{\pi (\mathbf{z}_{i})\left( 1+\alpha
t_{i}^{\gamma }e^{\mathbf{x}_{i}^\top \bm{\beta} } /\theta \right) ^{-\theta }}{(1-\pi (%
\mathbf{z}_{i}))+\pi (\mathbf{z}_{i})\left( 1+\alpha t_{i}^{\gamma }e^{ \mathbf{x}_{i}^\top
\bm{\beta} } /\theta \right) ^{-\theta }} \right\vert_{\left(\bm{\Lambda}^{(m)},\mathbf{D}\right) },   \label{p_i}
\end{equation}

\begin{equation}
a_{i}^{(m)}  \equiv  E( w_{i}| \bm{\Lambda}^{(m)},\mathbf{D}) 
 = \frac{\delta _{i}+\theta }{\theta +\alpha t_{i}^{\gamma }e^{\mathbf{x}_{i}^\top \bm{\beta} }}p_{i}+ \left.\frac{\delta _{i}+\theta }{\theta }(1-p_{i})\right\vert _{\left(
\bm{\Lambda}^{(m)},\mathbf{D}\right) } ,  \label{a_i}
\end{equation}

\begin{equation}
 b_{i}^{(m)}  \equiv E\left( \log w_{i}| \bm{\Lambda}^{(m)},\mathbf{D}\right) = %
\left[ \varphi (\delta _{i}+\theta)-\log \theta \right] (1-p_{i}) 
 + \left. \left[ \varphi (\delta _{i}+\theta )-\log (\theta +\alpha t_{i}^{\gamma }e^{\mathbf{x}_{i}^\top \bm{\beta} })\right] p_{i} \right\vert _{\left( \bm{\Lambda}^{(m)},\mathbf{D} \right) },  \label{b_i}
\end{equation}

\begin{equation}
c_{i}^{(m)} \equiv E\left( w_{i}y_{i}|\bm{\Lambda} ^{(m)},\mathbf{D}\right) =\left. 
\frac{\delta _{i}+\theta }{\theta +\alpha t_{i}^{\gamma }e^{\mathbf{x}_{i}^\top \bm{\beta} }}p_{i}\right\vert _{\left( \bm{\Lambda}^{(m)},\mathbf{D}\right) },
\label{c_i}
\end{equation}
where $\varphi (.)$ is the digamma function. In the M-step, the conditional expectation of the complete-data log-likelihood function is maximized with respect to the unknown parameters. For details about the derivation of expectations, please refer to the relevant calculations presented in Peng and Zhang \cite{peng2008estimation}.

\subsection{EM Algorithm for penMCFM}\label{sec:EMpenMCFM}

When the covariate vectors $\mathbf{x}$ and $\mathbf{z}$ are high-dimensional, variable selection is also a purpose of the analysis, together with estimation of the unknown parameters and regression coefficients. To this aim, we consider a penalized version of the
complete-data likelihood function, where a penalty term is added for the coefficients  $\bm{\beta}_{p}$ and $\mathbf{b}_{p},$ corresponding to the penalized part of the high-dimensional covariates $\mathbf{x}$ and $\mathbf{z}$. Then, the penalized complete-data log-likelihood function takes the form
\begin{equation*}
 l_{c}^{pen}(\bm{\Lambda})  = l_{c_{1}}(\mathbf{b)} + 
l_{c_{2}}(\alpha ,\gamma ,\bm{\beta}) + l_{c_{3}}(\theta) 
 - n \sum_{j=1}^{P^p_1}P_{\bm{\lambda}_1 }(b_{p,j}) - n \sum_{l=1}^{P^p_2}P_{\bm{\lambda}_2}(\beta_{p,l}),
\end{equation*}%
where, from the notation introduced at the beginning of Section \ref{sec:EM}, $P^p_1$ and $P^p_2$ are the number of covariates included in the covariate vectors $\mathbf{z}_{p}$ and $\mathbf{x}_{p}$, respectively, and $P_{\bm{\lambda}_{i}}(\cdot)$ is a penalty function depending on a vector of penalty parameters $\bm{\lambda}_i,$ for $i=1,2$.

In the E-step, the conditional expectation of the penalized complete-data
log-likelihood function with respect to the latent variables $y_{i}$'s and $%
w_{i}$'s are the same as in equations \eqref{p_i}-\eqref{c_i}. Hence, $p_{i}^{(m)}, a_{i}^{(m)}, b_{i}^{(m)}$ and $c_{i}^{(m)}$ in \eqref{p_i}-\eqref{c_i} are used at the $m$-th iteration of the E-step in penMCFM. Then, the conditional expectation of the penalized complete-data
log-likelihood function at the $m$-th step of the algorithm is $El_{c}^{pen, (m)}(\bm{\Lambda}) = El_{c_{1}}^{(m)}(%
\mathbf{b})+El_{c_{2}}^{(m)}(\alpha ,\gamma ,\bm{\beta})+El_{c_{3}}^{(m)}(\theta),$ where 
\begin{equation}
El_{c_{1}}^{(m)}(\mathbf{b})  =\sum_{i=1}^{n}\left[ p_{i}^{(m)} \mathbf{z}_{i}^\top \mathbf{b} -%
\log \left( 1+e^{\mathbf{z}_{i}^\top \mathbf{b} }\right) \right] 
 - n\sum_{j=1}^{P^p_1}P_{\mathbf{\lambda}_1}(b_{p,j}),  \label{E_lc1}
\end{equation}

\begin{equation}
El^{(m)}_{c_{2}}(\alpha ,\gamma ,\bm{\beta} )  = \sum_{i=1}^{n}  \delta
_{i}\log \left( \alpha \gamma t_{i}^{\gamma -1}e^{\mathbf{x}_{i}^\top \bm{\beta} }\right)  
 -  \sum_{i=1}^{n} c_{i}^{(m)}\alpha t_{i}^{\gamma }e^{\mathbf{x}_{i}^\top \bm{\beta} } 
 - n\sum_{l=1}^{P^p_2}P_{\mathbf{\lambda}_2}(\beta_{p,l}),
\label{E_lc2}
\end{equation}

\begin{equation}
El^{(m)}_{c_{3}}(\theta)  = \sum_{i=1}^{n} \left[  (\delta _{i}+\theta -1) b_{i}^{(m)} -a_{i}^{(m)}\theta \right]  
 + n (\theta \log \theta -\log\Gamma (\theta) ).
\label{E_lc3}
\end{equation}

We focus on the multi-step adaptive elastic-net penalty described by Xiao
and Xu \cite{xiao2015multi}, which aims to achieve increased sparsity while concurrently reducing false positives in the variable selection process, and we adapt it to the case of penMCFM. When applying such
elastic-net penalization for the high-dimensional covariates corresponding to $\mathbf{b}_{p}$ and $\bm{\beta}_p$
in (\ref{E_lc1}) and (\ref{E_lc2}), the components of the penalized complete-data log-likelihood function at the $m$th iteration of the E-step become

\begin{equation*}
 El_{c_{1}}^{(m)}(\mathbf{b)}  = 
 \frac{1}{n}\sum_{i=1}^{n}\left[ p_{i}^{(m)}\mathbf{z}_{i}^\top \mathbf{b} %
- \log \left( 1+e^{\mathbf{z}_{i}^\top \mathbf{b} }\right) \right] - \lambda_{1,Enet} \left[ \frac{(1-\alpha _{Enet})}{2}\sum_{j=1}^{P^p_1}b_{p,_{j}}^{2}+%
\alpha _{Enet}\sum_{j=1}^{P^p_1}w_{j}\left\vert b_{p,j}\right\vert \right] ,
\end{equation*}%

\begin{equation*}
 El^{(m)}_{c_{2}}(\alpha ,\gamma ,\bm{\beta} ) = 
 \frac{1}{n} \sum_{i=1}^{n} \left[ \delta
_{i}\log \left( \alpha \gamma t_{i}^{\gamma -1}e^{\mathbf{x}_{i}^\top \bm{\beta}  }\right) -c_{i}^{(m)}\alpha t_{i}^{\gamma }e^{ \mathbf{x}_{i}^\top \bm{\beta}} \right]  
 - \lambda_{2,Enet}\left[ \frac{(1-\alpha _{Enet})}{2}\sum_{l=1}^{P^p_2}\beta_{p,_{l}}^{2}+%
\alpha _{Enet}\sum_{l=1}^{P^p_2}w_{l}\left\vert \beta_{p,l}\right\vert \right] ,
\end{equation*}%
where $\lambda_{1,Enet}$ and $\lambda_{2,Enet}$ are the tuning parameters controlling the amount of penalty, $\alpha _{Enet}\in %
\left[ 0,1\right] $ is a higher level hyperparameter, and $w_{j}$ is the
data-driven weighting parameter corresponding to the $j-$th variable in the model (referred to as ``adaptive weights'' in the following). The tuning parameter values can be identified through cross-validation, and choices related to the tuning of these parameters are discussed in the following section. We update the adaptive weights $k$ times as 
\begin{equation*}
w_{j}^{(k)}=\frac{1}{\left\vert \widehat{b}_{p,j}^{(k-1)}\right\vert },\text{}j=1,...,P_1, 
\end{equation*}%
and 
\begin{equation*}w_{l}^{(k)}=\frac{1}{\left\vert \widehat{\beta}_{p,l}^{(k-1)}\right\vert },\text{}l=1,...,P_2,
\end{equation*}%
for $\mathbf{b}_p$ and $\bm{\beta}_p$.
When $k=1$, the adaptive weights equal $1$ for all variables, and this case is equivalent to
the regular elastic-net method. When $k=2,$ then the adaptive elastic-net method is used. The estimates of $\mathbf{b}_{p}$ and $\bm{\beta}_p$ at the $m$th iteration of the E-step are obtained by minimizing the negative penalized complete-data log-likelihood functions, that is respectively:

\begin{equation}
 -El_{c_{1}}^{(m)}(\mathbf{b)} = 
 -\frac{1}{n}\sum_{i=1}^{n}\left[ p_{i}^{(m)}\mathbf{z}_{i}^\top \mathbf{b} %
- \log \left( 1+e^{\mathbf{z}_{i}^\top \mathbf{b} }\right) \right] +\lambda_{1,Enet}  
\left[ \frac{(1-\alpha_{Enet})}{2}%
\sum_{j=1}^{P^p_1}b_{p,_{j}}^{2}+\alpha_{Enet}\sum_{j=1}^{P^p_1}w_{j}^{(k)}\left\vert b_{p,j}\right\vert \right] ,
\label{E_lc1_R}
\end{equation}

\begin{align}
 -El^{(m)}_{c_{2}}(\alpha ,\gamma ,\bm{\beta} ) &= 
 -\frac{1}{n} \sum_{i=1}^{n}\left[ \delta
_{i}\log \left( \alpha \gamma t_{i}^{\gamma -1}e^{\mathbf{x}_{i}^\top \bm{\beta}  }\right) -c_{i}^{(m)}\alpha t_{i}^{\gamma }e^{\mathbf{x}_{i}^\top \bm{\beta} }\right] \notag \\
& + \lambda_{2,Enet} 
 \left[ \frac{(1-\alpha _{Enet})}{2}\sum_{l=1}^{P^p_2}\beta_{p,_{l}}^{2}+%
\alpha _{Enet}\sum_{l=1}^{P^p_2}w_{l}^{(k)}\left\vert \beta_{p,l}\right\vert \right] .
\label{E_lc2_R}
\end{align}%

At the $m$th iteration of the M-step, the maximization of the $El_{c}^{pen, (m)}(\bm{\Lambda})$ is
equivalent to maximizing the $El^{(m)}_{c_{1}}(\mathbf{b)}$, $El^{(m)}_{c_{2}}(\alpha
,\gamma ,\bm{\beta)}$ and $El^{(m)}_{c_{3}}(\theta)$ separately to
obtain updated parameter estimates $\bm{\Lambda}^{(m+1)}$.

\subsection{Implementation of the penMCFM Algorithm}\label{sec:EMpenMCFMdetails}

At the $m$th iteration of the E-step, $El_{c_{1}}^{(m)}(\mathbf{b)}$ in (\ref{E_lc1}) has the same
structure as the penalized log-likelihood function of a logistic regression model, except for the parameters $p_{i}^{(m)}$. As evident from the definition in (\ref{p_i}), $p_{i}^{(m)}$ is not constrained to binary values, but can take any value in the interval $\left[ 0,1\right]$. Notably, it is important to note that these response values $p_{i}^{(m)}$ are not directly observable within this framework, in contrast to the typical situation in penalized logistic regression problems. Consequently, optimizing $El_{c_{1}}^{(m)}(\mathbf{b)}$ poses more challenges compared to solving a conventional penalization problem. This challenge has been recently highlighted by Bussy et al. \cite{bussy2019c} and Fu et al. \cite{fu2022controlled}, and implies the impossibility to directly employ the \texttt{glmnet} R package \cite{glmnet} to address the high-dimensional component of our problem.
The second term, $El_{c_{2}}^{(m)}(\alpha ,\gamma ,\bm{\beta)}$ in (\ref{E_lc2}), can be considered as the log-likelihood function of the penalized Weibull regression model with $h(t_{i})=\alpha \gamma t_{i}^{\gamma -1}e^{\mathbf{x}_{i}^\top \bm{\beta} +\log c_{i}^{(m)}}$ and $S(t_{i})=\exp \left( -\alpha t_{i}^{\gamma }e^{%
\mathbf{x}_{i}^\top \bm{\beta} +\log c_{i}^{(m)}}\right).$ Here $c_{i}^{(m)}$ is a constant that is evaluated by using (\ref{c_i}) before the optimization steps of the algorithm. We employ the \texttt{lbfgs} R package \cite{lbfgs}, which provides the functionalities to utilize $L_1$ penalties for the minimization of the cost functions outlined in (\ref{E_lc1_R}) and (\ref{E_lc2_R}). 
Finally the third term, $El_{c_{3}}^{(m)}(\theta \mathbf{)}$ in (\ref{E_lc3}), is maximized to obtain the estimate of the frailty distribution parameter $\theta $ (or its reciprocal, i.e., the variance).

We give a pseudo-code description of our EM algorithm for penMCFM in Algorithm \ref{EMalgorithm_1}. Initialization is carried out as follows: we use Cox regression estimates for $\bm{\beta}_{u}$, moment estimates for the Weibull distribution parameters $\alpha$ and $\gamma$, and we set $\bm{\beta}_p = \mathbf{b}_0=\mathbf{b}_u=\mathbf{b}_p=0$ and $\theta=1$. 
Generally, it is assumed that the tuning parameters for both penalties are common, i.e. $\bm \lambda_{1,Enet}= \bm \lambda_{2,Enet}=\bm \lambda_{Enet}$ for the sake of simplicity. We have followed this approach in our simulations as well. We then generate a sequence of $\bm \lambda_{Enet}$ based on the penalized Cox regression of $\mathbf{t},\: \bm{\delta},\: \mathbf{x}_u, \: \mathbf{x}_p$ using the \texttt{glmnet} R package \cite{glmnet}. 

 \begin{algorithm}
 
\caption{EM algorithm for penMCFM} \label{EMalgorithm_1}
\KwData { $\bm \delta,\: \mathbf{t, \: x_u, \: x_p, \: z_u, \: z_p}$}
\textbf{Initialization:} $\bm{\beta}_{u}=\bm{\widehat{\beta}_{Cox}},\: \bm{\beta}_{p}=0,\: \mathbf{b}=0,\: \alpha=\widehat{\alpha}_{ME},\\ \gamma=\widehat{\gamma}_{ME},\: \theta=1$\

\textbf{Determine:} $\bm \lambda_{Enet}$ based on $\mathbf{t}, \: \bm \delta,\: \mathbf{x_u, \: x_p}$ 

\KwIn {$\bm \lambda_{Enet},\: \alpha_{Enet},\: \alpha,\: \gamma,\: \theta,\: \bm{\beta_u},\: \bm{\beta_p},\: \mathbf{b},\: \epsilon$ }
\KwOut { $\widehat{\alpha},\: \widehat{\gamma},\: \widehat{\theta},\: \widehat{\bm{\beta}},\: \widehat{\mathbf{b}} $ }

\For{$i \gets 1$  \KwTo $length(\bm \lambda_{Enet})$} {

compute the adaptive weights $w_{j}^{(k)}$ and corresponding estimates of $(\alpha, \gamma, \theta, \bm{\beta}, \mathbf{b})$ given $\lambda_{Enet}^{i}\:, \alpha_{Enet}$ \;

\Repeat{ \scalebox{0.8}{$\left|El_{c_{1}}^{(m)}(\mathbf{b}^{(m)})- El_{c_{1}}^{(m-1)}(\mathbf{b}^{(m-1)})\right|<\epsilon$}, \\
\scalebox{0.8}{$\left| El_{c_{2}}^{(m)}(\alpha^{(m)} ,\gamma^{(m)} ,\bm{\beta}^{(m)}) - El_{c_{2}}^{(m-1)}(\alpha^{(m-1)} ,\gamma^{(m-1)} ,\bm{\beta}^{(m-1)}) \right|< \epsilon$},\\
\scalebox{0.8}{$\left|El_{c_{3}}^{(m)}(\theta^{(m)})- El_{c_{3}}^{(m-1)}(\theta^{(m-1)})\right|<\epsilon$ } } {

$m\gets ++$

\textbf{E-step:}

compute $p_{i}^{(m)}$, $a_{i}^{(m)}$, $b_{i}^{(m)}$ and $c_{i}^{(m)}$ from (\ref{p_i})-(\ref{c_i})\;

\textbf{M-step:}

update $\mathbf{b_p}^{(m)}$ minimizing (\ref{E_lc1_R}) 

update $\bm{\beta_p}^{(m)}$ minimizing (\ref{E_lc2_R})  

update $b_{0}^{(m)}$, $\mathbf{b_u}^{(m)}$ minimizing (\ref{E_lc1_R}) given $\mathbf{b_p}^{(m)}$\;

update $\alpha^{(m)}$, $\gamma^{(m)}$ and $\bm{\beta_u}^{(m)}$ minimizing (\ref{E_lc2_R}) given $c_i^{(m)}$, $\bm{\beta_p}^{(m)}$

update $\bm{\theta}^{(m)}$ minimizing $-El^{(m)}_{c_{3}}(\theta)$ in  (\ref{E_lc3}) given $a_i^{(m)}$, $b_i^{(m)}$
} }
 \end{algorithm}

\subsection{GMIFS method for penMCFM}\label{sec:GMIFS}

The GMIFS method is a variable selection technique that iteratively enhances the predictive power of a model by progressively adding variables, thus it can be adapted to obtain parameter estimations in penMCFM. Unlike traditional forward selection methods, GMIFS enforces a monotonic constraint on the coefficients of the added variables, ensuring that they either remain constant or increase with each iteration. By sequentially introducing variables with controlled increases in their coefficients, GMIFS strikes a balance between model sparsity and predictive accuracy. This method was first introduced by Hastie et al. \cite{hastie2007stagewise} for solving penalized least squares regression problems. Since then, Makowski and Archer \cite{makowski2015generalized}  and Hou and Archer \cite{hou2015regularization} provided extensions to Poisson regression and ordinal responses in the high dimensional setting, Yu et al. \cite{yu2018nonlinear} to  variable selection for nonlinear regression, and Fu et al. \cite{fu2022controlled} to the Weibull MCM. 

In the simulation studies, with the aim of comparing our EM-based method for penMCFM not only with the penalized Weibull MC model introduced in \cite{fu2022controlled}, but also with another variable selection algorithm, 
the GMIFS method is adapted so to be able to target our observed likelihood function in (\ref{l_observed}). For a comprehensive introduction to this method, readers are encouraged to refer to Hastie et al. \cite{hastie2007stagewise} and Fu et al. \cite{fu2022controlled}, and for further details on the implementation, to check our code details at \url{https://github.com/fatihki/penMCFM}.

\section{Simulation Study}\label{sec:Simulation}

In this section, we describe the results of extensive simulation studies that we have conducted to evaluate the performance of our proposed model. The evaluation of model performance involves assessing both the accuracy of the model predictions, and its capability to correctly identify covariates with true non-zero coefficients. We examine different scenarios involving varying censoring rates, cure rates, correlation among covariates, and number of nonzero coefficients. 

For the sake of performance comparisons in the simulation studies, besides our proposed penMCFM EM-based algorithm (detailed in Sections \ref{sec:EMpenMCFM}-\ref{sec:EMpenMCFMdetails}), which we refer to as penMCFM(EM), we also implemented a version of the penMCFM algorithm where we incorporate the GMIFS method (detailed in Section \ref{sec:GMIFS}), named penMCFM(GMIFS), and the GMIFS-based penalized Weibull MC model from Fu et al. \cite{fu2022controlled}, named MCM(GMIFS). We are also interested in exploring the performance of a survival model that does not account for a cure fraction. To this aim, the Cox PH model with lasso penalty is employed by using the \texttt{glmnet} \cite{glmnet} R package based on the observed data $(\bm \delta,\mathbf{t}, \mathbf{x_u}, \mathbf{x_p})$, where the tuning parameter $\lambda$ that gives the most regularized model such that the cross-validated error is within one standard error of the minimum is chosen. This $\lambda$ value is referred to as ``lambda.1se'' in \cite{glmnet}, and thus we refer to this method as penCox.1se. Note that this method only considers the latency part of our model, namely the one related to the $\bm{\beta}_p$ regression coefficients.  

The R code for implementing the proposed estimation algorithms and for replicating the simulation studies is available at \url{https://github.com/fatihki/penMCFM}.

\subsection{Data generation under the Weibull MCFM}

In order to generate random samples from a Weibull MCFM with right-censoring, cure rate and covariates, we follow the procedure outlined in Asano et al. \cite{asano2014assessing}. This method can be regarded as a modification of the inverse sampling technique, particularly suited for situations where the data generating process is assumed to have an improper survival function. From the population survival function $S_{pop}(t|\mathbf{x,z})$ in (\ref{S_pop_frailty}) we can derive the inverse function for the cdf of the population as
$$F_{pop}^{-1}(u|\mathbf{x,z}) = \left\{ \theta \alpha^{-1} e^{\mathbf{- \mathbf{x}^\top \bm{\beta}  }} \left( \left[ 1- u / \pi (\mathbf{z}) \right] ^{-1/\theta }-1\right)\right\} ^{1/\gamma }$$
where $0\leq u\leq (1-\pi (\mathbf{z}))$. The pseudo-code description of a random sample generation for the Weibull MCFM model is detailed in Algorithm \ref{data_gen_algorithm}.

\begin{algorithm}
\caption{Random sample generation from Weibull MCFM}\label{data_gen_algorithm}
\KwIn{$n,\: \alpha,\: \gamma,\: \theta,\: \bm{\beta},\: \mathbf{b,\: x,\: z},\: \lambda _{c}$}
\KwOut { $\mathbf{t}, \bm{\delta}$ }
\For{$i \gets 1$  \KwTo $n$} {

generate $u_i \sim U\left( 0,1\right)$

compute $\pi (\mathbf{z}_i)=e^{ \mathbf{z}_{i}^\top \mathbf{b} }/(1+e^{\mathbf{z}_{i}^\top \mathbf{b} })$ 

generate random survival time $t_{i}^{\ast }$:

\eIf {$u_{i}<\pi (\mathbf{z}_{i})$}{
        $t_{i}^{\ast }=F_{pop}^{-1}(u_{i}|\mathbf{x}_{i}\mathbf{,z}_{i})$     
        }{
        $t_{i}^{\ast }=\infty$
    }
generate censoring time: $c_i \sim Exp(\lambda _{c})$

determine survival time and censoring indicator: \\ 
$t_{i}=\min (c_{i},t_{i}^{\ast })$ and $\delta_{i}=I(t_{i}^{\ast }\leq c_{i})$
    }
\end{algorithm}

\subsection{Simulation Design}\label{sec:sim-design}

We carry out a large-scale simulation study to test the novel high-dimensional setup of our model. Our data generating process can be described as follows: we set the number of observations to $n=500$, and the penalized covariates sizes as $P^p_{1}=P^p_{2}=P=1000$. The sparsity levels of $\bm{\beta}_p$ and $\mathbf{b}_p$ are set to be $s=20,$ meaning that $s$ is the number of non-zero coefficients (thus, 2\% sparsity in this setting). We randomly split the dataset into a $80 \%$ training and a $20 \% $ testing  set to assess performance, and implement a $4$-fold cross-validation on the training set to tune the model parameters. The parameter $\lambda _{Enet}$ is tuned over a grid $\bm \lambda_{Enet} $ of $50$ values on a log scale in the interval $\left[\lambda _{Enet}^{\min }, \lambda_{Enet}^{\max }\right],$  using the same sequence of $\bm \lambda$ values as used for penalized Cox regression in the \texttt{glmnet} R package. The $\alpha _{Enet}$ parameter is chosen among the values $0.1,\: 0.5,\: 0.9$ and $1$.

A categorical variable with three levels is used as unique unpenalized covariate $\mathbf{Z}_{u}\in\mathbb{R}^{n \times P^u_1}$ with $P^u_1=1$, generated with weight probabilities $0.4,\: 0.35,\: 0.25$. The unpenalized covariates associated to the survival part of the model $\mathbf{X}_{u} \in \mathbb{R}^{n \times P^u_2}$ are i.i.d. standard normal, and $P^u_2=10$. The penalized covariates $\mathbf{Z}_p$ and $\mathbf{X}_p$ are assumed common, and generated from a $P$-dimensional Gaussian distribution $MVN(0,\mathbf{\Sigma }),$ where $\mathbf{\Sigma}\in \mathbb{R}^{P \times P}$ is a block diagonal matrix with block size $50$, and with the correlation between any pair of covariates within the same block defined by $corr(\mathbf{Z}_{p,i},\mathbf{Z}_{p,j})=corr(\mathbf{X}_{p,i},\mathbf{X}_{p,j})=\rho^{\left\vert i-j\right\vert }$, $\forall i,j=1,\ldots, P$; the tested values for $\rho$ are $0,\: 0.2,\: 0.5$.
The unpenalized regression coefficients are set as $(b_0, \mathbf{b}_u)= (-2,-1,1),$ while $\bm{\beta}_u \in \mathbb{R}^{n \times P^u_2}$ is randomly generated from a uniform distribution on the interval $[-3,3]$. For what concerns the penalized regression coefficients $\bm{\beta}_p,\: \mathbf{b}_p \in \mathbb{R}^{P},$ we consider five different settings with respect to their nonzero coefficients. In each setting, $s$ components of both $\bm{\beta}_p$ and $\mathbf{b}_p$ are nonzero and take the same value $v,$ which we will refer to as the ``signal''. The signals are occurring at roughly equally-spaced indices between $1$ and $P$, while the rest of the coefficients is equal to $0$. 
Hence, the non-zero coefficients are distributed equally in each block, and specifically one non-zero coefficient per block. Five different signal values are considered, specifically $\mathbf{v} = (0.5,\: 1,\: 1.5,\: 2,\: 2.5)$, to test the model when the signal in the data becomes smaller and smaller.

Finally, we set the Weibull distribution parameters as $(\alpha, \gamma)=(1.25, 2.5)$, the frailty distribution parameter $\theta=0.5,$ and $\lambda_c = 0.5$ for generating the censoring times. After generating all model coefficients and covariates according to these assumptions, we generate the data using Algorithm \ref{data_gen_algorithm}. This process is repeated $100$ times, and the average results across all runs of the EM on the $100$ datasets so obtained are subsequently reported. Notably, the average censoring (cure) rates in our simulated samples are approximately $ 85 (76) \%,\: 76 (70)\%,\: 70 (62) \%,\: 66 (59) \%,\: 63 (57) \%$ for the respective $v$ values chosen in $\mathbf{v}.$

\subsection{Performance evaluation}\label{sec:sim-measures}

To compare the performances of the various models considered in our simulation studies, we consider several performance metrics. For selecting the tuning parameter $\lambda_{Enet}$, we use $4$-fold cross-validation on a revised version of the \emph{concordance index} defined in (\ref{C_stat_cure}) below, and also used in \cite{fu2022controlled}.
Harrell's concordance index (or C-statistic, or C-index) is widely used as a
measure of performance in fitted survival models for censored data. The C-statistic for a
standard survival model is the proportion of concordant pairs out of the
total number of possible evaluation pairs, given by
$$
\widehat{C} =
\frac{\sum_{i=1}^{n}\sum_{j=1,i\neq j}^{n}I (\mathbf{x}_{i}^\top \widehat{\bm{\beta}} >\mathbf{x}_{j}^\top \widehat{\bm{\beta}} )
I_{i,j} }{ \sum_{i=1}^{n}\sum_{j=1,i\neq j}^{n}I_{i,j} },
$$ 
where $I_{i,j}=I[ t_{i}<t_{j},\delta_{i}=1 ]+I[ t_{i}=t_{j}, \delta_{i}=1, \delta_{j}=0 ]$, and $\widehat{\bm{\beta}}$ is the estimated vector of coefficients. Assano and Hirakawa \cite{asano2017assessing} noted that the value of $%
\widehat{\bm{\beta}}\mathbf{x}_{i}$ is the same for both cured patients
and censored uncured patients, and therefore proposed a modified C-statistics that considers this issue by introducing a cure status weighting for each patient. The weight is defined as $1$ for the uncured
patients who experienced the event, $0$ for the cured patients, and is equal to the estimate
of the uncured probability $\widehat{\pi}(\mathbf{z})$ for the censored patients.
Then, the C-statistic is defined by Assano and Hirakawa \cite{asano2017assessing} as
\begin{equation}
 \widehat{C}_{Cure}  = 
 \frac{\sum_{i=1}^{n}\sum_{j=1,i\neq j}^{n}I (\mathbf{x}_{i}^\top \widehat{\bm{\beta}} >\mathbf{x}_{j}^\top \widehat{\bm{\beta}} )%
\left\{ \vartheta_{j}y_{j}+(1-\vartheta_{j})\widehat{\pi }(\mathbf{z}%
_{j})\right\} I_{i,j}}{\sum_{i=1}^{n}\sum_{j=1,i\neq j}^{n}\left\{
\vartheta_{j}y_{j}+(1-\vartheta_{j})\widehat{\pi }(\mathbf{z}_{j})\right\} I_{i,j}},
\label{C_stat_cure}
\end{equation}
where $\vartheta_{i}$ is an indicator function taking values $\vartheta_{i}=1$ if $y_{i}=0$ or $1$, and $%
\vartheta_{i}=0 $ if $y_{i}$ is missing, and where $\widehat{\pi }(\mathbf{z}_{j})=e^{%
\mathbf{z}_{j}^\top \widehat{\mathbf{b}} } /(1+e^{ \mathbf{z}_{j}^\top \widehat{\mathbf{b}} } )$. Notice that $\widehat{C}_{Cure}$ in (\ref{C_stat_cure}) reduces to $\widehat{C}$ when $v_{i}=1$ and $y_{i}=1$ for all patients.

Moreover, we evaluate the performance of the variable selection part of the model by calculating the associated \emph{sensitivity}, i.e., the percentage of non-zero coefficients accurately estimated as non-zeros, \emph{specificity}, i.e., the percentage of zero
coefficients accurately estimated as zero, and \emph{False Positive Rate (FPR)}, i.e., the proportion of wrongly selected zero coefficients among those estimated as non-zero coefficients, which is equal to the proportion of mistakenly selected irrelevant variables among those identified as significant. We also evaluate the model prediction performance by considering the \emph{Relative Model Error (RME)}, and the model estimation performance by considering the \emph{estimation ERRor (ERR)}. For the vector of regression coefficients $\bm{\beta}$, these measures are defined as

\begin{eqnarray}
  \text{Sensitivity} & = & \sum_{l=1}^{P}I(\beta_{l}\neq 0\cap \widehat{\beta}_{l}\neq 0) / \sum_{l=1}^{P}I(\beta_{l}\neq 0) \nonumber \\
  \text{Specificity} & = & \sum_{l=1}^{P}I(\beta_{l}\neq 0\cap \widehat{\beta}_{l}=0) / \sum_{l=1}^{P}I(\beta_{l}=0) \nonumber \\
  \text{FPR} & = & \sum_{l=1}^{P}I(\beta_{l} = 0\cap \widehat{\beta}_{l}\neq 0)/\sum_{l=1}^{P}I(\beta_{l} = 0) \nonumber \\
  \text{RME} & = & (\widehat{\bm{\beta}}-\bm{\beta})^{T}\bm{\Sigma }(\widehat{\bm{\beta}}-\bm{\beta}) / (\widehat{\bm{\beta}}^{\ast }-\bm{\beta})^{T}%
\bm{\Sigma}(\widehat{\bm{\beta}}^{\ast }-\bm{\beta})  \nonumber \\
  \text{ERR} & = & (\widehat{\bm{\beta}}-\bm{\beta})^{T}(\widehat{\bm{\beta}}-\bm{\beta}) / (\widehat{\bm{\beta}}^{\ast }-\bm{\beta})^{T}(\widehat{\bm{\beta}}^{\ast }-\bm{\beta}) \nonumber 
\end{eqnarray}
where $\bm{\Sigma}$ is the covariance matrix of the covariates, $\bm{\beta}$ is the true vector of coefficients, and $\widehat{\bm{\beta}}^{\ast} $ is the oracle estimate of the coefficients derived from the model where only the true signals were included and the coefficients for the other covariates were forced to be zero.

Furthermore, according to the model, the probability of being cured is sample-specific: this means that, even when two patients share identical clinical characteristics, their unique genomic profiles can influence the cure probability in different ways. Recall that the uncured probability for the $i$th subject is computed by using the formula $\pi (%
\mathbf{z}_{i})=e^{\mathbf{z}_{i}^\top \mathbf{b} } /(1+e^{ \mathbf{z}_{i}^\top \mathbf{b} } )$, for given $%
\mathbf{b}$ coefficients and for the $i$th subject covariates $\mathbf{z}_{i}$, $i=1,...,n$.
Therefore, the \emph{average true uncured probability} can be computed as $\sum_{k=1}^{M}\left[
\sum_{i=1}^{n}\pi^{k}(\mathbf{z}_{i})/n\right] /M$ over $M$ Monte
Carlo runs of the EM algorithm, and over the $n$ subjects in the sample, where $\pi^{k}(\mathbf{z}%
_{i})$ is the true uncured probability for the $i$th subject in the $k$th Monte
Carlo simulation, with $M$ being the total number of simulated datasets. Consequently, we can also evaluate the accuracy of model estimates in terms of
bias and mean squared error (MSE) of the estimated uncured probability $%
\widehat{\pi }(\mathbf{z})$, which are calculated as 
\begin{equation*}
Bias\left( \widehat{\pi }(\mathbf{z})\right) =\frac{1}{M}\sum_{k=1}^{M}\left[
\frac{1}{n}\sum_{i=1}^{n}\left\{ \widehat{\pi }^{k}(\mathbf{z}_{i})-\pi
^{k}(\mathbf{z}_{i})\right\} \right],
\end{equation*}%
and
\begin{equation*}
MSE\left( \widehat{\pi }(\mathbf{z})\right) =\frac{1}{M}\sum_{k=1}^{M}\left[ 
\frac{1}{n}\sum_{i=1}^{n}\left\{ \widehat{\pi }^{k}(\mathbf{z}_{i})-\pi
^{k}(\mathbf{z}_{i})\right\} ^{2}\right],
\end{equation*}%
where $\pi ^{k}(\mathbf{z}_{i})$ and $\widehat{\pi }^{k}(\mathbf{z}_{i})$
are respectively the true uncured probability and its estimate for the $i$-th subject and
the $k$-th Monte Carlo run, similarly to what reported in Pal et al. \cite{pal2023new}.

\subsection{Simulation Results}

\begin{figure}[H]
\centering
\includegraphics[width=0.95\textwidth]{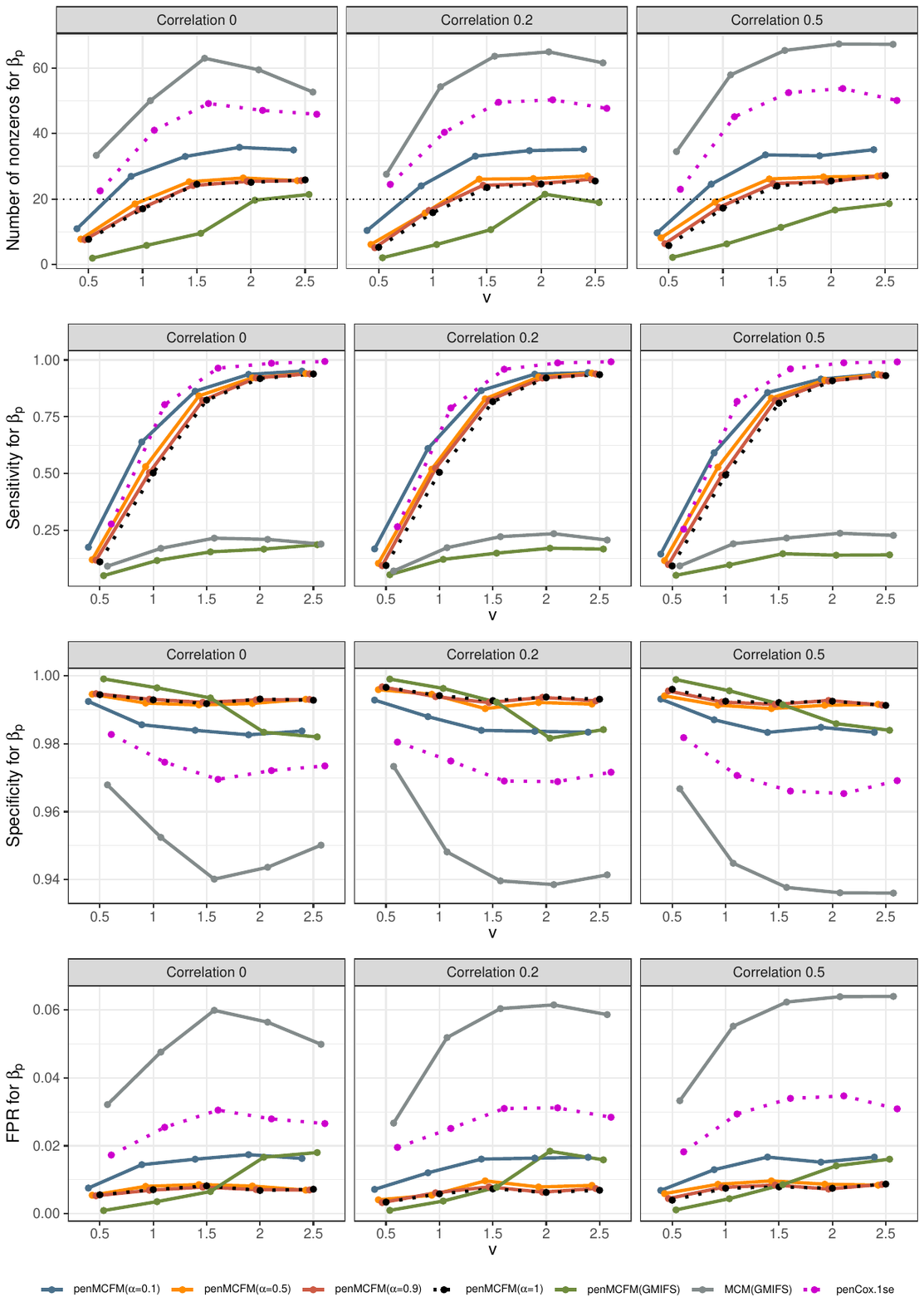}
\caption{Results of simulation studies. Variable selection performance in inference for $\bm{\beta}_{p}$.}
\label{betap_performance_plot}
\end{figure}

\begin{figure}[H]
\centering
\includegraphics[width=0.95\textwidth]{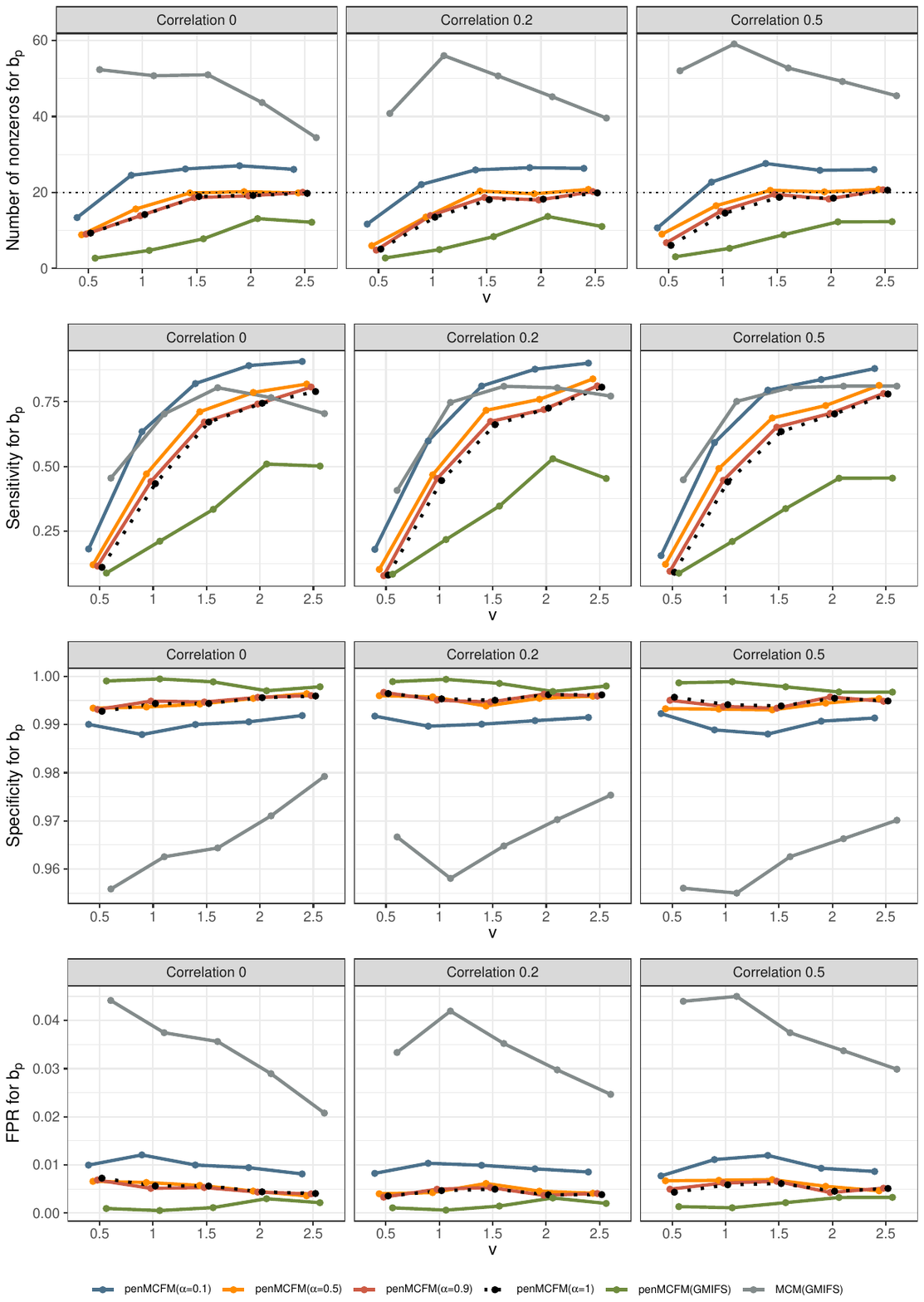}
\caption{Results of simulation studies. Variable selection performance in inference for $\textbf{b}_{p}$.}
\label{bp_performance_plot}
\end{figure}

\begin{figure}[H]
\centering
\includegraphics[width=0.915\textwidth]{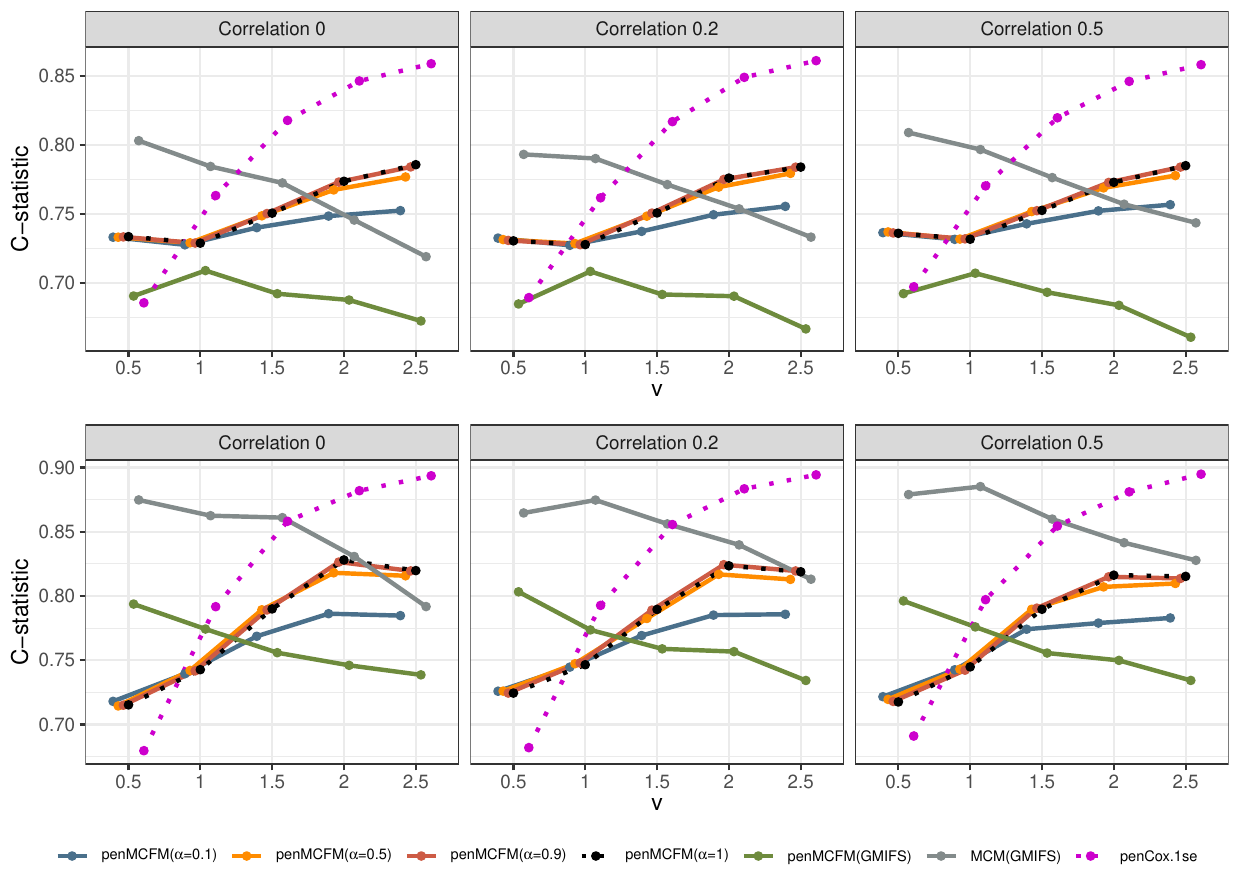}
\caption{Results of simulation studies. C-statistics plot based on the train (top row) and test (bottom row) data.}
\label{Cstat_train_test_plot}
\end{figure}

\begin{figure}[H]
\centering
\includegraphics[width=0.915\textwidth]{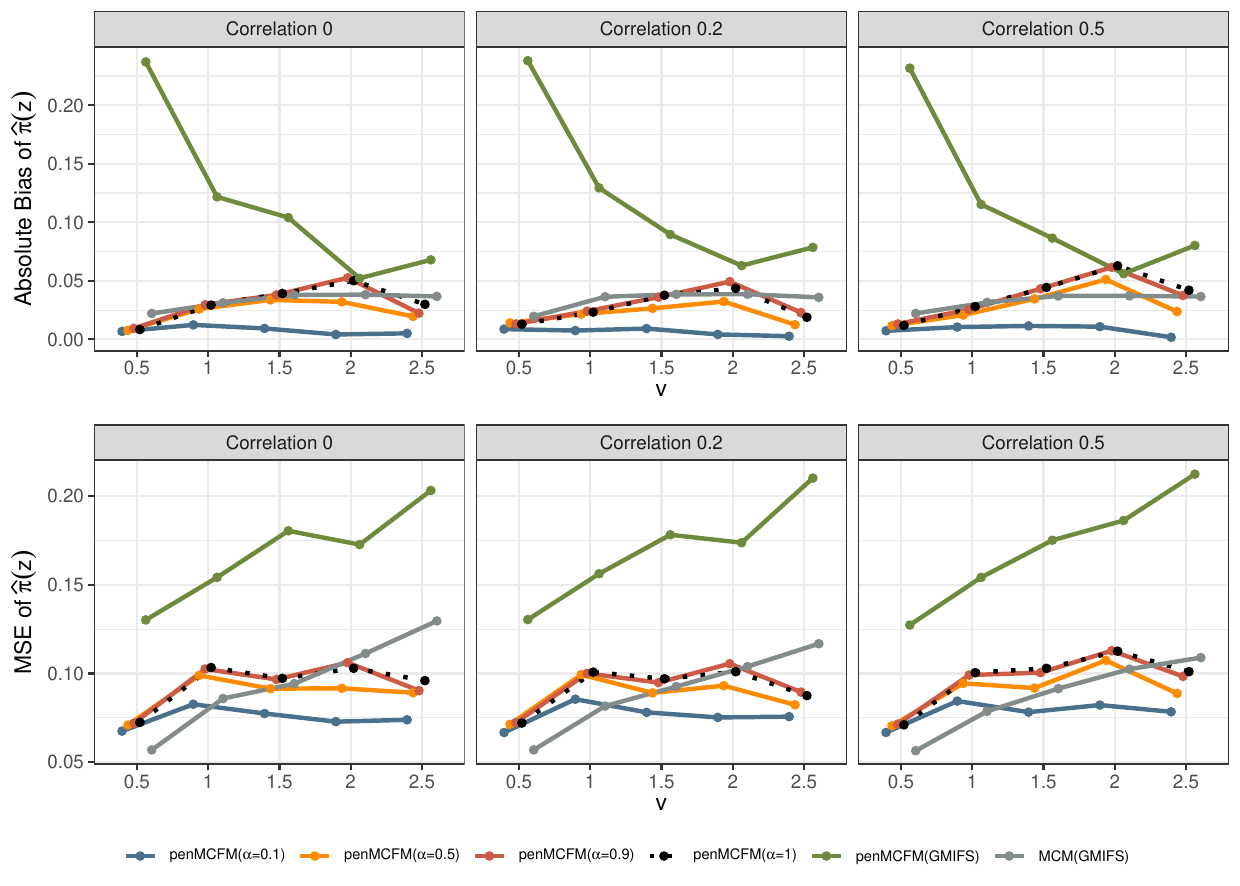}
\caption{Results of simulation studies. Absolute bias and MSE plots for the uncured rate estimates.}
\label{uncured_performance_plot}
\end{figure}

We here report the simulation results for each scenario specified in the simulation design described in Section \ref{sec:sim-design}, over $M=100$ simulated datasets, when using the different performance measures as introduced in Section \ref{sec:sim-measures}. We plot the number of non-zero regression coefficients, and the sensitivity, specificity, and FPR, to show the performance of the variable selection for both sets of regression coefficients $\bm{\beta}_p$ and $\textbf{b}_p$ in Figures \ref{betap_performance_plot} and \ref{bp_performance_plot}, respectively. The C-statistics plots for both training and testing data are shown in Figure \ref{Cstat_train_test_plot}. Here the regular C-statistic, namely $\widehat{C}$, is used for the penCox.1se method (which does not include a cured part), while $\widehat{C}_{Cure}$ is used for all other methods. The performance of the uncured rate estimates is also shown in Figure \ref{uncured_performance_plot}, where the absolute Bias and MSE are reported. Finally, we present results that illustrate the prediction and estimation performance of the models, as well as the cure rates, by reporting the RME, ERR, $Bias\left( \widehat{\pi }(\mathbf{z})\right)$ and $MSE\left( \widehat{\pi }(\mathbf{z})\right) $ in Table \ref{table:Correlation_0}, for the case in which the data are generated with $\rho = 0$. Corresponding tables for the cases of $\rho = 0.2$ and $0.5$ are included in Tables S$1$ and S$2$ of the Supplementary Material.
 
Each panel in the figures shows the average of a given metric over $M$ repetitions for the respective methods under consideration. In the table, mean results of the different methods are listed along with the standard deviation (in parenthesis). It is worth noting that the GMIFS method is independent of the $\alpha_{Enet}$ parameter, while penCox.1se only includes the lasso case for the penalized Cox regression, and thus results associated to these two methods remain unchanged regardless of the $\alpha_{Enet}$ values.

To assess the accuracy of variable selection, it is vital to consider sensitivity, specificity, and the FPR plots in combination. Figures \ref{betap_performance_plot} and \ref{bp_performance_plot} reveal that the proposed penMCFM(EM) method exhibits very similar average results when choosing $\alpha_{Enet}$ equal to either values $ 0.5,\: 0.9,\: 1$ across all metrics, and moreover, these figures also show that penMCFM(EM) is the method providing best overall results across all simulated scenarios and considered metrics. 
Notably, when looking specifically at Figure \ref{betap_performance_plot}, the penMCFM(GMIFS) method consistently selects the fewest variables, often below the true value of $20$, while conversely the MCM model version, MCM(GMIFS), chooses the largest number of variables compared to other methods. These behaviors obviously reflect in the often suboptimal sensitivity, specificity and FPR values attained by these two methods. The penCox.1se, which utilizes the penalized Cox regression with a lasso penalty, ranks as the second method with regards to the number of selected variables in Figure \ref{betap_performance_plot}, thus showing large sensitivity but suboptimal specificity and FPR. 
If we compare the variable selection performance of our proposed penMCFM(EM) method when choosing $\alpha_{Enet} = 1$ and penCox.1se, they show similar sensitivity plots although penMCFM(EM) selects less variables than penCox.1se. Moreover, penMCFM(EM) attains better performance across $\alpha_{Enet}$ values than penCox.1se in terms of specificity and FPR across all $\mathbf{v}$ and $\rho$ settings. 
The MCM (GMIFS) method demonstrates better sensitivity performance for $\mathbf{b}_p$ (Figure \ref{bp_performance_plot}) as compared to $\bm{\beta}_p$ (Figure \ref{betap_performance_plot}).
However, similarly to what shown for $\bm{\beta}_p$ in Figure \ref{betap_performance_plot}, penMCFM(EM) consistently outperforms other methods across all $\alpha_{Enet}$ values also for estimating $\mathbf{b}_p$ (Figure \ref{bp_performance_plot}), as 
it selects the fewest variables and frequently approaches the true value of $20$ more closely than other methods, while maintaining large sensitivity and specificity, and exhibiting low FPR. 

The overall performance of the methods in terms of variable selection generally deteriorates when the signal becomes weaker, specifically when $v = 0.5$, and it improves as the signal magnitude increases, as observed in both figures.  The effect of the signal parameter $v$ is apparent in the increasing variable selection performance particularly in the number of selected variables and sensitivity. On the other hand, while the other methods seem to have slightly deteriorating performance when the correlation among variables is increasing, the impact of the correlation parameter $\rho$ on the performance of penMCFM(EM) cannot be prominently discerned.
We therefore conclude that the penMCFM(EM) method, especially when choosing $\alpha_{Enet} = 0.5,\: 0.9,$ or 1, attains better variable selection performance than other methods in terms of the true number of selected variables, high sensitivity and specificity, and low FPR. 

From inspection of Figure \ref{Cstat_train_test_plot}, we note that performance in terms of C-statistics is generally slightly better on the testing set than on the training set for all methods, as also previously observed in a similar model setting by Fu et al. \cite{fu2022controlled}. As the signal magnitude increases, the penMCFM(EM) and penCox.1se methods as expected tend to attain larger C-statistics values for both the train and test data. penCox.1se shows generally the best performance except for the cases $v= 0.5$ and $1$, when performance deteriorates more than for other methods. penMCFM(EM) generally shows larger C-statistic values when choosing $\alpha_{Enet} = 0.5,\: 0.9,\: 1 $  as compared to $\alpha_{Enet} = 0.1$, and with respect to penMCFM(GMIFS) (except for $v= 0.5,\: 1$ on the test data). 
Even if the C-statistics values attained by MCM(GMIFS) tend to be larger than those attained by penMCFM(EM), the performance of the two methods becomes more and more similar as the signal magnitude increases, both on the training and on the testing datasets. It is also worth noting that, when calculating the C-statistics for a cure model, $\widehat{C}_{Cure}$ in (\ref{C_stat_cure}), both coefficient vectors $\bm{\beta}_p$ and $\textbf{b}_p$ are considered in the calculations, thus making it plausible that this contributes to the larger values attained by MCM(GMIFS) as compared to  penMCFM(EM).
Nonetheless, the MCM(GMIFS) method consistently exhibits the highest count of nonzero variables for both $\bm \beta_p$ and $\textbf{b}_p$ in all scenarios (Figures \ref{betap_performance_plot} and \ref{bp_performance_plot}, respectively), thus showing a worse variable selection performance.
To summarize, we can conclude that generally all the methods show acceptable and comparable C-statistics values for almost all considered scenarios. We thus give more importance to the evaluation of the variable selection performance, which discriminates more among the considered methods, and which impacts in a much more crucial way on the reliable interpretation of the model outputs.

Finally, Figure \ref{uncured_performance_plot} shows that penMCFM(EM) when choosing $\alpha_{Enet} = 1$ and $0.5$ generally gives the lowest first and second average absolute bias and MSE values of $\widehat{\pi }(\mathbf{z})$, respectively. penMCFM(GMIFS) exhibits the least favorable performance among all methods. Additionally, it is worth noting that the penMCFM(EM) method showcases consistent behavior across varying correlation sizes for each $\alpha_{Enet}$ value.

From Table \ref{table:Correlation_0}, it can be observed that our proposed method penMCFM(EM) outperforms other methods in terms of the RME and ERR metrics, except for the most difficult case when $v = 0.5$. In almost all scenarios, penMCFM(EM) with $\alpha_{Enet} = 0.1 $ shows the best prediction and estimation performance for the regression coefficients in the cure part, and therefore we observe similar performance for the estimation of $\pi(\mathbf{z})$.

\section{Application to RNA-Seq data from TCGA-BRCA}\label{sec:Application}

In this section we utilize publicly available omics data from The Cancer Genome Atlas (TCGA) project\footnote{See: \url{http://cancergenome.nih.gov}} for showcasing the use of the proposed penMCFM model on a real case application. The overall survival time, demographic and gene expression data from primary invasive BReast CAncer patients in the TCGA database (TCGA-BRCA) were obtained from the Genomic Data Commons Data Portal data release v33.0. We retrieve the RNA-Seq data from the primary tumor of TCGA-BRCA patients, together with the accompanying metadata comprising survival outcomes, as well as clinical and demographic variables. We also use the BCR Biotab files to gather some additional clinical variables related to hormone status, such as estrogen receptor (ER), human epidermal growth receptor 2 (HER2) and progesterone receptor (PR), for the same TCGA-BRCA patients.

The original data set comprises $1111$ samples with $60660$ gene expression features, and $87$ clinical and demographic variables. We specifically consider protein-coding genes, and after eliminating duplicate genes we are left with $19938$ RNA-seq features. We then remove observations which show a survival time of less than one month, leaving $1017$ samples before arrangement of the clinical variables. We use years as unit for the survival time variable in this analysis. We apply the DESeq2 normalization approach for RNA-seq data as implemented in the R/Bioconductor package \texttt{DESeq2} \cite{love2014moderated} before further statistical analysis, as suggested by Zhao et al. \cite{zhao2021tpm, zhao2024tutorial}. 

For data preprocessing, we reduce the $19938$ RNA-seq features using the variance filter method implemented in the \texttt{M3C} R package \cite{john2020m3c}, thus reducing the features considered to the top $2000$ RNA-seq, which explain around $41 \% $ of the variation of the whole data.
Since prior knowledge of BRCA plays a crucial role in the identification of potential prognostic biomarkers, we consolidate genes of interest from five distinct sources of prior knowledge as in Li and Liu \cite{li2021detecting, li2023biomarker}. This selection includes: 
\begin{itemize}
    \item the $147$ genes from the KEGG breast cancer pathway \cite{kanehisa2021kegg},
    \item the $519$ genes related to BRCA from the top ranked gene ontology (GO) terms sorted in gene ontology annotations (GOA) \cite{gene2021gene},
    \item the $70$ genes known as the ``MammaPrint'' BRCA signatures \cite{cardoso201670},
    \item the $128$ genes collected by the online consensus survival analysis web server for breast cancers (OSbrca) previously published BRCA biomarkers \cite{yan2019osbrca},
    \item the $10$ BRCA prognosis signatures selected by the scPrognosis method using single-cell RNA sequencing (scRNA-seq) data \cite{li2020novel}.
\end{itemize}
When we take the intersection of these sources with our already reduced set of $19938$ RNA-seq features, and combine them with the top $2000$ features, we obtain a total of $2650$ RNA-seq features. Lastly, we filter the samples based on the clinical variables listed in Table \ref{table:TCGACovariates}, resulting in a dataset of $828$ observations and around $89\%$ censoring rate. The remaining portion of the RNA-seq data, which includes $189$ observations and has around $70\%$ censoring rate, will be used for validating the identified genes. Even if the censoring rate is slightly lower on this validation set (as compared to the training and testing sets), filtering on clinical variables makes this validation dataset at most homogeneous to the training and testing sets with respect to covariates, which is most crucial in calculating the Prognostic Risk Score.

We consider two real data scenarios for the choice of the penalized covariates $\mathbf{X}_{p}$ and $\mathbf{Z}_{p}$, which are the same for the cure and the survival part: Scenario $1$, where the penalized covariates only include the top $2000$ RNA-seq features after the variance filtering, and Scenario $2$, which is built upon the larger set of $2650$ RNA-seq features, combining the same top $2000$ RNA-seq features as in Scenario $1$ with the known genes related to BRCA. In both scenarios, we employ the same set of unpenalized covariates $\mathbf{X}_{u}$ and $\mathbf{Z}_{u}$. Unpenalized covariates in $\mathbf{X}_{u}$ include the following clinical variables: \textit{ER, PR, HER2, PT, RT, Age}, while $\mathbf{Z}_{u}$ includes \textit{PAM50, MS, LN, TS, PS, ER, PR, HER2} (as in De Bin et al. \cite{de2014investigating}, Volkmann et al. \cite{volkmann2019plea} and  Li and Liu \cite{li2021detecting}).

\begin{table}[H]
	\centering
 \begin{threeparttable}
	\renewcommand{\arraystretch}{0.90}
 \caption{Simulation studies results for $\mathbf{b}_p,\: \bm{\beta}_p$ and $\pi(\mathbf{z})$ when $\rho=0$}
 \label{table:Correlation_0}
	\begin{tabular}{ccccccrc} \hline
		\toprule
		\multicolumn{2}{c}{} & \multicolumn{2}{c@{\quad}}{$\bm{\beta}_p$} 
		& \multicolumn{2}{c@{\quad}}{$\mathbf{b}_p$} 
		& \multicolumn{2}{c@{\quad}}{$\pi(\mathbf{z})$} \\ \cmidrule(r){3-4} \cmidrule(r){5-6} \cmidrule(l){7-8}
		    $v$ & Method & RME(SD) &ERR (SD) & RME(SD) &ERR (SD) &  Bias & MSE  \\  \hline
		 0.5 & A1 & 1.248 (0.268) & 1.248 (0.268) &0.491 (0.938) & 0.491  (0.938) &  \textbf{0.007} & 0.067\\ 	
          & A2 & 1.560 (1.246) & 1.560 (1.246) & 0.515 (0.977) & 0.515 (0.977) & 0.008 & 0.071 \\ 
          & A3 & 1.747 (1.746) & 1.747 (1.746) & 0.553 (1.211) & 0.553 (1.211) & 0.009 & 0.072 \\ 
          & A4 & 1.821 (1.912) & 1.821 (1.912) & 0.559 (1.227) & 0.559 (1.227) & 0.008 & 0.072 \\ 
		  & B & 1.193 (0.056) & 1.193 (0.056) & 0.496 (0.936) & 0.496 (0.936) & 0.237 & 0.130 \\
		  & C & 1.502 (0.322) & 1.502 (0.322) & \textbf{0.461} (0.881) & \textbf{0.461} (0.881) & -0.022 & \textbf{0.057} \\ 
		  & D & \textbf{1.134} (0.282) & \textbf{1.134} (0.282)  & & & & \\
        &  & & & & & & \\
		 1 & A1 & \textbf{0.801} (0.113) & \textbf{0.801} (0.113) & \textbf{1.365} (2.165) & \textbf{1.365}  (2.165) & \textbf{0.012} & \textbf{0.083} \\
        & A2 & 0.807 (0.149) & 0.807 (0.149) & 1.525 (2.578) & 1.525 (2.578) & 0.026 & 0.099 \\
        & A3 & 0.809 (0.153) & 0.809 (0.153) & 1.536 (2.604) & 1.536 (2.604) & 0.030 & 0.103 \\ 
        & A4 & 0.811 (0.153) & 0.811 (0.153) & 1.527 (2.563) & 1.527 (2.563) & 0.029 & 0.103 \\ 
		  & B& 1.056 (0.028)& 1.056 (0.028) & 1.809 (2.816) & 1.809 (2.816) & 0.122 & 0.154\\
		  & C & 1.075 (0.045) & 1.075 (0.045) & 1.404 (2.192) &1.404 (2.192) & -0.031 & 0.086\\
		  & D & 0.851 (0.085) & 0.851 (0.085) & & & & \\
        &  & & & & & & \\
		 1.5 & A1 & 0.729 (0.089) & 0.729 (0.089) & \textbf{2.855} (5.976) & \textbf{2.855} (5.976)& \textbf{0.009} & \textbf{0.077}\\
         & A2 & 0.644 (0.125)& 0.644 (0.125) & 2.953 (6.416) & 2.953 (6.416) & 0.034 & 0.091 \\
         & A3 & 0.635 (0.140) & 0.635 (0.140)   & 2.985 (6.421) & 2.985 (6.421) & 0.038 & 0.097 \\
         & A4 & \textbf{0.631} (0.137) & \textbf{0.631} (0.137)  & 2.989 (6.465) & 2.989 (6.465) & 0.039 & 0.097 \\
		  & B & 1.021 (0.019) & 1.021 (0.019) & 3.873 (8.061) & 3.873 (8.061) & 0.104 & 0.180\\
		  & C & 1.019 (0.026)& 1.019 (0.026) & 3.177 (6.822) & 3.177 (6.822)& -0.038 & 0.094\\
		  & D & 0.807 (0.048) & 0.807 (0.048)  & & & & \\
       &  & & & & & & \\
		 2 &A1 & 0.748  (0.056)& 0.748 (0.056) & \textbf{3.267} (7.496) & \textbf{3.267} (7.496)& \textbf{0.004} & \textbf{0.073} \\ 
       & A2 & 0.669 (0.093) & 0.669 (0.093) & 3.290 (7.573) & 3.290 (7.573) & 0.032 & 0.092 \\
       & A3 & 0.638 (0.096) & 0.638 (0.096) & 3.430 (8.095) & 3.430 (8.095) & 0.053 & 0.106 \\ 
        & A4 & \textbf{0.636} (0.096) & \textbf{0.636} (0.096) & 3.431 (8.154) & 3.430 (8.154) & 0.050 & 0.103 \\ 
		  & B & 1.012  (0.014)& 1.012 (0.014) & 4.033 (8.926)& 4.033 (8.926) & 0.052 & 0.173 \\ 
		  & C & 1.013 (0.017)& 1.013 (0.017) & 3.645 (8.309) & 3.645 (8.309) & -0.038 & 0.111 \\
		  & D & 0.837 (0.037) & 0.837 (0.037) & & & & \\
      &  & & & & & & \\
		 2.5 & A1 & 0.804 (0.047) & 0.804 (0.047) & 4.657 (11.521) & 4.657 (11.521) & \textbf{-0.005} & \textbf{0.074} \\ 
        & A2 & 0.740 (0.069) & 0.740 (0.069) & 4.603 (10.956) & 4.603 (10.956) & 0.020 & 0.089 \\
        & A3 & 0.718 (0.077) & 0.718 (0.077) & \textbf{4.458} (10.339) & \textbf{4.458} (10.339) & 0.022 & 0.090 \\
        & A4 & \textbf{0.715} (0.080) & \textbf{0.715} (0.080) & 4.632 (11.091) & 4.632 (11.091) & 0.030 & 0.096 \\ 
		  & B & 1.012 (0.017)& 1.012 (0.017) & 5.560 (13.218) & 5.560  (13.218) & 0.068 & 0.203 \\
		  & C & 1.012 (0.012)& 1.012 (0.012)  & 5.202 (12.640) &5.202 (12.640) & -0.037 & 0.130 \\ 
		  & D & 0.872 (0.024) & 0.872 (0.024) & & & & \\ \hline
		
  	\end{tabular}
   \begin{tablenotes}
     \item \footnotesize Method A1-A4: penMCFM (EM) for $\alpha_{Enet}=0.1,0.5,0.9,1$, B: penMCFM (GMIFS), C: MCM (GMIFS), \\D: penCox.1se; The best result appears in bold.
    \end{tablenotes}
 \end{threeparttable}
\end{table}

\subsection{Identification of biomarkers}

To identify the relevant features (biomarkers), we initially partition the dataset randomly into a $80 \%$ training and a $20 \% $ testing set. In order to select the tuning parameter $\lambda_{Enet}$, we employ a $4$-fold cross-validation on the training dataset to obtain the optimal $\lambda_{Enet}$ that maximizes the C-statistic ($\widehat{C}$ for the penCox.1se method and $\widehat{C}_{Cure}$ for all other methods). We additionally tune the $\alpha_{Enet}$ parameter, by considering values in the set $\{0.1, 0.3, 0.5, 0.7, 0.9, 1\}$, as the one maximizing the C-statistic value in the testing dataset when employing the $\lambda_{Enet}$ parameter selected from the training dataset. We then repeat this random splitting (and subsequent parameter tuning) $20$ times to get several new training and testing datasets, as a part of our experiments. It is crucial to emphasize the importance of such approach to ensure feature selection robustness, as the selected features are consolidated as the union of the genes associated with non-zero coefficients in all $20$ experiments.

\begin{figure}[H]
     \centering    
     \begin{subfigure}[b]{0.95\textwidth}
         \centering
         \includegraphics[width=\textwidth]{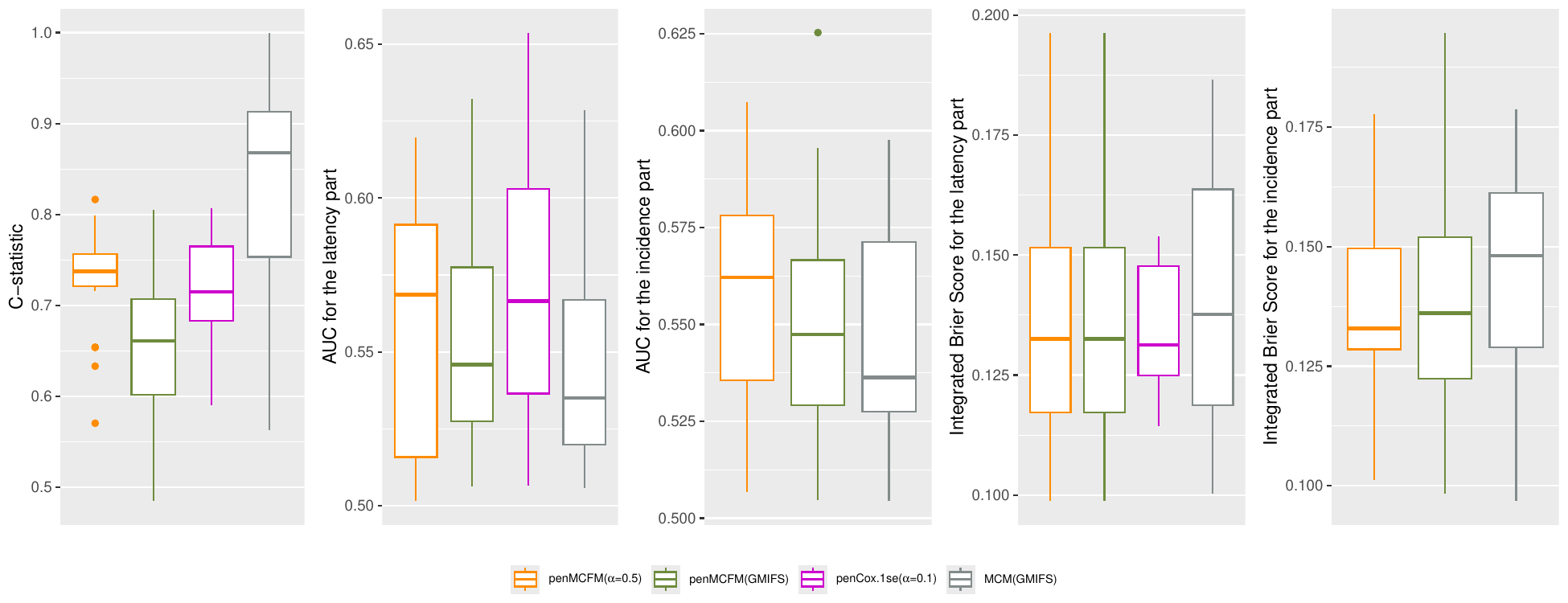}
         \caption{Scenario $1$}
         \label{Cstat_test_plot_Scenario_1}
     \end{subfigure}
     \hfill
     \begin{subfigure}[b]{0.95\textwidth}
         \centering
         \includegraphics[width=\textwidth]{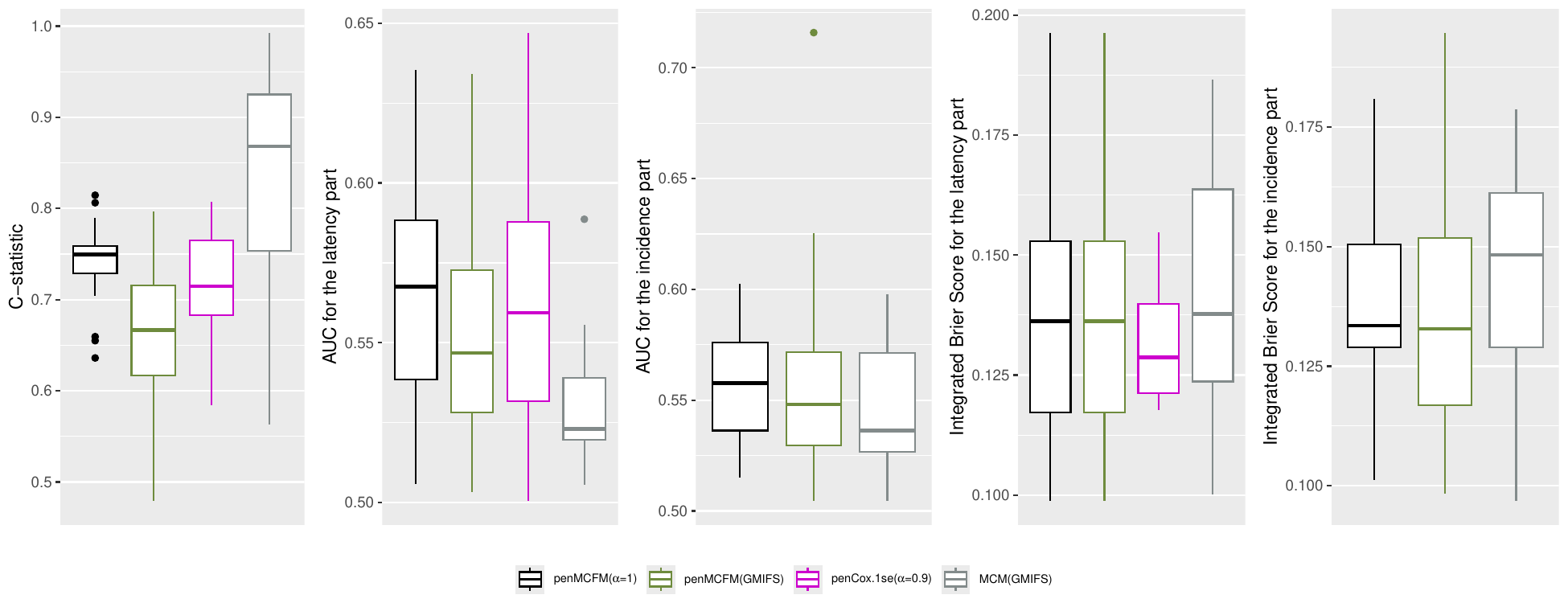}
         \caption{Scenario $2$}
         \label{Cstat_test_plot_Scenario_2}
     \end{subfigure}
     \caption{Results of the analysis of the TCGA-BRCA data. Boxplots of the C-statistic, AUC and IBS values obtained on the testing datasets over the $20$ repeated data-splitting processes}
     \label{Cstat_test_plot}
 \end{figure}

\begin{figure}[!h]
     \centering    
     \begin{subfigure}[b]{0.49\textwidth}
         \centering
         \includegraphics[width=\textwidth]{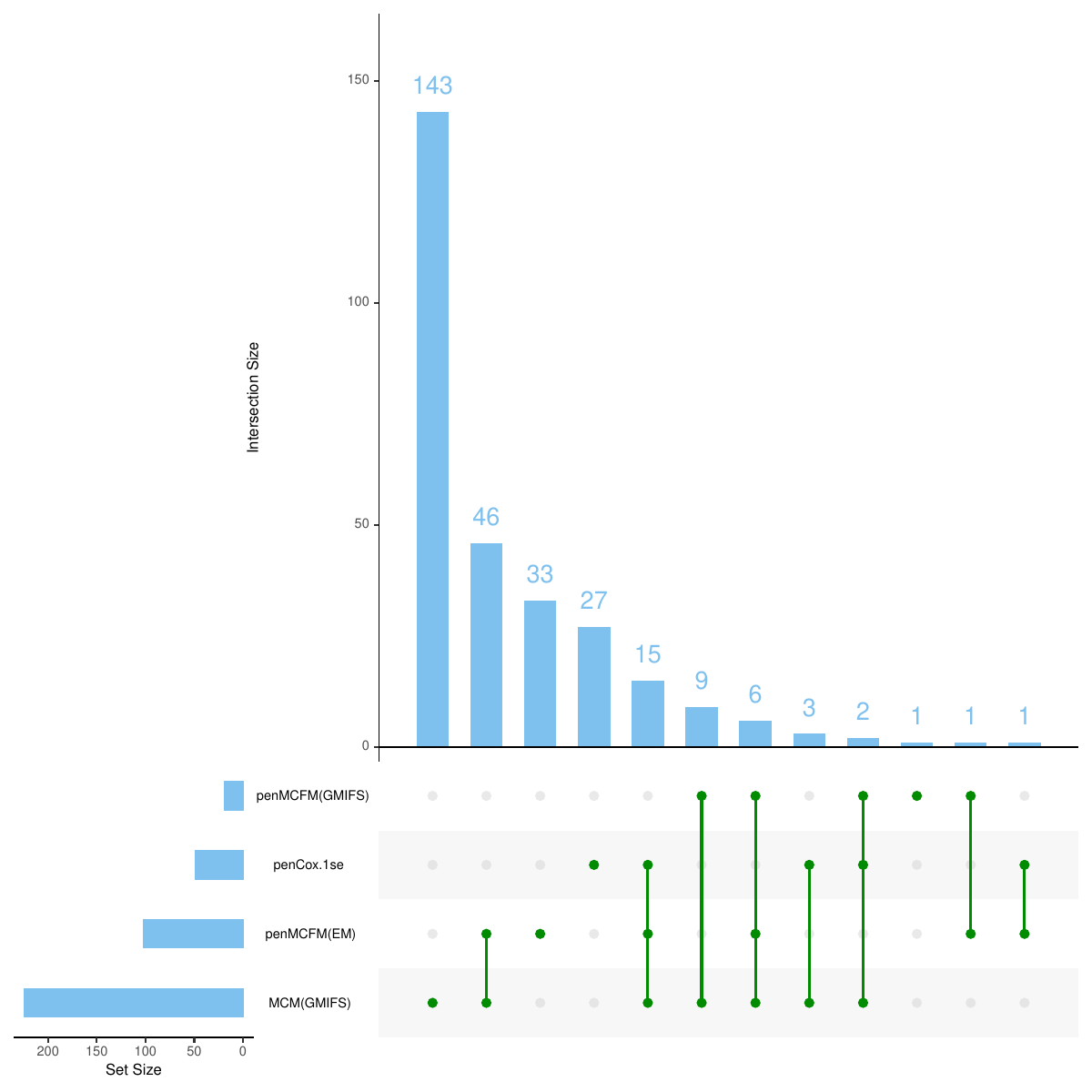}
         \caption{Scenario $1$}
         \label{upset_plot_Scenario_1}
     \end{subfigure}
     \hfill
     \begin{subfigure}[b]{0.49\textwidth}
         \centering
         \includegraphics[width=\textwidth]{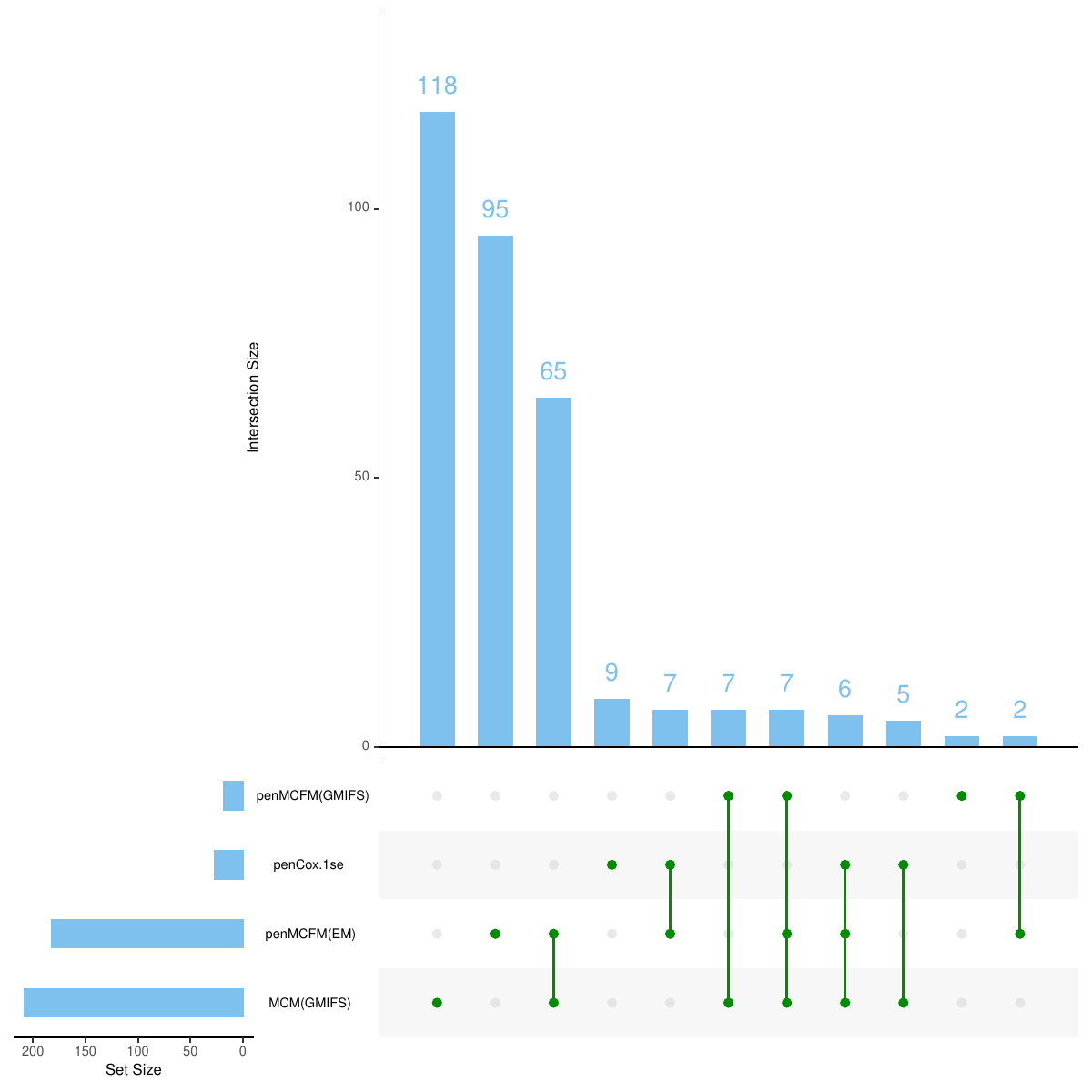}
         \caption{Scenario $2$}
         \label{upset_plot_Scenario_2}
     \end{subfigure}
     \caption{Results of the analysis of the TCGA-BRCA data. Overlap of the selected gene sets (nonzero $\bm{\beta}_p$ coefficients) among the four methods: the blue barplots report the frequencies of intersections among the methods, while the bottom green lines report which methods are considered for the overlap}
     \label{upset_plots_betap}
 \end{figure}

We employ our methods in both Scenarios 1 and 2 as detailed above. The $\alpha_{Enet}$ tuning parameter is determined based on the average values of the C-statistic over the $20$ repeats for the penMCFM(EM) and penCox.1se methods, the only ones allowing this parameter. The results obtained using the optimally selected $\alpha_{Enet}$ parameters can be seen in Figure \ref{Cstat_test_plot}, which shows the boxplots of the C-statistic for the different methods. The number of identified biomarker genes is different for the different methods: for instance, in Scenario $2$, the penMCFM(EM) method with $\alpha_{Enet} = 1$ selects $182$ genes for the latency part of the model, genes which correspondingly show nonzero values in $\bm{\beta}_p$, while MCM(GMIFS) selects $208$ genes. These identified sets of biomarker genes show great overlap across different methods, as shown in Figure \ref{upset_plots_betap}, which illustrates the overlap of the selected genes for the latency part of the model across the four methods. Interestingly, penMCFM(EM) generally does not select any features for the incidence part of the model, meaning that no coefficient in $\textbf{b}_p$ is estimated to be significantly different from zero, while the GMIFS method (when used with both models penMCFM and MCM) usually selects some features. A similar result was also observed in Fu et al. \cite{fu2022controlled} in an application to acute myeloid leukemia data. A summary of the overlap of the selected genes of these two methods for the incidence part of the model is also given in Figure S$1$ of the Supplementary Material. Furthermore, the selected sets of biomarker genes for the different methods, along with their frequencies of occurrence obtained over the $20$ repeats of each method, are provided at \url{https://github.com/fatihki/penMCFM} as excel files.

From inspection of Figure \ref{Cstat_test_plot} one can see that the proposed method penMCFM(EM) achieves the second-largest C-statistic values, whereas MCM(GMIFS) attains the largest C-statistic. It is worth noting however that the latter method selects a larger number of genes for both parts of the model compared to other methods, which may contribute to its greater C-statistic values. Moreover, it is also possible to consider other metrics, such as the area under the receiver operating characteristic (ROC) curve and the integrated Brier score, for comparing the different methods, for both the incidence and latency parts of the model. To calculate the area under the ROC curve (AUC), each method is used to estimate $5$-years survival for the subjects in the testing set, and compared to outcome on the corresponding time interval. The integrated Brier score (IBS) is a composite measure of discrimination and calibration, and lower values indicate better model calibration and predictive accuracy.
The procedures outlined by Zhao et al. \cite{zhao2024tutorial} were followed to compute both the AUC and IBS metrics. Our analysis leads to the conclusion that the penMCFM(EM) method, as depicted in Figures \ref{Cstat_test_plot_Scenario_1} and \ref{Cstat_test_plot_Scenario_2}, exhibits superior AUC and IBS performance compared to the alternative methods for both the incidence and latency components of the model.

For a fair comparison of the methods performance, we carry out more investigations for the identified biomarker genes via gene set enrichment analyses, and by constructing a prognostic risk score, as described in Sections \ref{sec:functionalEA} and \ref{sec:RiskScore}, respectively. Furthermore, we validate a subset of the top-identified genes derived from penMCFM(EM) when considering the intersection of results from both scenarios, by referring to the existing BRCA literature. This additional validation can be found in Table S$3$ of the Supplementary Material.

\begin{table}[H]
\renewcommand{\arraystretch}{1.23}
   \centering 
    \caption{Demographic and clinical characteristics of TCGA-BRCA patients}
    \label{table:TCGACovariates}
\begin{threeparttable}[b]
\begin{tabular}{cccccccc} \hline
   \textbf{Clinical outcomes} &  mean & range &   event rate & \\  \hline
     Overall survival (year) & 3.19 & $[0.08, 18.05]$   & 0.109 & & & \\ 
     & & &  & & \\
    \textbf{Clinical covariates} & \textbf{Levels}  & $n$ &  \textbf{Clinical covariates} & \textbf{Levels}  & $n$ &  \\  \hline
    PAM50 Subtypes & Basal &  144  & Pathological Stage (PS) &  Stage I & 151 \\
                     & Her2 & 58 &               & Stage II  & 478 \\
                     & LumA & 432 &              & Stage III  & 180\\
                     & LumB & 164 &              & Stage IV  & 14 \\
                     & Normal & 30 &              & Stage X  & 5 \\
   Lymph nodes (LN)& N0 & 402 &        Tumor size  (TS) & T1 & 221 \\
                    & N1 & 273&                 & T2 & 481 \\
                    & M2 & 87 &                 & T3 & 101 \\
                    & N3 & 56 &                  & T4 & 24 \\
                    & NX & 10 &                  & TX & 1 \\
   Metastases stage (MS) & M0 & 706  &   HER2 status & Equivocal & 172 \\
                    & M1 & 14 &                  & Positive & 143 \\ 
                    & MX & 108 &                  & Negative & 513 \\
  Pharmaceutical Therapy (PT)& Yes & 669 &  ER status & Positive & 648 \\
                         & No & 159 &              & Negative & 180 \\
 Radiation Therapy  (RT)  & Yes & 451 &     PR status &  Positive & 567 \\
                      & No & 377 &               & Negative & 261 \\
  
          Age (year)\tnote{1}  & 58.79  & $[26.6,90]$ &  &   &   & \\ \hline
\end{tabular}
   \begin{tablenotes}
     \item [1] mean and range
    \end{tablenotes}
\end{threeparttable}
\end{table}

\subsection{Functional enrichment analysis}\label{sec:functionalEA}

We perform both GO- and the Kyoto encyclopedia of genes and genomes (KEGG)- Enrichment Analyses (EA) on the union of the selected genes obtained in both scenarios to investigate the pathological implications of these identified biomarker genes. To this aim, we utilize the R/Bioconductor package \texttt{clusterProfiler} \cite{yu2012clusterprofiler, wu2021clusterprofiler} for all the analyses described in this section. First, we perform GO-EA on the genes identified from penMCFM(EM), penMCFM(GMIFS), MCM(GMIFS) and penCox.1se in Scenarios $1$ and $2$, and then we conduct the KEGG pathway EA (KEGG-EA). We only present here the results obtained when studying the identified genes from the latency part of the model in penMCFM(EM), while the results obtained with other methods (namely penMCFM(GMIFS), MCM(GMIFS), and penCox.1se) are provided in the Supplementary Materials. Moreover, the detailed results of enrichment analyses for each method are given in \url{https://github.com/fatihki/penMCFM} as excel files.

\begin{figure}[H]
     \centering
     \begin{subfigure}[b]{0.48\textwidth}
         \centering
         \includegraphics[width=\textwidth]{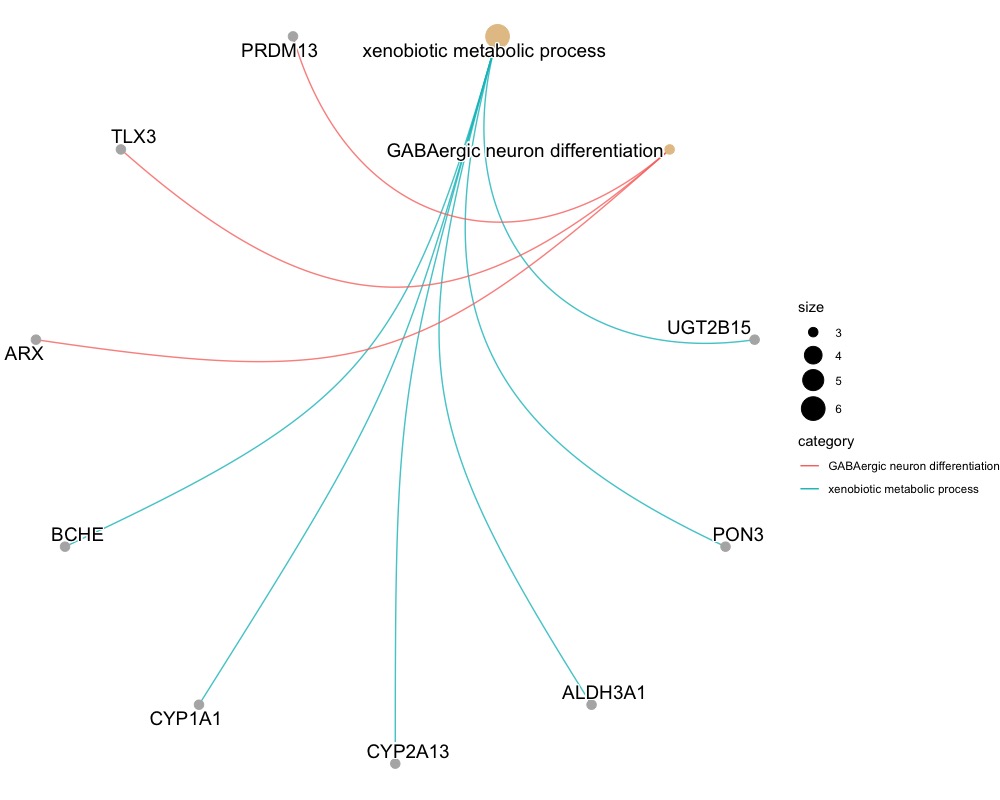}
         \caption{}
         \label{go_Scenario1}
     \end{subfigure}
          \begin{subfigure}[b]{0.51\textwidth}
         \centering
         \includegraphics[width=\textwidth]{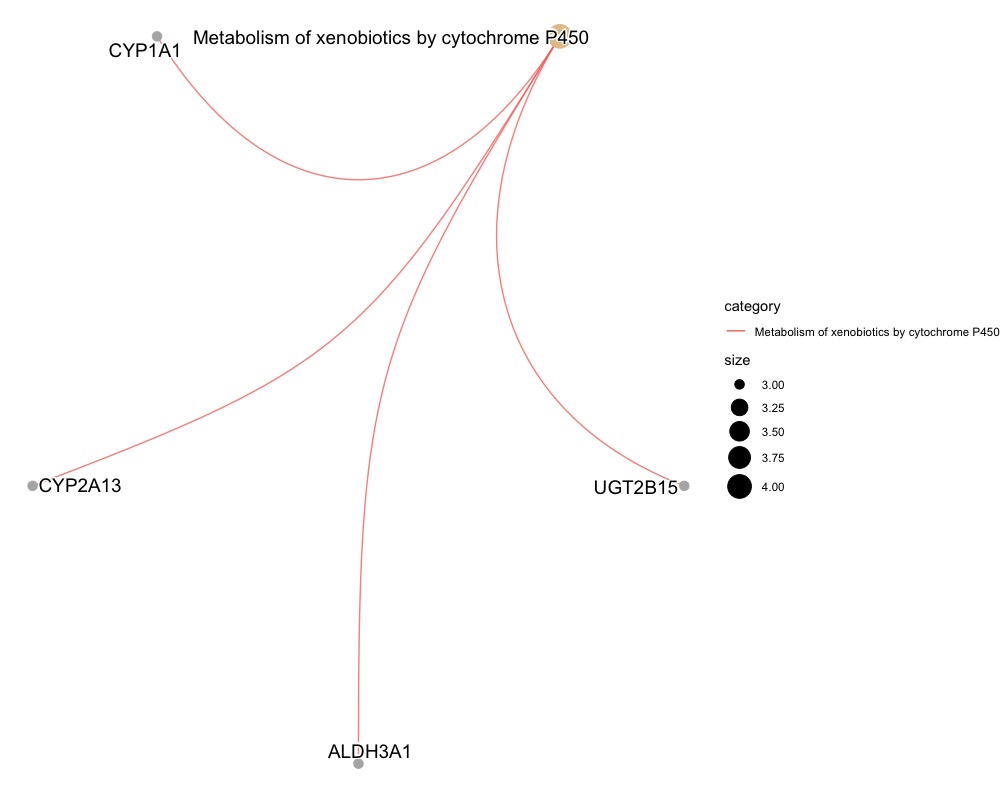}
         \caption{}
      \label{kegg_Scenario1}
     \end{subfigure}
      \caption{Results of the analysis of the TCGA-BRCA data. Gene network plots related to the significant GO and KEGG pathway terms obtained from EA for Scenario $1$: (a) GO-enriched terms (b) KEGG-enriched terms}
      \label{go_kegg_Scenario1}
\end{figure}

In Scenario $1$, only one biological process (BP) for the GO-EA is significant, the ``xenobiotic metabolic process'', and the association between this BP and similar breast cancer evolution processes is examined by Lee et al. \cite{lee2022interactions}. Another BP identified by GO-EA is ``GABAergic neuron differentiation'', with a p-value of $0.057$. For KEGG-EA, the only significantly enriched term is ``metabolism of xenobiotics by cytochrome P450''. The association between this enzyme expression and cancer risk, progression, metastasis and prognosis has been widely reported in basic and clinical studies (see the review study of Luo et al. \cite{luo2021cytochrome} for more details). Also the role of CYP1A1 and CYP2A13 genes in breast and lung cancers has been investigated extensively (see Sneha et al. \cite{sneha2021intratumoural} and references therein for more details). The gene network plots related to the significant GO and KEGG pathway terms are presented in Figures \ref{go_Scenario1} and \ref{kegg_Scenario1}.  The EA results for other methods, namely penMCFM(GMIFS) and MCM(GMIFS), are respectively shown in Figures S$2$-S$3$ in the Supplementary Materials. 
While the number of enriched terms in both GO- and KEGG-EA is limited for all the methods, some of the identified terms exhibit relevance to breast cancer.

\begin{figure}[H]
     \centering
     \begin{subfigure}[b]{0.49\textwidth}
         \centering
         \includegraphics[width=\textwidth]{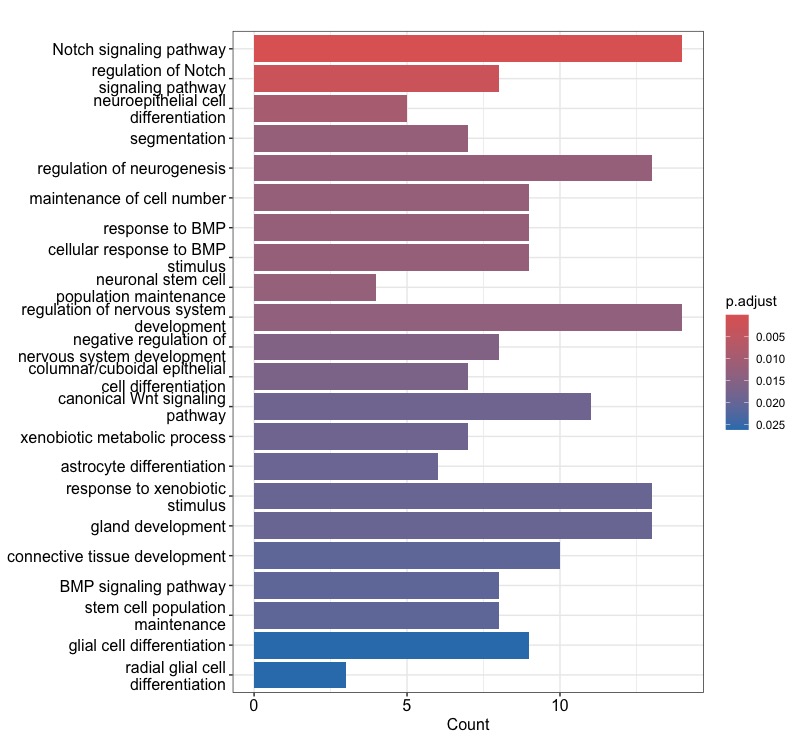}
         \caption{}
         \label{go_Scenario2a}
     \end{subfigure}
          \begin{subfigure}[b]{0.50\textwidth}
         \centering
         \includegraphics[width=\textwidth]{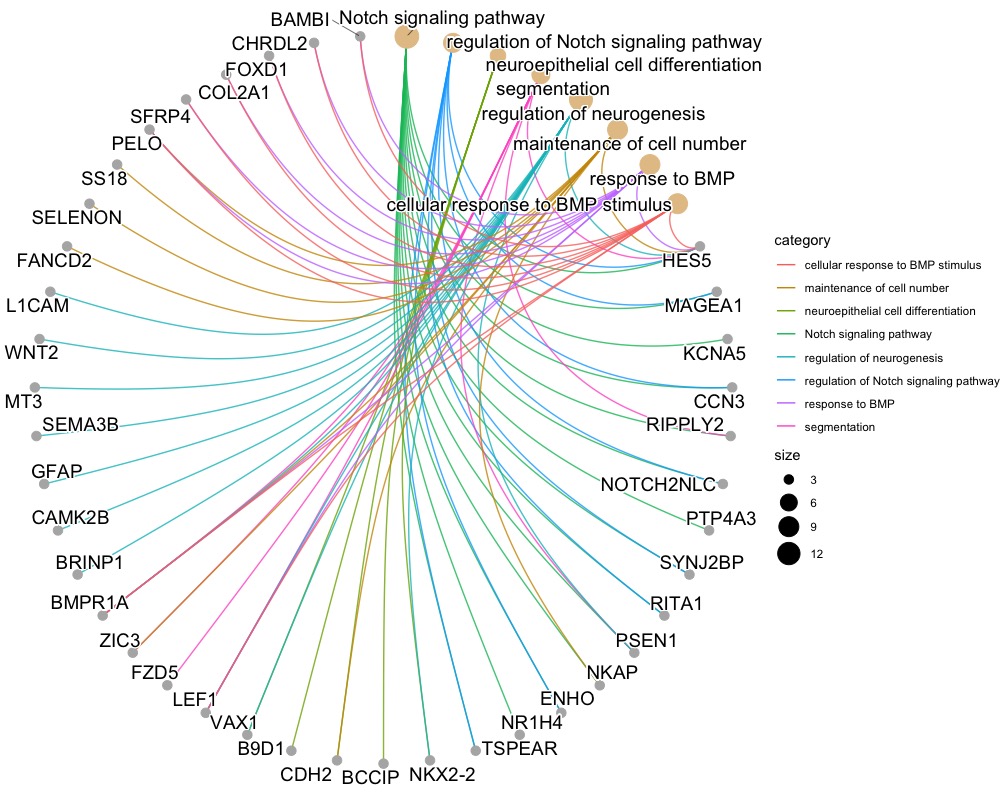}
         \caption{}
         \label{go_Scenario2b}
     \end{subfigure}
      \caption{Results of the analysis of the TCGA-BRCA data. GO-EA for Scenario $2$: (a) Barplot of significantly enriched terms (b) Network plot of enriched GO terms and related selected genes}
    \label{go_Scenario2}
\end{figure}

\begin{figure}[H]
     \centering    
     \begin{subfigure}[b]{0.48\textwidth}
         \centering
         \includegraphics[width=\textwidth]{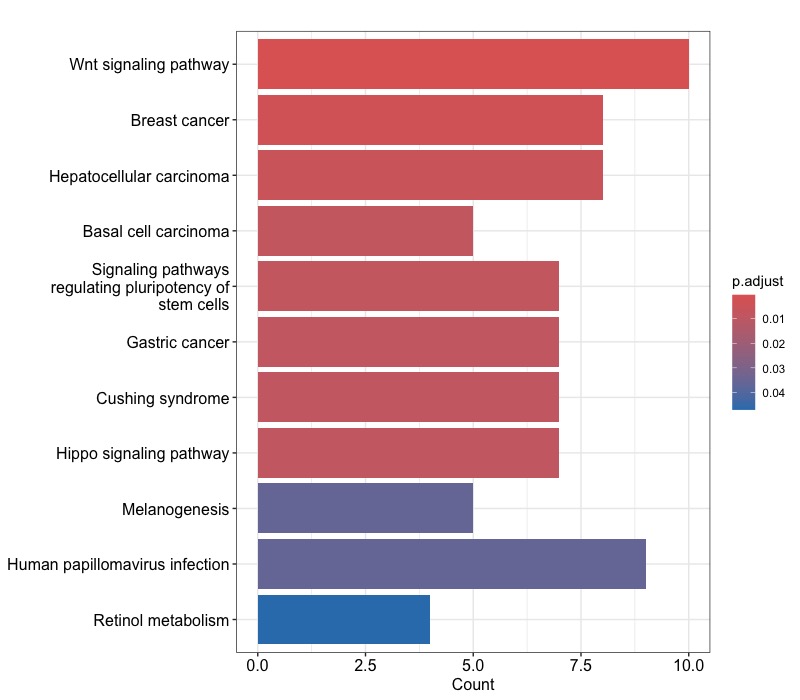}
         \caption{}
         \label{kegg_Scenario2a}
     \end{subfigure}
          \hfill
     \begin{subfigure}[b]{0.51\textwidth}
         \centering
         \includegraphics[width=\textwidth]{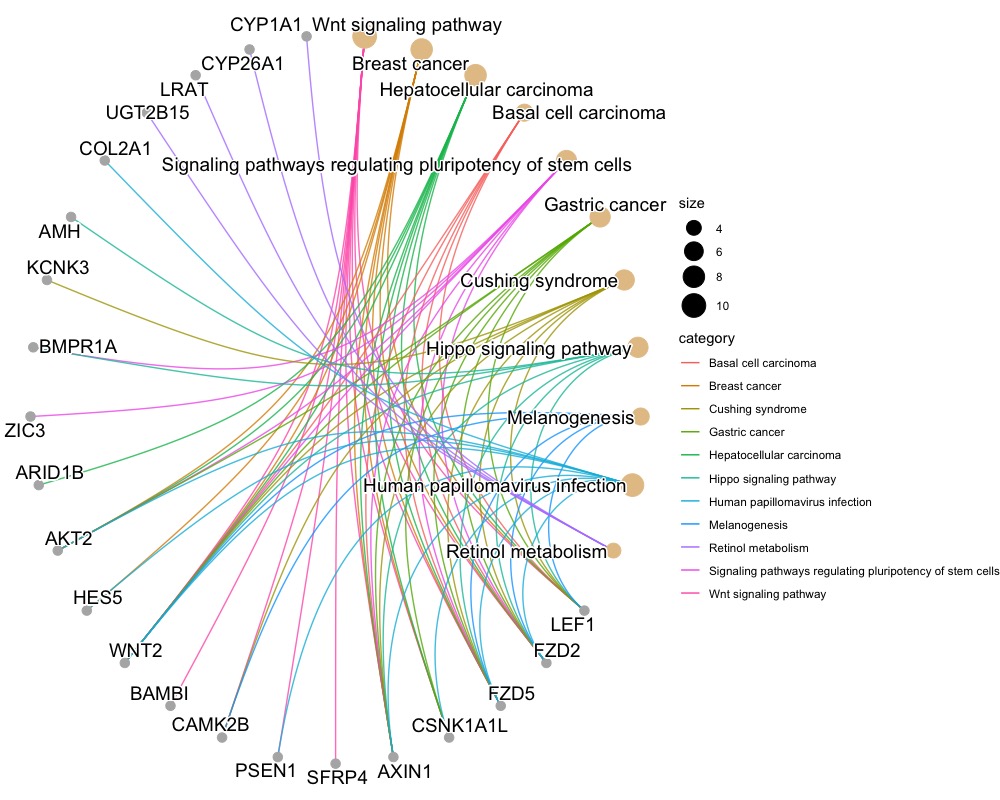}
         \caption{}
         \label{kegg_Scenario2b}
     \end{subfigure}
     \caption{Results of the analysis of the TCGA-BRCA data. KEGG-EA for Scenario $2$: (a) Barplot of significantly enriched terms (b) Network plot of enriched KEGG pathway terms and related selected genes}
    \label{kegg_Scenario2}
 \end{figure}

In Scenario $2$, we detect totally $28$ enriched BP terms via GO-EA, including some important BPs associated with breast cancer, such as ``Notch signaling pathway'' \cite{yousefi2022notch}, ``canonical Wnt signaling pathway'' \cite{xu2020wnt, abreu2022wnt}, and ``BMP signaling pathway'' \cite{ehata2022bone}. The barplot for some of the enriched terms in GO-EA, and the gene network plot related to the top-significant GO terms, are presented in Figures \ref{go_Scenario2a} and \ref{go_Scenario2b} respectively. Concerning results of KEGG-EA in this Scenario, Figure \ref{kegg_Scenario2a} shows that the ``Wnt signaling'' and ``breast cancer pathways'' are two of the most significant KEGG pathways, as confirmed by the growing number of studies in the literature demonstrating that Wnt signaling involves the proliferation, metastasis, immune microenvironment regulation, stemless maintenance, therapeutic resistance, and phenotype shaping of breast cancer (see Xu et al. \cite{xu2020wnt} and Abreu de Oliveira et al. \cite{abreu2022wnt}, and references therein,  for more details). The gene network plot related to the significant KEGG-EA pathway terms is presented in Figure \ref{kegg_Scenario2b}. Moreover, the enriched KEGG pathway ``hsa05224: Breast cancer'' is reported in Figure S$5$ in the Supplementary Materials to more clearly show the associated biological meaning.
We observe a greater number of enriched terms in the EA for all methods in Scenario $2$ compared to Scenario $1$. Notably, the ``breast cancer'' pathway is identified as one of the most enriched terms in at least one part of the model structure for all methods.

\subsection{Prognostic risk score construction and validation of identified genes}\label{sec:RiskScore}

We establish a prognostic risk score (PRS) to demonstrate the prognostic power of our model and methods. The PRS is constructed similarly to in Li and Liu \cite{li2021detecting}, by using the  average estimates of the coefficients corresponding to the selected genes over the $20$ repeats. Precisely, the PRS is defined as 
\begin{equation}
    PRS = \widehat{\theta}_1 \mathbf{x}_1 + \widehat{\theta}_2 \mathbf{x}_2 + ... +  \widehat{\theta}_{P^*} \mathbf{x}_{P^*}
\end{equation}
where $P^*$ is the total number of selected genes for the considered method, $\widehat{\theta}_i$ is the average estimate of the coefficient corresponding to the $i$-th gene, and $\mathbf{x}_i$ is the expression value of the $i$-th gene in the data. We dichotomize the prognostic scores based on the median value and subsequently employ a log-rank test to compare the survival curves between the two groups of patients.

\begin{figure}[H]
     \centering    
     \begin{subfigure}[b]{0.49\textwidth}
         \centering
         \includegraphics[width=\textwidth]{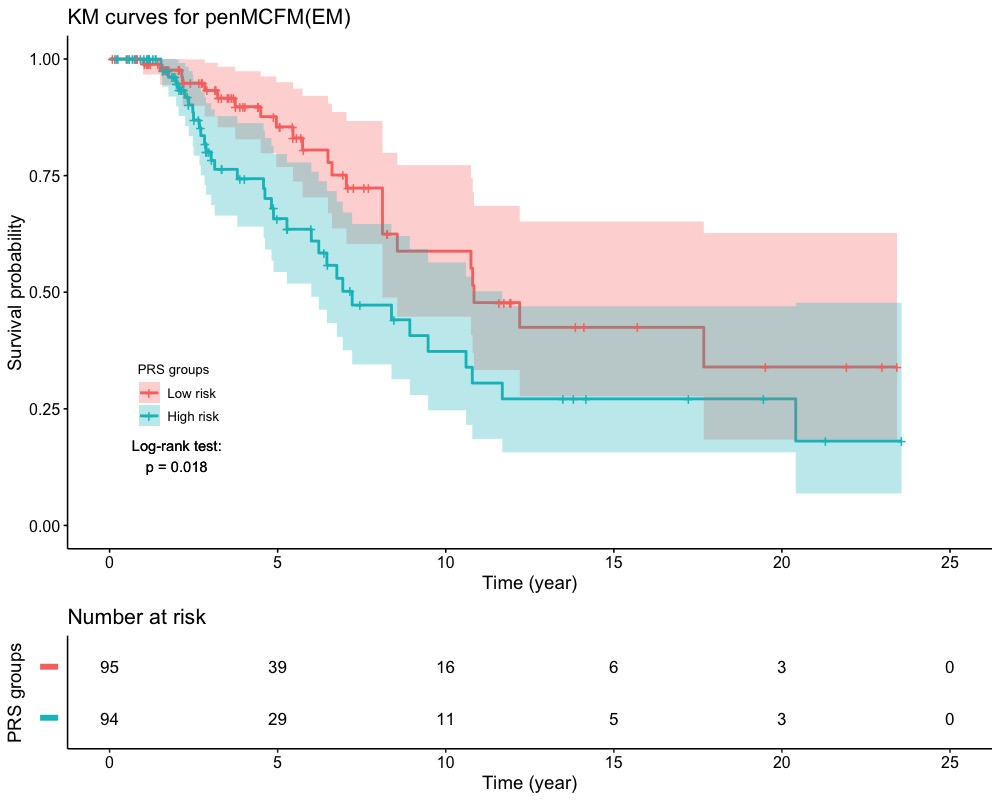}
         \caption{}
         \label{}
     \end{subfigure}
        \hfill
     \begin{subfigure}[b]{0.49\textwidth}
         \centering
         \includegraphics[width=\textwidth]{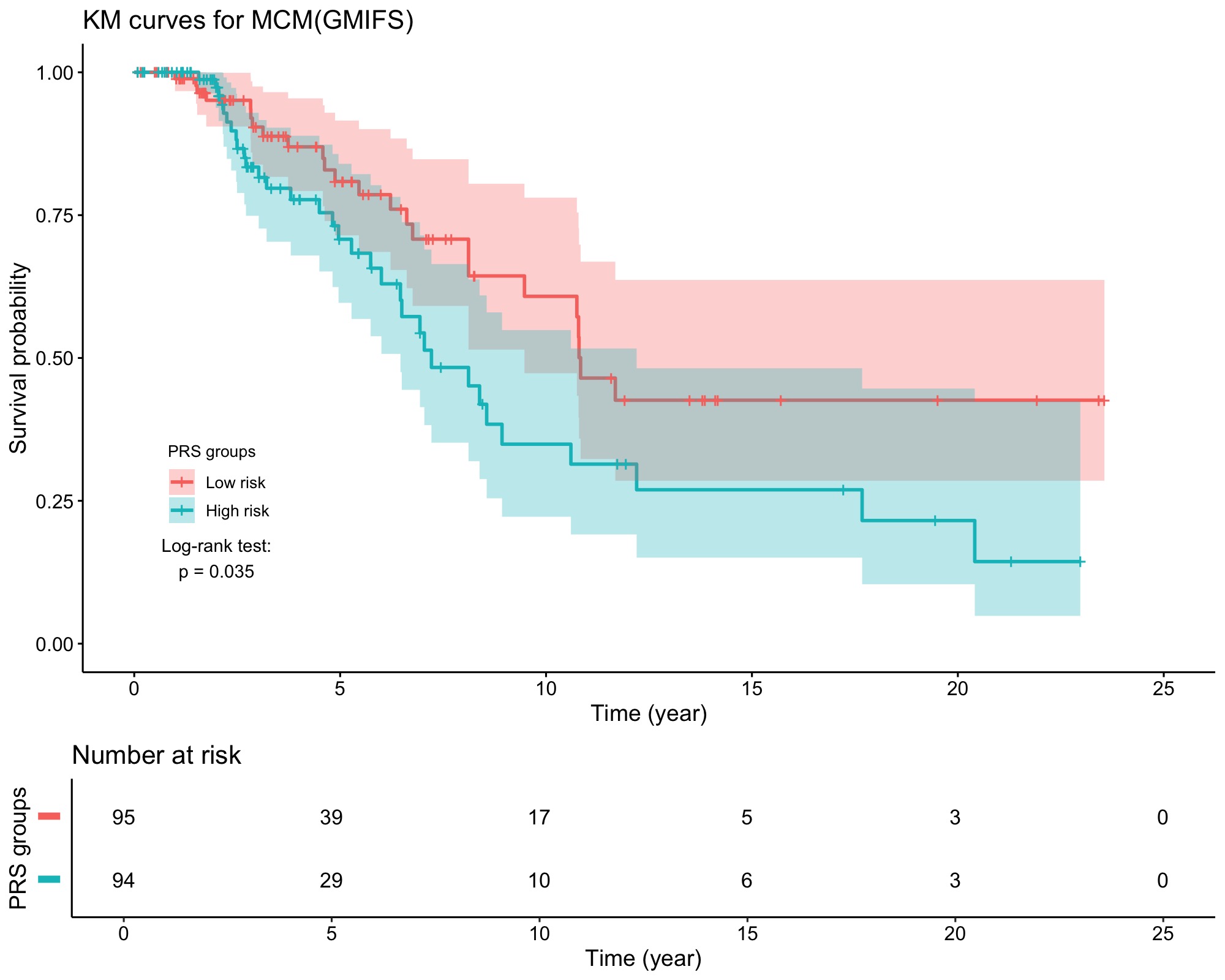}
         \caption{}
         \label{}
     \end{subfigure}
     \caption{Results of the analysis of the TCGA-BRCA data. KM curves for the TCGA-BRCA patients in the validation dataset when dichotomized into two groups by the median PRS  using the average results of $20$ method repeats in Scenario $2$: (a) penMCFM(EM) and (b) MCM(GMIFS)}
    \label{KM_plots_wrt_PRS_validation_data}
 \end{figure}

We use the validation dataset, which includes $189$ observations with their RNA-seq features and survival time, to validate the selected genes with the proposed PRS. We assign the subjects in the validation dataset to a high and low risk group using the median of the obtained PRS values. This procedure is applied when using the selected genes obtained with penMCFM(EM), penMCFM(GMIFS), MCM(GMIFS) and penCox.1se in Scenario $2$. The KM curves of these subgroups are presented in Figure \ref{KM_plots_wrt_PRS_validation_data} for penMCFM(EM) and MCM(GMIFS):  
in the figure it can be observed that the log-rank test shows significant differences between the survival distributions corresponding to the high- and low-risk groups as obtained by the PRS calculations carried out on the outputs of both penMCFM(EM) and MCM(GMIFS) methods (with p-values $0.018$ and $0.035$, respectively).
Since the $p$-values of the log-rank tests for penMCFM(GMIFS) and penCox.1se are not significant, these KM plots are only shown in Figures S$14$(a)-(b) in the Supplementary Materials.

Moreover, the heatmaps of the expression values of the selected genes, based on the penMCFM(EM) and MCM(GMIFS) methods, in the validation dataset are also presented in Figure S$15$ in the Supplementary Materials. In these heatmaps, only the genes which are selected in at least $2$ out of $20$ re-run of the methods are shown, for visualization purposes. From inspection of these figures it can be observed that the penMCFM(EM) method gives much more visible clusters of genes showing similar expression behaviours than those based on the MCM(GMIFS) method.

\section{Discussion and Conclusion}\label{sec:Conclusion}

In this study, we have introduced a novel method that adapts the mixture cure frailty model to account for high-dimensional covariates in both the incidence and the latency components of the model. The method extends the classical MCFM to higher dimensions using a multi-step adaptive elastic net penalty term inside the EM algorithm, therefore the name penMCFM. The proposed penMCFM allows performing variable selection in the MCFM model in high-dimensional settings where the number of covariates is significantly larger than the sample size. A comparative analysis with alternative survival models, including the penalized mixture cure model and penalized Cox regression, was also carried out. We also applied the proposed model along with alternative approaches to analyze RNAseq data from BRCA samples from TCGA. Finally, we conducted functional enrichment analyses of the identified biomarkers to show the disease mechanisms associated with BRCA, and then validated these biomarkers using an independent validation dataset to demonstrate their effectiveness and usefulness.

Moreover, we conducted a comprehensive investigation comparing our primary methodology, penMCFM(EM), with alternative methods using various metrics including C-statistic, AUC, and IBS, alongside functional enrichment analysis and the construction of prognostic risk scores, on various real data scenarios. Our analysis revealed that penMCFM(EM) and MCM(GMIFS) methods demonstrated comparable performances under certain circumstances, while penMCFM(EM) exhibited superior performance in others. Notably, despite similarities in performance observed in certain scenarios, penMCFM(EM) consistently selected fewer genes than MCM(GMIFS), its main competitor. Moreover, through rigorous simulation studies, penMCFM(EM) consistently outperformed alternative methods across different evaluation metrics. Consequently, we conclude that penMCFM(EM) represents a compelling alternative, offering advantages across diverse data generating scenarios. Additionally, our future research agenda entails applying this method to other cancer datasets, and conducting further comparative analyses against relevant methods.

A few practical issues regarding the proposed algorithm deserve some discussion. One notable computational challenge in our algorithm pertains to the efficient selection of the tuning parameters, specifically $\lambda_{1,Enet}$ and $\lambda_{2,Enet}$. While in penMCFM it is assumed that these parameters are identical for both penalties, alternative methods (such as grid search, random search, or Latin hypercube sampling) could be employed to identify optimal different values for $\lambda_{1,Enet}$ and $\lambda_{2,Enet}$, however at a much higher computational cost. In one experimental setting on simulated data, using Latin hypercube sampling with $100$ pairs for tuning $\lambda_{1,Enet}$ and $\lambda_{2,Enet}$ did not improve the results as compared to using the same value for the two parameters, leading us to omit this approach.
Moreover, when selecting the optimal tuning parameters  $\lambda_{Enet}$ and $\alpha_{Enet}$ via cross-validation, we opted for optimizing the C-statistics, i.e.  $\widehat{C}_{Cure}$ and $\widehat{C}$, instead of the commonly used Akaike Information Criterion (AIC) and Bayesian Information Criterion (BIC). While AIC and BIC were initially included in the cross-validation tuning process, it was observed that these criteria tended to select only a limited number of variables. Consequently, these results are not presented in the study.

Some potential directions for future research include exploring a semi-parametric approach as an alternative to the current use of parametric distributions in the latency part. Additionally, investigating a Bayesian approach for variable selection could be a fruitful direction for further investigation. Finally, to the best of our knowledge, the literature on cure frailty models for recurrent events in high-dimensional settings has not flourished to the same extent as that based on survival data. This could be another potentially fruitful direction of future research.

\section*{Authors contributions}
FK, DMS and VV jointly contributed to the method conception and development. FK was responsible for the implementation, for all data analyses, and for drafting the paper. VV contributed to the paper drafting and to the interpretation of all results. Both DMS and VV revised the paper for submission.
  
\section*{Acknowledgements}

This project has received funding from the European Union’s Horizon 2020 research and innovation programme under the Marie Skłodowska-Curie grant agreement No 801133.
Computer simulations were performed in HPC solutions provided by the Oslo Centre for Epidemiology and Biostatistics, at the University of Oslo.
The results published in the real data study are  based upon data generated by the TCGA Research Network: https://www.cancer.gov/tcga.

The authors would also like to thank Dr. Zhi Zhao for fruitful discussions.

\bibliographystyle{unsrt}
\bibliography{all_references_arXiv}

\begin{thebibliography}{100}

\bibitem{price2001modelling}
Dionne~L Price and Amita~K Manatunga.
\newblock Modelling survival data with a cured fraction using frailty models.
\newblock {\em Statistics in Medicine}, 20(9-10):1515--1527, 2001.

\bibitem{boag1949maximum}
John~W Boag.
\newblock Maximum likelihood estimates of the proportion of patients cured by
  cancer therapy.
\newblock {\em Journal of the Royal Statistical Society. Series B
  (Methodological)}, 11(1):15--53, 1949.

\bibitem{berkson1952survival}
Joseph Berkson and Robert~P Gage.
\newblock Survival curve for cancer patients following treatment.
\newblock {\em Journal of the American Statistical Association},
  47(259):501--515, 1952.

\bibitem{peng2008estimation}
Yingwei Peng and Jiajia Zhang.
\newblock Estimation method of the semiparametric mixture cure gamma frailty
  model.
\newblock {\em Statistics in Medicine}, 27(25):5177--5194, 2008.

\bibitem{sy2000estimation}
Judy~P Sy and Jeremy~MG Taylor.
\newblock Estimation in a cox proportional hazards cure model.
\newblock {\em Biometrics}, 56(1):227--236, 2000.

\bibitem{peng2000nonparametric}
Yingwei Peng and Keith~BG Dear.
\newblock A nonparametric mixture model for cure rate estimation.
\newblock {\em Biometrics}, 56(1):237--243, 2000.

\bibitem{dempster1977maximum}
Arthur~P Dempster, Nan~M Laird, and Donald~B Rubin.
\newblock Maximum likelihood from incomplete data via the {EM} algorithm.
\newblock {\em Journal of the Royal Statistical Society: Series B
  (Methodological)}, 39(1):1--22, 1977.

\bibitem{peng2008identifiability}
Yingwei Peng and Jiajia Zhang.
\newblock Identifiability of a mixture cure frailty model.
\newblock {\em Statistics \& Probability Letters}, 78(16):2604--2608, 2008.

\bibitem{cai2012smcure}
Chao Cai, Yubo Zou, Yingwei Peng, and Jiajia Zhang.
\newblock smcure: An r-package for estimating semiparametric mixture cure
  models.
\newblock {\em Computer Methods and Programs in Biomedicine},
  108(3):1255--1260, 2012.

\bibitem{smcure}
Chao Cai, Yubo Zou, Yingwei Peng, Jiajia Zhang, and Maintainer~Chao Cai.
\newblock smcure: An r-package for estimating semiparametric ph and aft mixture
  cure models.
\newblock {\em https://cran.r-project.org/web/packages/smcure/smcure.pdf},
  2.1., 2022.

\bibitem{rondeau2013cure}
Virginie Rondeau, Emmanuel Schaffner, Fabien Corbiere, Juan~R Gonzalez, and
  Simone Mathoulin-P{\'e}lissier.
\newblock Cure frailty models for survival data: application to recurrences for
  breast cancer and to hospital readmissions for colorectal cancer.
\newblock {\em Statistical Methods in Medical Research}, 22(3):243--260, 2013.

\bibitem{tibshirani1996regression}
Robert Tibshirani.
\newblock Regression shrinkage and selection via the lasso.
\newblock {\em Journal of the Royal Statistical Society Series B: Statistical
  Methodology}, 58(1):267--288, 1996.

\bibitem{zou2005regularization}
Hui Zou and Trevor Hastie.
\newblock Regularization and variable selection via the elastic net.
\newblock {\em Journal of the Royal Statistical Society: Series B (Statistical
  Methodology)}, 67(2):301--320, 2005.

\bibitem{zou2009adaptive}
Hui Zou and Hao~Helen Zhang.
\newblock On the adaptive elastic-net with a diverging number of parameters.
\newblock {\em Annals of Statistics}, 37(4):1733--1751, 2009.

\bibitem{friedman2010regularization}
Jerome Friedman, Trevor Hastie, and Rob Tibshirani.
\newblock Regularization paths for generalized linear models via coordinate
  descent.
\newblock {\em Journal of Statistical Software}, 33(1):1--22, 2010.

\bibitem{glmnet}
Jerome Friedman, Trevor Hastie, Rob Tibshirani, Balasubramanian Narasimhan,
  Kenneth Tay, Noah Simon, Junyang Qian, and James Yang.
\newblock glmnet: Lasso and elastic-net regularized generalized linear models.
\newblock {\em https://cran.r-project.org/web/packages/glmnet/index.html},
  4.1-7, 2023.

\bibitem{simon2011regularization}
Noah Simon, Jerome Friedman, Trevor Hastie, and Rob Tibshirani.
\newblock Regularization paths for cox’s proportional hazards model via
  coordinate descent.
\newblock {\em Journal of Statistical Software}, 39(5):1--13, 2011.

\bibitem{tay2023elastic}
J~Kenneth Tay, Balasubramanian Narasimhan, and Trevor Hastie.
\newblock Elastic net regularization paths for all generalized linear models.
\newblock {\em Journal of Statistical Software}, 106(1):1--31, 2023.

\bibitem{buhlmann2008discussion}
Peter B{\"u}hlmann and Lukas Meier.
\newblock Discussion: One-step sparse estimates in nonconcave penalized
  likelihood models.
\newblock {\em The Annals of Statistics}, 36:1534--1541, 2008.

\bibitem{xiao2015multi}
Nan Xiao and Qing-Song Xu.
\newblock Multi-step adaptive elastic-net: reducing false positives in
  high-dimensional variable selection.
\newblock {\em Journal of Statistical Computation and Simulation},
  85(18):3755--3765, 2015.

\bibitem{msaenet}
Nan Xiao.
\newblock msaenet: Multi-step adaptive estimation methods for sparse
  regressions.
\newblock {\em https://cran.r-project.org/web/packages/msaenet/msaenet.pdf},
  3.1, 2022.

\bibitem{buhlmann2011statistics}
Peter B{\"u}hlmann and Sara Van De~Geer.
\newblock {\em Statistics for high-dimensional data: methods, theory and
  applications}.
\newblock Springer Berlin, Heidelberg, 2011.

\bibitem{hastie2015statistical}
Trevor Hastie, Robert Tibshirani, and Martin Wainwright.
\newblock {\em Statistical learning with sparsity: the lasso and
  generalizations}.
\newblock CRC press, 2015.

\bibitem{james2021statistical}
Gareth James, Daniela Witten, Trevor Hastie, Robert Tibshirani, Gareth James,
  Daniela Witten, Trevor Hastie, and Robert Tibshirani.
\newblock {\em An Introduction to Statistical Learning}.
\newblock Springer New York, NY, 2021.

\bibitem{liu2012variable}
Xiang Liu, Yingwei Peng, Dongsheng Tu, and Hua Liang.
\newblock Variable selection in semiparametric cure models based on penalized
  likelihood, with application to breast cancer clinical trials.
\newblock {\em Statistics in Medicine}, 31(24):2882--2891, 2012.

\bibitem{Fan2017promoting}
Xinyan Fan, Mengque Liu, Kuangnan Fang, Yuan Huang, and Shuangge Ma.
\newblock Promoting structural effects of covariates in the cure rate model
  with penalization.
\newblock {\em Statistical Methods in Medical Research}, 26(5):2078--2092,
  2017.

\bibitem{masud2018variable}
Abdullah Masud, Wanzhu Tu, and Zhangsheng Yu.
\newblock Variable selection for mixture and promotion time cure rate models.
\newblock {\em Statistical Methods in Medical Research}, 27(7):2185--2199,
  2018.

\bibitem{beretta2019variable}
Alessandro Beretta and C{\'e}dric Heuchenne.
\newblock Variable selection in proportional hazards cure model with
  time-varying covariates, application to us bank failures.
\newblock {\em Journal of Applied Statistics}, 46(9):1529--1549, 2019.

\bibitem{penPHcure}
Alessandro Beretta and Cédric Heuchenne.
\newblock Variable selection in ph cure model with time-varying covariates.
\newblock {\em
  https://cran.r-project.org/web/packages/penPHcure/penPHcure.pdf}, 1.0.2,
  2022.

\bibitem{sun2019variable}
Liuquan Sun, Shuwei Li, Lianming Wang, and Xinyuan Song.
\newblock Variable selection in semiparametric nonmixture cure model with
  interval-censored failure time data: An application to the prostate cancer
  screening study.
\newblock {\em Statistics in Medicine}, 38(16):3026--3039, 2019.

\bibitem{bussy2019c}
Simon Bussy, Agathe Guilloux, St{\'e}phane Ga{\"\i}ffas, and Anne-Sophie
  Jannot.
\newblock C-mix: A high-dimensional mixture model for censored durations, with
  applications to genetic data.
\newblock {\em Statistical Methods in Medical Research}, 28(5):1523--1539,
  2019.

\bibitem{shi2020promoting}
Xingjie Shi, Shuangge Ma, and Yuan Huang.
\newblock Promoting sign consistency in the cure model estimation and
  selection.
\newblock {\em Statistical Methods in Medical Research}, 29(1):15--28, 2020.

\bibitem{xie2021mixture}
Yujing Xie and Zhangsheng Yu.
\newblock Mixture cure rate models with neural network estimated nonparametric
  components.
\newblock {\em Computational Statistics}, 36(4):2467--2489, 2021.

\bibitem{xu2021variable}
Yang Xu, Shishun Zhao, Tao Hu, and Jianguo Sun.
\newblock Variable selection for generalized odds rate mixture cure models with
  interval-censored failure time data.
\newblock {\em Computational Statistics \& Data Analysis}, 156:1--17, 2021.

\bibitem{fu2022controlled}
Han Fu, Deedra Nicolet, Krzysztof Mr{\'o}zek, Richard~M Stone, Ann-Kathrin
  Eisfeld, John~C Byrd, and Kellie~J Archer.
\newblock Controlled variable selection in weibull mixture cure models for
  high-dimensional data.
\newblock {\em Statistics in Medicine}, 41(22):4340--4366, 2022.

\bibitem{hastie2007stagewise}
Trevor Hastie, Jonathan Taylor, Robert Tibshirani, and Guenther Walther.
\newblock Forward stagewise regression and the monotone lasso.
\newblock {\em Electronic Journal of Statistics}, 1:1--29, 2007.

\bibitem{farewell1982use}
Vern~T Farewell.
\newblock The use of mixture models for the analysis of survival data with
  long-term survivors.
\newblock {\em Biometrics}, pages 1041--1046, 1982.

\bibitem{lbfgs}
Antonio Coppola and Brandon M.~Stewart.
\newblock lbfgs: Efficient l-bfgs and owl-qn optimization in r.
\newblock {\em https://cran.r-project.org/web/packages/lbfgs/lbfgs.pdf,
  https://cran.r-project.org/web/packages/lbfgs/vignettes/Vignette.pdf},
  1.2.1.2, 2022.

\bibitem{makowski2015generalized}
Mateusz Makowski and Kellie~J Archer.
\newblock Generalized monotone incremental forward stagewise method for
  modeling count data: Application predicting micronuclei frequency.
\newblock {\em Cancer Informatics}, 14:97--105, 2015.

\bibitem{hou2015regularization}
Jiayi Hou and Kellie~J Archer.
\newblock Regularization method for predicting an ordinal response using
  longitudinal high-dimensional genomic data.
\newblock {\em Statistical Applications in Genetics and Molecular Biology},
  14(1):93--111, 2015.

\bibitem{yu2018nonlinear}
Tianwei Yu.
\newblock Nonlinear variable selection with continuous outcome: A fully
  nonparametric incremental forward stagewise approach.
\newblock {\em Statistical Analysis and Data Mining: The ASA Data Science
  Journal}, 11(4):188--197, 2018.

\bibitem{asano2014assessing}
Junichi Asano, Akihiro Hirakawa, and Chikuma Hamada.
\newblock Assessing the prediction accuracy of cure in the cox proportional
  hazards cure model: an application to breast cancer data.
\newblock {\em Pharmaceutical Statistics}, 13(6):357--363, 2014.

\bibitem{asano2017assessing}
Junichi Asano and Akihiro Hirakawa.
\newblock Assessing the prediction accuracy of a cure model for censored
  survival data with long-term survivors: application to breast cancer data.
\newblock {\em Journal of Biopharmaceutical Statistics}, 27(6):918--932, 2017.

\bibitem{pal2023new}
Suvra Pal, Yingwei Peng, and Wisdom Aselisewine.
\newblock A new approach to modeling the cure rate in the presence of interval
  censored data.
\newblock {\em Computational Statistics}, pages 1--27, 2023.

\bibitem{love2014moderated}
Michael~I Love, Wolfgang Huber, and Simon Anders.
\newblock Moderated estimation of fold change and dispersion for rna-seq data
  with deseq2.
\newblock {\em Genome Biology}, 15(12):1--21, 2014.

\bibitem{zhao2021tpm}
Yingdong Zhao, Ming-Chung Li, Mariam~M Konat{\'e}, Li~Chen, Biswajit Das, Chris
  Karlovich, P~Mickey Williams, Yvonne~A Evrard, James~H Doroshow, and Lisa~M
  McShane.
\newblock Tpm, fpkm, or normalized counts? a comparative study of
  quantification measures for the analysis of rna-seq data from the nci
  patient-derived models repository.
\newblock {\em Journal of Translational Medicine}, 19(1):1--15, 2021.

\bibitem{zhao2024tutorial}
Zhi Zhao, John Zobolas, Manuela Zucknick, and Tero Aittokallio.
\newblock Tutorial on survival modeling with applications to omics data.
\newblock {\em Bioinformatics}, pages 1--15, 2024.

\bibitem{john2020m3c}
Christopher~R John, David Watson, Dominic Russ, Katriona Goldmann, Michael
  Ehrenstein, Costantino Pitzalis, Myles Lewis, and Michael Barnes.
\newblock M3c: Monte carlo reference-based consensus clustering.
\newblock {\em Scientific Reports}, 10(1 (1816)):1--14, 2020.

\bibitem{li2021detecting}
Lingyu Li and Zhi-Ping Liu.
\newblock Detecting prognostic biomarkers of breast cancer by regularized cox
  proportional hazards models.
\newblock {\em Journal of Translational Medicine}, 19:1--20, 2021.

\bibitem{li2023biomarker}
Lingyu Li and Zhi-Ping Liu.
\newblock Biomarker discovery from high-throughput data by connected
  network-constrained support vector machine.
\newblock {\em Expert Systems with Applications}, 226 (120179):1--12, 2023.

\bibitem{kanehisa2021kegg}
Minoru Kanehisa, Miho Furumichi, Yoko Sato, Mari Ishiguro-Watanabe, and Mao
  Tanabe.
\newblock Kegg: integrating viruses and cellular organisms.
\newblock {\em Nucleic Acids Research}, 49(D1):D545--D551, 2021.

\bibitem{gene2021gene}
The Gene~Ontology Consortium.
\newblock The gene ontology resource: enriching a gold mine.
\newblock {\em Nucleic Acids Research}, 49(D1):D325--D334, 2021.

\bibitem{cardoso201670}
Fatima Cardoso, Laura~J van’t Veer, Jan Bogaerts, Leen Slaets, Giuseppe
  Viale, Suzette Delaloge, Jean-Yves Pierga, Etienne Brain, Sylvain Causeret,
  Mauro DeLorenzi, et~al.
\newblock 70-gene signature as an aid to treatment decisions in early-stage
  breast cancer.
\newblock {\em New England Journal of Medicine}, 375(8):717--729, 2016.

\bibitem{yan2019osbrca}
Zhongyi Yan, Qiang Wang, Xiaoxiao Sun, Bingbing Ban, Zhendong Lu, Yifang Dang,
  Longxiang Xie, Lu~Zhang, Yongqiang Li, Wan Zhu, et~al.
\newblock Osbrca: a web server for breast cancer prognostic biomarker
  investigation with massive data from tens of cohorts.
\newblock {\em Frontiers in Oncology}, 9(1349):1--8, 2019.

\bibitem{li2020novel}
Xiaomei Li, Lin Liu, Gregory~J Goodall, Andreas Schreiber, Taosheng Xu, Jiuyong
  Li, and Thuc~D Le.
\newblock A novel single-cell based method for breast cancer prognosis.
\newblock {\em PLoS Computational Biology}, 16(8(e1008133)):1--20, 2020.

\bibitem{de2014investigating}
Riccardo De~Bin, Willi Sauerbrei, and Anne-Laure Boulesteix.
\newblock Investigating the prediction ability of survival models based on both
  clinical and omics data: two case studies.
\newblock {\em Statistics in Medicine}, 33(30):5310--5329, 2014.

\bibitem{volkmann2019plea}
Alexander Volkmann, Riccardo De~Bin, Willi Sauerbrei, and Anne-Laure
  Boulesteix.
\newblock A plea for taking all available clinical information into account
  when assessing the predictive value of omics data.
\newblock {\em BMC medical research methodology}, 19:1--15, 2019.

\bibitem{yu2012clusterprofiler}
Guangchuang Yu, Li-Gen Wang, Yanyan Han, and Qing-Yu He.
\newblock clusterprofiler: an r package for comparing biological themes among
  gene clusters.
\newblock {\em Omics: A Journal of Integrative Biology}, 16(5):284--287, 2012.

\bibitem{wu2021clusterprofiler}
Tianzhi Wu, Erqiang Hu, Shuangbin Xu, Meijun Chen, Pingfan Guo, Zehan Dai,
  Tingze Feng, Lang Zhou, Wenli Tang, LI~Zhan, et~al.
\newblock clusterprofiler 4.0: A universal enrichment tool for interpreting
  omics data.
\newblock {\em The Innovation}, 2(3-100141), 2021.

\bibitem{lee2022interactions}
Derrick~G Lee, Johanna~M Schuetz, Agnes~S Lai, Igor Burstyn, Angela
  Brooks-Wilson, Kristan~J Aronson, and John~J Spinelli.
\newblock Interactions between exposure to polycyclic aromatic hydrocarbons and
  xenobiotic metabolism genes, and risk of breast cancer.
\newblock {\em Breast Cancer}, 29:38--49, 2022.

\bibitem{luo2021cytochrome}
Bin Luo, Dandan Yan, Honglin Yan, and Jingping Yuan.
\newblock Cytochrome p450: Implications for human breast cancer (review).
\newblock {\em Oncology Letters}, 22(1):1--9, 2021.

\bibitem{sneha2021intratumoural}
Smarakan Sneha, Simon~C Baker, Andrew Green, Sarah Storr, Radhika Aiyappa,
  Stewart Martin, and Klaus Pors.
\newblock Intratumoural cytochrome p450 expression in breast cancer: impact on
  standard of care treatment and new efforts to develop tumour-selective
  therapies.
\newblock {\em Biomedicines}, 9(3(290)):1--22, 2021.

\bibitem{yousefi2022notch}
Hassan Yousefi, Afshin Bahramy, Narges Zafari, Mahsa~Rostamian Delavar, Khoa
  Nguyen, Atousa Haghi, Tahmineh Kandelouei, Cecilia Vittori, Parham Jazireian,
  Sajad Maleki, et~al.
\newblock Notch signaling pathway: a comprehensive prognostic and gene
  expression profile analysis in breast cancer.
\newblock {\em BMC Cancer}, 22(1):1282, 2022.

\bibitem{xu2020wnt}
Xiufang Xu, Miaofeng Zhang, Faying Xu, and Shaojie Jiang.
\newblock Wnt signaling in breast cancer: biological mechanisms, challenges and
  opportunities.
\newblock {\em Molecular cancer}, 19:1--35, 2020.

\bibitem{abreu2022wnt}
Willy~Antoni Abreu~de Oliveira, Youssef El~Laithy, Alejandra Bruna, Daniela
  Annibali, and Frederic Lluis.
\newblock Wnt signaling in the breast: From development to disease.
\newblock {\em Frontiers in Cell and Developmental Biology}, 10( 884467):1--19,
  2022.

\bibitem{ehata2022bone}
Shogo Ehata and Kohei Miyazono.
\newblock Bone morphogenetic protein signaling in cancer; some topics in the
  recent 10 years.
\newblock {\em Frontiers in Cell and Developmental Biology}, 10:883523, 2022.

\bibitem{ascierto2012signature}
Maria~Libera Ascierto, Maciej Kmieciak, Michael~O Idowu, Rose Manjili, Yingdong
  Zhao, Margaret Grimes, Catherine Dumur, Ena Wang, Viswanathan Ramakrishnan,
  Xiang-Yang Wang, et~al.
\newblock A signature of immune function genes associated with recurrence-free
  survival in breast cancer patients.
\newblock {\em Breast Cancer Research and Treatment}, 131:871--880, 2012.

\bibitem{lee2021novel}
Hannah Lee, Mi~Jeong Kwon, Beom-Mo Koo, Hee~Geon Park, Jinil Han, and Young~Kee
  Shin.
\newblock A novel immune prognostic index for stratification of high-risk
  patients with early breast cancer.
\newblock {\em Scientific Reports}, 11(1):128, 2021.

\bibitem{liang2015molecular}
Feng Liang, Hongzhu Qu, Qiang Lin, Yadong Yang, Xiuyan Ruan, Bo~Zhang, Yi~Liu,
  Chengze Yu, Hongyan Zhang, Xiangdong Fang, et~al.
\newblock Molecular biomarkers screened by next-generation rna sequencing for
  non-sentinel lymph node status prediction in breast cancer patients with
  metastatic sentinel lymph nodes.
\newblock {\em World Journal of Surgical Oncology}, 13:1--10, 2015.

\bibitem{xia2021igll5}
Zhi-Nan Xia, Xing-Yuan Wang, Li-Cheng Cai, Wen-Gang Jian, and Cheng Zhang.
\newblock Igll5 is correlated with tumor-infiltrating immune cells in clear
  cell renal cell carcinoma.
\newblock {\em FEBS Open Bio}, 11(3):898--910, 2021.

\bibitem{lee2019gabrq}
Dongjun Lee, Mihyang Ha, Chae~Mi Hong, Jayoung Kim, Su~Min Park, Dongsu Park,
  Dong~Hyun Sohn, Ho~Jin Shin, Hak-Sun Yu, Chi~Dae Kim, et~al.
\newblock Gabrq expression is a potential prognostic marker for patients with
  clear cell renal cell carcinoma.
\newblock {\em Oncology Letters}, 18(6):5731--5738, 2019.

\bibitem{yan2020distinct}
Ling Yan, Yi-Zhen Gong, Meng-Nan Shao, Guo-Tian Ruan, Hai-Lun Xie, Xi-Wen Liao,
  Xiang-Kun Wang, Quan-Fa Han, Xin Zhou, Li-Cheng Zhu, et~al.
\newblock Distinct diagnostic and prognostic values of $\gamma$-aminobutyric
  acid type a receptor family genes in patients with colon adenocarcinoma.
\newblock {\em Oncology Letters}, 20(1):275--291, 2020.

\bibitem{li2023delta}
Na~Li, Xiang Xu, Dan Liu, Jiaxin Gao, Ying Gao, Xufeng Wu, Huiming Sheng, Qun
  Li, and Jun Mi.
\newblock The delta subunit of the gabaa receptor is necessary for the
  gpt2-promoted breast cancer metastasis.
\newblock {\em Theranostics}, 13(4):1355, 2023.

\bibitem{doberstein2014l1cam}
Kai Doberstein, Karin Milde-Langosch, Niko~P Bretz, Uwe Schirmer, Ayelet
  Harari, Isabell Witzel, Alon Ben-Arie, Michael Hubalek, Elisabeth
  M{\"u}ller-Holzner, Susanne Reinold, et~al.
\newblock L1cam is expressed in triple-negative breast cancers and is inversely
  correlated with androgen receptor.
\newblock {\em BMC cancer}, 14(1):1--13, 2014.

\bibitem{altevogt2016l1cam}
Peter Altevogt, Kai Doberstein, and Mina Fogel.
\newblock L1cam in human cancer.
\newblock {\em International Journal of Cancer}, 138(7):1565--1576, 2016.

\bibitem{barron2022gene}
Carlos~A Barr{\'o}n-Gallardo, Mariel Garcia-Chagoll{\'a}n, Andres~J
  Mor{\'a}n-Mendoza, Raul Delgadillo-Cristerna, Mar{\'\i}a~G
  Mart{\'\i}nez-Silva, Mar{\'\i}a~M Villase{\~n}or-Garc{\'\i}a, Adriana
  Aguilar-Lemarroy, and Luis~F Jave-Su{\'a}rez.
\newblock A gene expression signature in her2+ breast cancer patients related
  to neoadjuvant chemotherapy resistance, overall survival, and disease-free
  survival.
\newblock {\em Frontiers in Genetics}, 13:991706, 2022.

\bibitem{meng2021screening}
Di~Meng, Tongjun Liu, Feng Ma, and Mingguo Wang.
\newblock Screening the key genes of prognostic value in the microenvironment
  for head and neck squamous cell carcinoma.
\newblock {\em Medicine}, 100(4):1--8, 2021.

\bibitem{lei2022correlation}
Ping Lei, Hongmei Wang, Liting Yu, Cong Xu, Haojie Sun, Yihan Lyu, Lianqin Li,
  and Dao-Lai Zhang.
\newblock A correlation study of adhesion g protein-coupled receptors as
  potential therapeutic targets in uterine corpus endometrial cancer.
\newblock {\em International Immunopharmacology}, 108:108743, 2022.

\bibitem{singh2022genome}
Yashbir Singh, Naidu Subbarao, Abhinav Jaimini, Quincy~A Hathaway, Amina
  Kunovac, Bradley Erickson, Vishnu Swarup, and Himanshu~Narayan Singh.
\newblock Genome-wide expression reveals potential biomarkers in breast cancer
  bone metastasis.
\newblock {\em Journal of Integrative Bioinformatics}, 19(3):20210041, 2022.

\bibitem{gao2019prognostic}
Chundi Gao, Jing Zhuang, Chao Zhou, Ke~Ma, Minzhang Zhao, Cun Liu, Lijuan Liu,
  Huayao Li, Fubin Feng, and Changgang Sun.
\newblock Prognostic value of aberrantly expressed methylation gene profiles in
  lung squamous cell carcinoma: A study based on the cancer genome atlas.
\newblock {\em Journal of Cellular Physiology}, 234(5):6519--6528, 2019.

\bibitem{li2018combined}
Xiaohong Li, Eric~C Rouchka, Guy~N Brock, Jun Yan, Timothy~E O’Toole, David~A
  Tieri, and Nigel~GF Cooper.
\newblock A combined approach with gene-wise normalization improves the
  analysis of rna-seq data in human breast cancer subtypes.
\newblock {\em PloS One}, 13(8):e0201813, 2018.

\bibitem{zhang2019identification}
Xiuzhi Zhang, Chunyan Kang, Ningning Li, Xiaoli Liu, Jinzhong Zhang, Fenglan
  Gao, and Liping Dai.
\newblock Identification of special key genes for alcohol-related
  hepatocellular carcinoma through bioinformatic analysis.
\newblock {\em PeerJ}, 7:e6375, 2019.

\bibitem{rao2017g}
Angad Rao and Deron~R Herr.
\newblock G protein-coupled receptor gpr19 regulates e-cadherin expression and
  invasion of breast cancer cells.
\newblock {\em Biochimica et Biophysica Acta (BBA)-Molecular Cell Research},
  1864(7):1318--1327, 2017.

\bibitem{jia2023low}
Linghui Jia, Liting Liao, Yongshuai Jiang, Xiangyu Hu, Guotao Lu, Weiming Xiao,
  Weijuan Gong, Xiaoqin Jia, and Jia Xiaoqin.
\newblock Low-dose adropin stimulates inflammasome activation of macrophage via
  mitochondrial ros involved in colorectal cancer progression.
\newblock {\em BMC Cancer}, 23(1):1042, 2023.

\bibitem{weber2018olfactory}
Lea Weber, D{\'e}sir{\'e}e Ma{\ss}berg, Christian Becker, Janine Altm{\"u}ller,
  Burkhard Ubrig, Gabriele Bonatz, Gerhard W{\"o}lk, Stathis Philippou, Andrea
  Tannapfel, Hanns Hatt, et~al.
\newblock Olfactory receptors as biomarkers in human breast carcinoma tissues.
\newblock {\em Frontiers in Oncology}, 8:33, 2018.

\bibitem{masjedi2019olfactory}
Shirin Masjedi, Laurence~J Zwiebel, and Todd~D Giorgio.
\newblock Olfactory receptor gene abundance in invasive breast carcinoma.
\newblock {\em Scientific Reports}, 9(1):13736, 2019.

\bibitem{hong2007inactivation}
Kyeong-Man Hong, Sei-Hoon Yang, Sinchita~Roy Chowdhuri, Audrey Player, Megan
  Hames, Junya Fukuoka, Daoud Meerzaman, Tatiana Dracheva, Zhifu Sun, Ping
  Yang, et~al.
\newblock Inactivation of llc1 gene in nonsmall cell lung cancer.
\newblock {\em International Journal of Cancer}, 120(11):2353--2358, 2007.

\bibitem{shih2018identification}
Andrew~J Shih, Andrew Menzin, Jill Whyte, John Lovecchio, Anthony Liew, Houman
  Khalili, Tawfiqul Bhuiya, Peter~K Gregersen, and Annette~T Lee.
\newblock Identification of grade and origin specific cell populations in
  serous epithelial ovarian cancer by single cell rna-seq.
\newblock {\em PLoS One}, 13(11):e0206785, 2018.

\bibitem{bose2022computing}
Banabithi Bose, Matthew Moravec, and Serdar Bozdag.
\newblock Computing microrna-gene interaction networks in pan-cancer using
  mirdriver.
\newblock {\em Scientific Reports}, 12(1):3717, 2022.

\bibitem{furrer2022association}
Daniela Furrer, Dzevka Dragic, Sue-Ling Chang, Fr{\'e}d{\'e}ric Fournier,
  Arnaud Droit, Simon Jacob, and Caroline Diorio.
\newblock Association between genome-wide epigenetic and genetic alterations in
  breast cancer tissue and response to her2-targeted therapies in her2-positive
  breast cancer patients: new findings and a systematic review.
\newblock {\em Cancer Drug Resistance}, 5(4):995, 2022.

\bibitem{brophy2017gene}
Patrick~D Brophy, Maria Rasmussen, Mrutyunjaya Parida, Greg Bonde, Benjamin~W
  Darbro, Xiaojing Hong, Jason~C Clarke, Kevin~A Peterson, James Denegre,
  Michael Schneider, et~al.
\newblock A gene implicated in activation of retinoic acid receptor targets is
  a novel renal agenesis gene in humans.
\newblock {\em Genetics}, 207(1):215--228, 2017.

\bibitem{dong2023greb1l}
Ke~Dong, Chenchen Geng, Xiaohong Zhan, Zhi Sun, Qian Pu, Peng Li, Haiyun Song,
  Guanghui Zhao, and Haidong Gao.
\newblock Greb1l overexpression is associated with good clinical outcomes in
  breast cancer.
\newblock {\em European Journal of Medical Research}, 28(1):510, 2023.

\bibitem{o2016bmp2}
Hannah~L O’Neill, Amy~P Cassidy, Olivia~B Harris, and John~W Cassidy.
\newblock Bmp2/bmpr1a is linked to tumour progression in dedifferentiated
  liposarcomas.
\newblock {\em PeerJ}, 4:e1957, 2016.

\bibitem{pickup2015deletion}
Michael~W Pickup, Laura~D Hover, Yan Guo, Agnieszka~E Gorska, Anna Chytil,
  Sergey~V Novitskiy, Harold~L Moses, and Philip Owens.
\newblock Deletion of the bmp receptor bmpr1a impairs mammary tumor formation
  and metastasis.
\newblock {\em Oncotarget}, 6(26):22890, 2015.

\bibitem{hermawan2023bioinformatics}
Adam Hermawan and Herwandhani Putri.
\newblock Bioinformatics analysis of the genetic and epigenetic alterations of
  bone morphogenetic protein receptors in metastatic breast cancer.
\newblock {\em Biochemical Genetics}, pages 1--27, 2023.

\bibitem{li2022increased}
Xintong Li, Yuandong Xie, Shuoyao Su, Zhe Liu, Jia Zhao, and Dezhong Wen.
\newblock Increased expression of tspear in colorectal cancer predicts poor
  prognosis.
\newblock 2022.

\bibitem{dhakal2021divergent}
A~Dhakal, N~Mladkova, and DM~Blakaj.
\newblock Divergent role of intermediate filaments in clinical outcomes of
  hpv-positive and hpv-negative head and neck squamous cell carcinoma.
\newblock {\em International Journal of Radiation Oncology, Biology, Physics},
  111(3):e368, 2021.

\bibitem{takan2023light}
I{\c{s}}{\i}l Takan, G{\"o}khan Karak{\"u}lah, Aikaterini Louka, and Athanasia
  Pavlopoulou.
\newblock “in the light of evolution:” keratins as exceptional tumor
  biomarkers.
\newblock {\em PeerJ}, 11:e15099, 2023.

\bibitem{zhao2021identifying}
Hongcan Zhao, Danli Sheng, Ze~Qian, Sunyi Ye, Jianzhong Chen, and Zhe Tang.
\newblock Identifying gng4 might play an important role in colorectal cancer
  tmb.
\newblock {\em Cancer Biomarkers}, 32(4):435--450, 2021.

\bibitem{mao2021identification}
Xiao-hong Mao, Qiang Ye, Guo-bing Zhang, Jin-ying Jiang, Hong-ying Zhao,
  Yan-fei Shao, Zi-qi Ye, Zi-xue Xuan, and Ping Huang.
\newblock Identification of differentially methylated genes as diagnostic and
  prognostic biomarkers of breast cancer.
\newblock {\em World journal of surgical oncology}, 19:1--11, 2021.

\bibitem{duan2022g}
Lianhui Duan, Xuefei Liu, Ziwei Luo, Chen Zhang, Chun Wu, Weiping Mu, Zhixiang
  Zuo, Xiaoqing Pei, and Tian Shao.
\newblock G-protein subunit gamma 4 as a potential biomarker for predicting the
  response of chemotherapy and immunotherapy in bladder cancer.
\newblock {\em Genes}, 13(4):693, 2022.

\bibitem{chou2012suppression}
Ruey-Hwang Chou, Hui-Chin Wen, Wei-Guang Liang, Sheng-Chieh Lin, Hsiao-Wei
  Yuan, Cheng-Wen Wu, and Wun-Shaing~Wayne Chang.
\newblock Suppression of the invasion and migration of cancer cells by serpinb
  family genes and their derived peptides.
\newblock {\em Oncology Reports}, 27(1):238--245, 2012.

\bibitem{wei2023serpinb7}
Hua-Fang Wei, Rui-Feng Zhang, Yue-Chen Zhao, and Xian-Shuang Tong.
\newblock Serpinb7 as a prognostic biomarker in cervical cancer: Association
  with immune infiltration and facilitation of the malignant phenotype.
\newblock {\em Heliyon}, 9(9), 2023.

\bibitem{ou2023serpine1}
Junwen Ou, Qiulin Liao, Yanping Du, Wentao Xi, Qiong Meng, Kexin Li, Qichun
  Cai, and Clifford~LK Pang.
\newblock Serpine1 and serpinb7 as potential biomarkers for intravenous vitamin
  c treatment in non-small-cell lung cancer.
\newblock {\em Free Radical Biology and Medicine}, 209:96--107, 2023.

\bibitem{hu2008cancer}
Jianhua Hu.
\newblock Cancer outlier detection based on likelihood ratio test.
\newblock {\em Bioinformatics}, 24(19):2193--2199, 2008.

\bibitem{tang2019signature}
Zhengyi Tang, Ganguan Wei, Longcheng Zhang, and Zhiwen Xu.
\newblock Signature micrornas and long noncoding rnas in laryngeal cancer
  recurrence identified using a competing endogenous rna network.
\newblock {\em Molecular Medicine Reports}, 19(6):4806--4818, 2019.

\bibitem{nwosu2022variable}
Gerald Nwosu, Shilpa~B Reddy, Heather Rose~Mead Riordan, and Jing-Qiong Kang.
\newblock Variable expression of gabaa receptor subunit gamma 2 mutation in a
  nuclear family displaying developmental and encephalopathic phenotype.
\newblock {\em International Journal of Molecular Sciences}, 23(17):9683, 2022.

\bibitem{elgaaen2012znf385b}
Bente~Vilming Elgaaen, Ole~Kristoffer Olstad, Leiv Sandvik, Elin {\O}degaard,
  Torill Sauer, Anne~Cathrine Staff, and Kaare~M Gautvik.
\newblock Znf385b and vegfa are strongly differentially expressed in serous
  ovarian carcinomas and correlate with survival.
\newblock {\em PloS One}, 7:e46317, 2012.

\bibitem{yan2021downregulated}
Ning Yan, Cong Liu, Fang Tian, Ling Wang, Yimin Wang, Zhaoying Yang, Yan Jiao,
  Miao He, et~al.
\newblock Downregulated mrna expression of znf385b is an independent predictor
  of breast cancer.
\newblock {\em International Journal of Genomics}, 2021, 2021.

\bibitem{zhong2022identification}
Zhenhua Zhong, Wenqiang Jiang, Jing Zhang, Zhanwen Li, and Fengfeng Fan.
\newblock Identification and validation of a novel 16-gene prognostic signature
  for patients with breast cancer.
\newblock {\em Scientific Reports}, 12(1):12349, 2022.

\bibitem{zhong2020identification}
Guansheng Zhong, Weiyang Lou, Qinyan Shen, Kun Yu, and Yajuan Zheng.
\newblock Identification of key genes as potential biomarkers for
  triple-negative breast cancer using integrating genomics analysis.
\newblock {\em Molecular Medicine Reports}, 21(2):557--566, 2020.

\bibitem{chen2021multiomics}
Chen Chen, Dan Gao, Jinlong Huo, Rui Qu, Youming Guo, Xiaochi Hu, and Libo Luo.
\newblock Multiomics analysis reveals ct83 is the most specific gene for triple
  negative breast cancer and its hypomethylation is oncogenic in breast cancer.
\newblock {\em Scientific Reports}, 11(1):12172, 2021.

\bibitem{li2022natural}
Qingyang Li, Wei Hu, Baoyi Liao, Chanchan Song, and Liangping Li.
\newblock Natural high-avidity t-cell receptor efficiently mediates regression
  of cancer/testis antigen 83 positive common solid cancers.
\newblock {\em Journal for Immunotherapy of Cancer}, 10(7):e004713, 2022.

\end{thebibliography}

\clearpage
\begin{center}
\textbf{\LARGE Supplementary Materials to the paper\\ ``A Weibull Mixture Cure Frailty Model for High-dimensional Covariates''}
\end{center}




\newcolumntype{L}[1]{>{\raggedright\arraybackslash}p{#1}}
\newcolumntype{C}[1]{>{\centering\arraybackslash}p{#1}}
\newcolumntype{R}[1]{>{\raggedleft\arraybackslash}p{#1}}

\setcounter{figure}{0}
\setcounter{table}{0}
\setcounter{page}{1}

\renewcommand{\thefigure}{S\arabic{figure}}
\renewcommand{\thetable}{S\arabic{table}}

\pagenumbering{Roman}

\section*{S-1 Additional tables for the Simulation Study}
\addcontentsline{toc}{section}{S-1 Additional tables for the Simulation Study}

In Tables \ref{table:Correlation_02} and \ref{table:Correlation_05}, we present the prediction performance for the regression coefficients and the uncured rate estimate performance for the simulated scenarios with correlation $\rho = 0.2$ and $0.5$.

\begin{table}[!h]
	\centering
 \begin{threeparttable}
	\renewcommand{\arraystretch}{0.90}
 \caption{Simulation studies results for $\mathbf{b}_p,\: \bm{\beta}_p$ and $\pi(\mathbf{z})$ when $\rho=0.2$}
  \label{table:Correlation_02}
	\begin{tabular}{ccccccrc} \hline
		\toprule
		\multicolumn{2}{c}{} & \multicolumn{2}{c@{\quad}}{$\bm{\beta}_p$} 
		& \multicolumn{2}{c@{\quad}}{$\mathbf{b}_p$} 
		& \multicolumn{2}{c@{\quad}}{$\pi(\mathbf{z})$} \\ \cmidrule(r){3-4} \cmidrule(r){5-6} \cmidrule(l){7-8}
            $v$ & Method & RME(SD) &ERR (SD) & RME(SD) &ERR (SD) &  Bias & MSE  \\  \hline
    0.5 & A1 & 1.225 (0.254) & 1.227 (0.258) & 0.320 (0.522) & 0.321 (0.522) & \textbf{0.009} & 0.067 \\ 
        & A2 & 1.340 (0.642) & 1.342 (0.647) & 0.330 (0.532) & 0.330 (0.532) & 0.014 & 0.071 \\
        & A3 & 1.415 (0.965) & 1.417 (0.968) & 0.331 (0.534) & 0.332 (0.534) & 0.013 & 0.072 \\
        & A4 & 1.432 (0.984) & 1.434 (0.986) & 0.332 (0.535) & 0.332 (0.535) & 0.013 & 0.072 \\ 
		  & B & 1.191 (0.050) & 1.191 (0.050)  &0.335 (0.535) & 0.335 (0.535)& 0.238 & 0.130 \\
		  & C & 1.428 (0.323) & 1.428 (0.321) &\textbf{0.303} (0.506) & \textbf{0.304} (0.506)& -0.020 & \textbf{0.057} \\ 
		  & D & \textbf{1.125} (0.125) & \textbf{1.127} (0.123) & & & & \\
     &  & & & & & & \\
		 1 & A1 & 0.817  (0.127) & 0.818 (0.126) & \textbf{1.385} (2.632) & \textbf{1.387} (2.633) & \textbf{0.007} & 0.085\\ 
        & A2 & 0.803 (0.147) & 0.804 (0.147)  & 1.513 (2.881) & 1.514 (2.881) & 0.022 & 0.099 \\
        & A3 & \textbf{0.785} (0.157) & \textbf{0.786} (0.156) & 1.517 (2.922) & 1.518 (2.922) & 0.024 & 0.100 \\ 
        & A4 & 0.789 (0.161) & 0.790 (0.160) & 1.512 (2.909) & 1.513 (2.909) & 0.023 & 0.101 \\ 
		  & B & 1.059 (0.028) & 1.059 (0.027)  & 1.816 (3.405) & 1.816 (3.405) & 0.129 & 0.156 \\ 
		  & C & 1.082 (0.051) & 1.084 (0.051)  & 1.411 (2.919) & 1.413 (2.920) & -0.036 & \textbf{0.082} \\ 
		  & D & 0.851 (0.087) & 0.852 (0.086) & & & & \\
     &  & & & & & & \\
		 1.5 & A1 & 0.727 (0.089) & 0.728 (0.089) & 2.040 (4.292)& 2.041 (4.292) & \textbf{0.009} & \textbf{0.078} \\ 
          & A2 & 0.654 (0.129) & 0.654 (0.129) & \textbf{2.033} (4.343) & \textbf{2.034} (4.343) & 0.026 & 0.089 \\ 
         & A3 & \textbf{0.634} (0.132) & \textbf{0.635 }(0.132) & 2.106 (4.651) & 2.107 (4.652) & 0.036 & 0.095 \\ 
         & A4 & 0.638 (0.133) & 0.638 (0.133)  & 2.152 (4.839) & 2.153 (4.839) & 0.038 & 0.097 \\ 
		  & B & 1.024 (0.023) & 1.024 (0.023)  & 2.681 (5.605) & 2.681 (5.605)& 0.090 & 0.178 \\ 
		  & C & 1.017 (0.027) & 1.018 (0.026) & 2.257 (4.878)& 2.258 (4.880) & -0.038 & 0.093 \\ 
		  & D & 0.808 (0.054) & 0.808 (0.054) & & & & \\ 
     &  & & & & & & \\
		 2 & A1 & 0.752 (0.066) & 0.753 (0.066) & \textbf{3.922} (9.992)& \textbf{3.924} (10.000) & \textbf{0.004} & \textbf{0.075} \\ 
        & A2 & 0.669 (0.094) & 0.669 (0.094)& 4.056 (10.551) & 4.056 (10.552) & 0.032 & 0.093 \\ 
        & A3 & 0.648 (0.104) & 0.648 (0.104) & 4.047 (10.497) & 4.048 (10.498) & 0.049 & 0.105 \\ 
        & A4 & \textbf{0.646} (0.103) & \textbf{0.646} (0.102) & 4.027 (10.440) & 4.028 (10.441) & 0.044 & 0.101 \\ 
		  & B & 1.012  (0.016) & 1.012 (0.016) & 4.934 (12.511)& 4.935 (12.513) & 0.063 & 0.174 \\ 
		  & C & 1.010  (0.017) & 1.011 (0.016) & 4.190 (10.397) & 4.192 (10.405) & -0.039 & 0.104 \\
		  & D & 0.831  (0.035) & 0.832 (0.035) & & & & \\
     &  & & & & & & \\
		 2.5 & A1 & 0.801 (0.052) & 0.802 (0.052) & \textbf{2.764} (7.826) & \textbf{2.764} (7.828) & \textbf{-0.003} & \textbf{0.076} \\
        & A2 & 0.737 (0.069) & 0.737 (0.069) & 2.790 (8.113) & 2.790 (8.113) & 0.012 & 0.082 \\
        & A3 & \textbf{0.715} (0.076) & \textbf{0.715} (0.076) & 2.815 (8.181) & 2.816 (8.181) & 0.023 & 0.089 \\
        & A4 & 0.717 (0.077) & 0.717 (0.077) & 2.805 (8.142) & 2.805 (8.142) & 0.019 & 0.087 \\ 
		  & B & 1.014  (0.012) & 1.014 (0.012) & 3.422 (9.565) & 3.422 (9.565) & 0.079 & 0.210 \\
		  & C & 1.011  (0.013) & 1.011 (0.013) & 3.064 (8.548) & 3.066 (8.554) & -0.036 & 0.117 \\ 
		  & D & 0.869  (0.026) & 0.869 (0.026) & & & & \\  \hline

  	\end{tabular}
   \begin{tablenotes}
    \item \footnotesize Method A1-A4: penMCFM (EM) for $\alpha_{Enet}=0.1,0.5,0.9,1$, B: penMCFM (GMIFS), C: MCM (GMIFS), D: penCox.1se; The best result appears in bold.
    \end{tablenotes}
 \end{threeparttable}
\end{table}

\begin{table}[!h]
	\centering
 \begin{threeparttable}
	\renewcommand{\arraystretch}{0.90}
 \caption{Simulation studies results for $\mathbf{b}_p,\: \bm{\beta}_p$ and $\pi(\mathbf{z})$ when $\rho=0.5$}
  \label{table:Correlation_05}
	\begin{tabular}{ccccccrc} \hline
		\toprule
		\multicolumn{2}{c}{} & \multicolumn{2}{c@{\quad}}{$\bm{\beta}_p$} 
		& \multicolumn{2}{c@{\quad}}{$\mathbf{b}_p$} 
		& \multicolumn{2}{c@{\quad}}{$\pi(\mathbf{z})$} \\ \cmidrule(r){3-4} \cmidrule(r){5-6} \cmidrule(l){7-8}
            $v$ & Method & RME(SD) &ERR (SD) & RME(SD) &ERR (SD) &  Bias & MSE  \\  \hline
    0.5 & A1 & 1.232 (0.244)& 1.255 (0.270) & 0.396 (0.906) & 0.402 (0.923) & \textbf{0.007} & 0.067 \\ 
        & A2 & 1.514 (0.950) & 1.549 (1.001)  & 0.421 (0.927) & 0.426 (0.934) & 0.012 & 0.070 \\
        & A3 & 1.556 (1.288) & 1.593 (1.352) & 0.431 (0.940)& 0.438 (0.954) & 0.013 & 0.071 \\
        & A4 & 1.454 (0.869) & 1.485 (0.918) & 0.420 (0.938) & 0.426 (0.950) & 0.012 & 0.071 \\
		  & B & \textbf{1.202} (0.049) & \textbf{1.203} (0.049)& 0.421 (0.959) & 0.421 (0.961) & 0.232 & 0.127 \\ 
		  & C & 1.528 (0.359)& 1.530 (0.349)& \textbf{0.366} (0.847) & \textbf{0.379} (0.875) & -0.022 & \textbf{0.056} \\ 
		   & D & 1.424 (1.859) & 1.429 (1.816) & & & & \\ 
      &  & & & & & & \\
	   1 & A1 & 0.812 (0.119) & 0.831 (0.111) & 1.401 (2.072) & 1.423 (2.103) & \textbf{0.010} & 0.084 \\ 
        & A2 & 0.798 (0.149) & 0.819 (0.145)& 1.432 (2.093) & 1.451 (2.124) & 0.021 & 0.094 \\ 
        & A3 & 0.791 (0.151) & 0.811 (0.145) & 1.460 (2.134) & 1.481 (2.165) & 0.026 & 0.099 \\ 
        & A4 & \textbf{0.788} (0.151) & \textbf{0.809} (0.145) & 1.457 (2.129) & 1.478 (2.162) & 0.028 & 0.100 \\ 
		  & B & 1.061 (0.030)& 1.062 (0.028) & 1.795 (2.638)& 1.797 (2.640) & 0.115 & 0.154 \\ 
		  & C & 1.064 (0.048)& 1.078 (0.048) & \textbf{1.332} (1.879) & \textbf{1.366} (1.942) & -0.032 & \textbf{0.079} \\
		  & D & 0.829 (0.083) & 0.841 (0.081) & & & & \\  
     &  & & & & & & \\
		 1.5 & A1 & 0.714 (0.080) & 0.726 (0.078) & 2.658 (6.115) & 2.675 (6.123) & \textbf{0.011} & \textbf{0.078} \\ 
         & A2 & 0.634 (0.108) & 0.644 (0.108) & \textbf{2.652} (5.995) & \textbf{2.668} (6.003) & 0.035 & 0.092 \\ 
         & A3 & \textbf{0.620} (0.119) & \textbf{0.629} (0.119) & 2.713 (5.951) & 2.727 (5.955) & 0.043 & 0.100 \\ 
         & A4 & 0.625 (0.124) & 0.634 (0.123) & 2.721 (5.941) & 2.735 (5.945) & 0.044 & 0.103 \\ 
		  & B & 1.022 (0.019)& 1.024 (0.018) & 3.467 (7.430) & 3.469 (7.430) & 0.086 & 0.175 \\ 
		  & C & 1.010 (0.029) & 1.020 (0.024)& 2.955 (7.087) & 2.972 (7.092) & -0.037 & 0.091 \\ 
		  & D & 0.800 (0.054) & 0.808 (0.052) & & & & \\ 
     &  & & & & & & \\
		 2 & A1 & 0.758 (0.062) & 0.766 (0.062) & 3.348 (6.477) & 3.367 (6.511) & \textbf{0.011} & \textbf{0.082} \\ 
        & A2 & 0.668 (0.087) & 0.674 (0.087)& \textbf{3.337} (6.545) & \textbf{3.353} (6.570) & 0.051 & 0.107 \\ 
        & A3 & 0.649 (0.086) & 0.654 (0.087) & 3.352 (6.573) & 3.366 (6.592) & 0.062 & 0.113 \\ 
        & A4 & \textbf{0.642} (0.089) & \textbf{0.647} (0.089) & 3.344 (6.556) & 3.358 (6.575) & 0.063 & 0.113 \\
		  & B & 1.014 (0.017) & 1.017 (0.016) & 4.113 (7.934) & 4.119 (7.943) & 0.056 & 0.186 \\ 
		  & C & 1.003 (0.020)& 1.011 (0.018)& 3.721 (7.153) & 3.745 (7.205) & -0.037 & 0.102 \\ 
		    & D & 0.828 (0.036) & 0.833 (0.035) & & & & \\ 
       &  & & & & & & \\
		 2.5 & A1 & 0.800 (0.051) & 0.808 (0.049) & 3.588 (8.465) & 3.606 (8.510)& \textbf{0.002} & \textbf{0.078} \\ 
          & A2 & 0.734 (0.058) & 0.738 (0.059) & 3.501 (8.299) & 3.516 (8.323) & 0.024 & 0.089 \\ 
         & A3 & 0.712 (0.072) & 0.716 (0.072) & 3.457 (8.289) & 3.473 (8.306) & 0.037 & 0.098 \\ 
         & A4 & \textbf{0.705} (0.072) & \textbf{0.709} (0.073) & \textbf{3.429} (8.211) & \textbf{3.445} (8.228) & 0.042 & 0.101 \\ 
		   & B & 1.014 (0.013) & 1.016 (0.011) & 4.476 (10.818) & 4.479 (10.820) & 0.080 & 0.212 \\ 
		   & C & 1.005 (0.015) & 1.011 (0.012)& 3.909 (9.191)& 3.931 (9.252) & -0.037 & 0.109 \\ 
		   & D & 0.865 (0.028) & 0.868 (0.027) & & & & \\  \hline
    
    	\end{tabular}
   \begin{tablenotes}
    \item \footnotesize Method A1-A4: penMCFM (EM) for $\alpha_{Enet}=0.1,0.5,0.9,1$, B: penMCFM (GMIFS), C: MCM (GMIFS), D: penCox.1se; The best result appears in bold.
    \end{tablenotes}
 \end{threeparttable}
\end{table}

\clearpage

\section*{S-2 Additional figures showing results of the analysis of TCGA-BRCA RNA-Seq data}
\addcontentsline{toc}{section}{S-2 Additional figures showing results of the analysis of TCGA-BRCA RNA-Seq data}

\begin{figure}[!h]
     \centering    
     \begin{subfigure}[b]{0.49\textwidth}
         \centering
         \includegraphics[width=\textwidth]{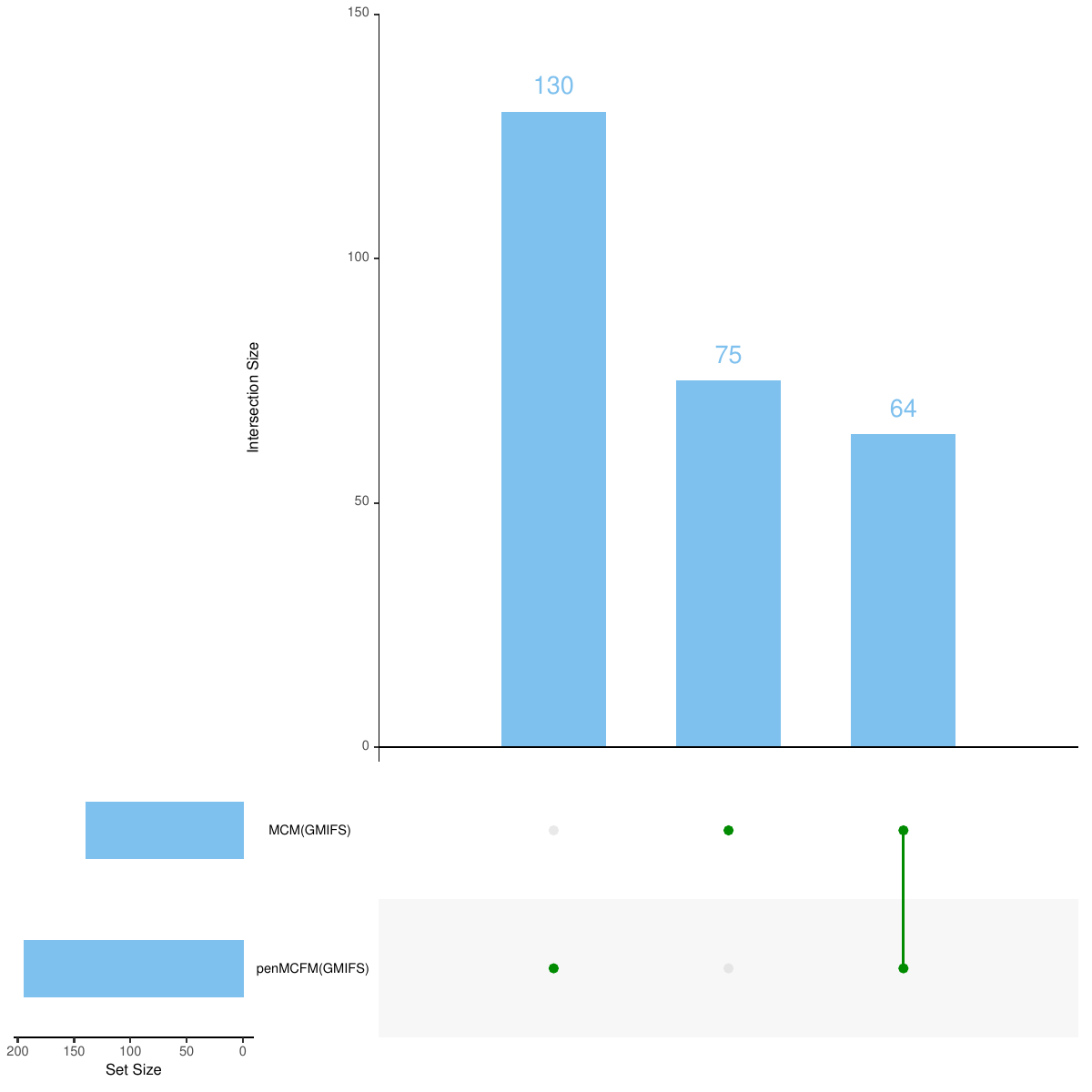}
         \caption{Scenario $1$}
         \label{upset_plot_Scenario_1}
     \end{subfigure}
     \hfill
     \begin{subfigure}[b]{0.49\textwidth}
         \centering
         \includegraphics[width=\textwidth]{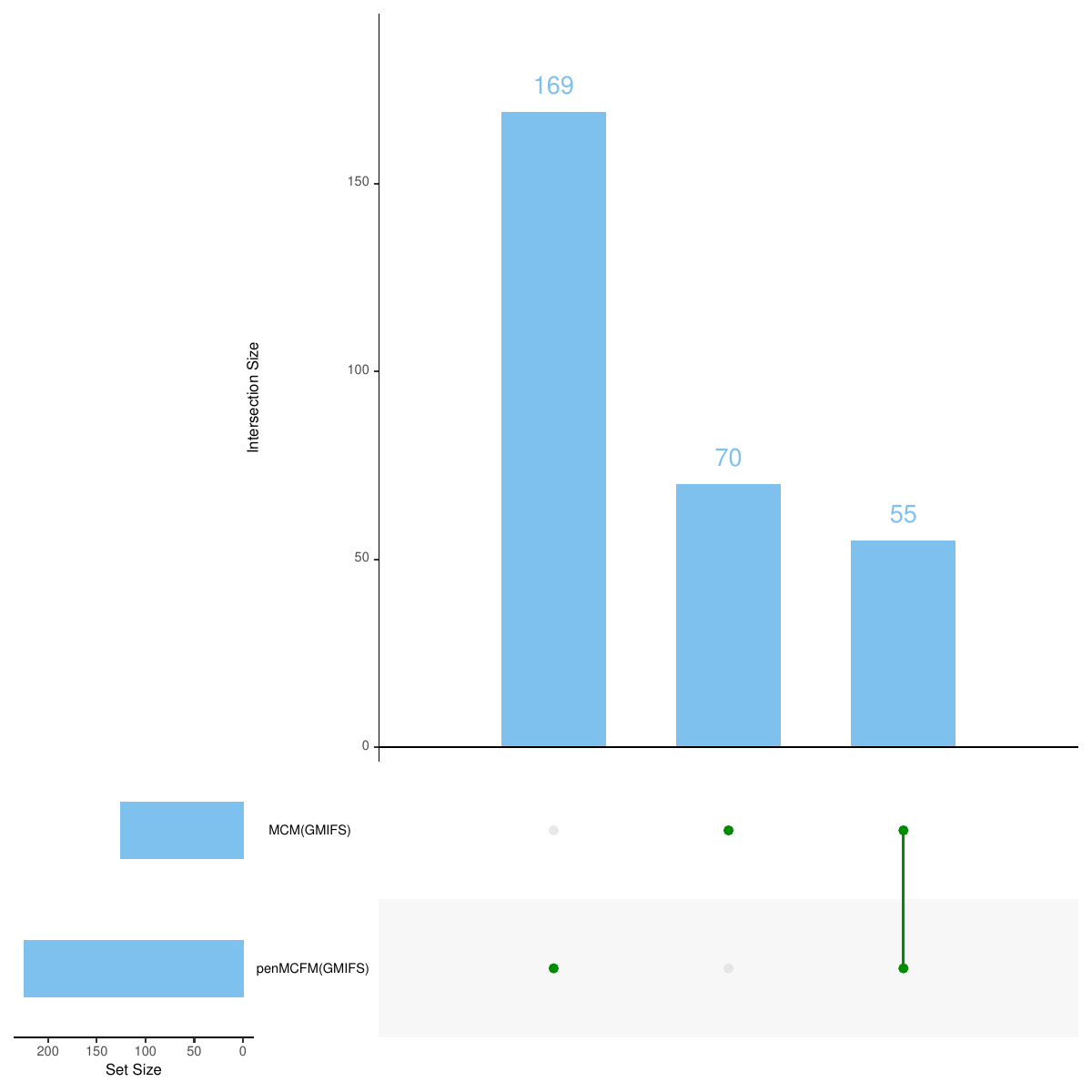}
         \caption{Scenario $2$}
         \label{upset_plot_Scenario_2}
     \end{subfigure}
     \caption{Results of the analysis of the TCGA-BRCA data. Overlap of the selected gene sets (nonzero $\mathbf{b}_p$ coefficients) between two GMIFS methods: the blue barplots report the frequencies of intersections between the methods, while the bottom green lines report which methods are considered for the overlap}
     \label{upset_plots_bp}
 \end{figure}

In Figure \ref{upset_plots_bp}, we present the overlap of the selected genes for the incidence part of the model across the two GMIFS methods.

\subsection*{S-2.1 Results for Scenario $1$}
\addcontentsline{toc}{subsection}{S-2.1 Results for Scenario $1$}

\subsubsection*{S-2.1.1 penMCFM(GMIFS): enrichment analysis results}

\begin{figure}[!h]
     \centering
     \begin{subfigure}[b]{0.49\textwidth}
         \centering
         \includegraphics[width=\textwidth]{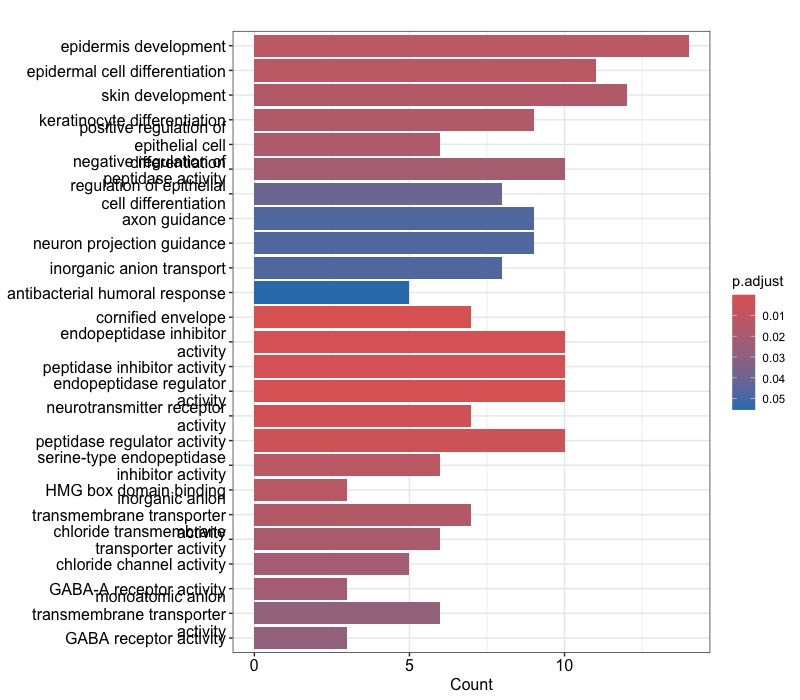}
         \caption{}
         \label{}
     \end{subfigure}
          \begin{subfigure}[b]{0.49\textwidth}
         \centering
         \includegraphics[width=\textwidth]{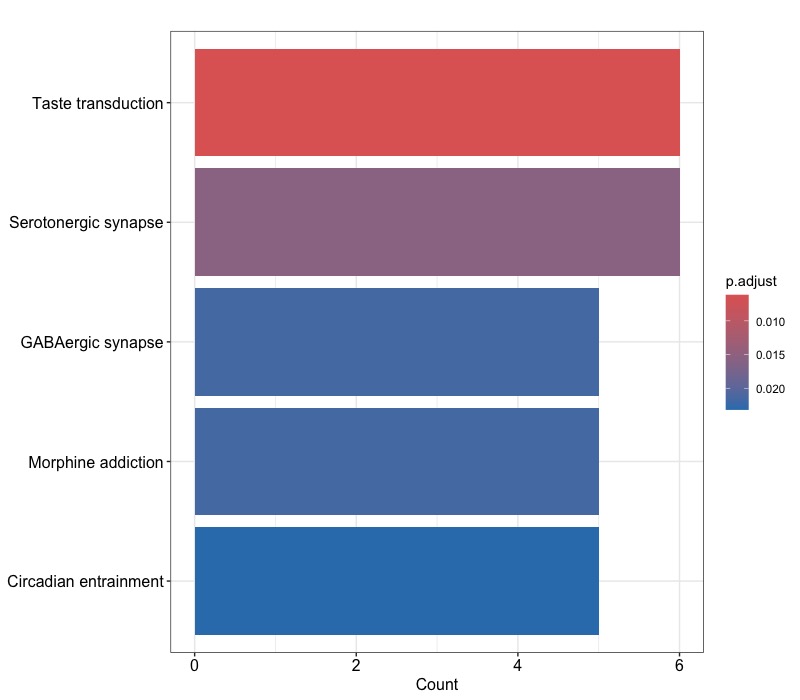}
         \caption{}
         \label{}
     \end{subfigure}
      \caption{Results of the analysis of the TCGA-BRCA data. GO-EA and KEGG-EA for the incidence part of the model based on penMCFM(GMIFS): (a) Barplot of significantly enriched GO terms (b) Barplot of significantly enriched KEGG terms}
    \label{bp_penMCFM_gmifs_Scenario1}
\end{figure}

In Figure \ref{bp_penMCFM_gmifs_Scenario1}, we present GO-EA and KEGG-EA for the incidence part of the model based on penMCFM(GMIFS). For the latency part, we have only one molecular function identified through GO-EA, namely the ``histone deacetylase binding (GO:0042826)''. On the other hand, no significant enrichment terms are identified in KEGG-EA.

\subsubsection*{S-2.1.2 MCM(GMIFS): enrichment analysis results}

\begin{figure}[!h]
     \begin{subfigure}[b]{0.49\textwidth}
     \centering
         \includegraphics[width=\textwidth]{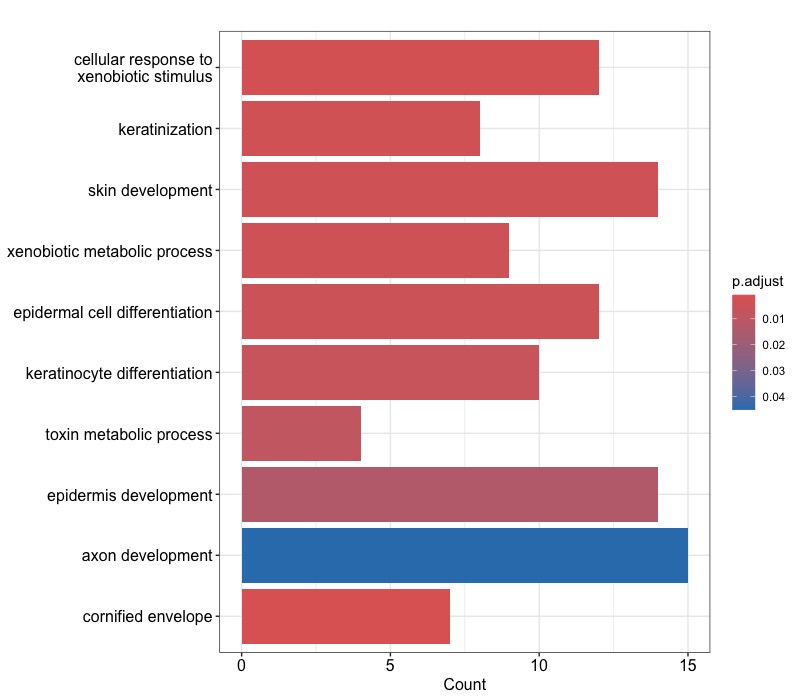}         
         \caption{}
         \label{}
     \end{subfigure}
          \begin{subfigure}[b]{0.49\textwidth}
         \centering
         \includegraphics[width=\textwidth]{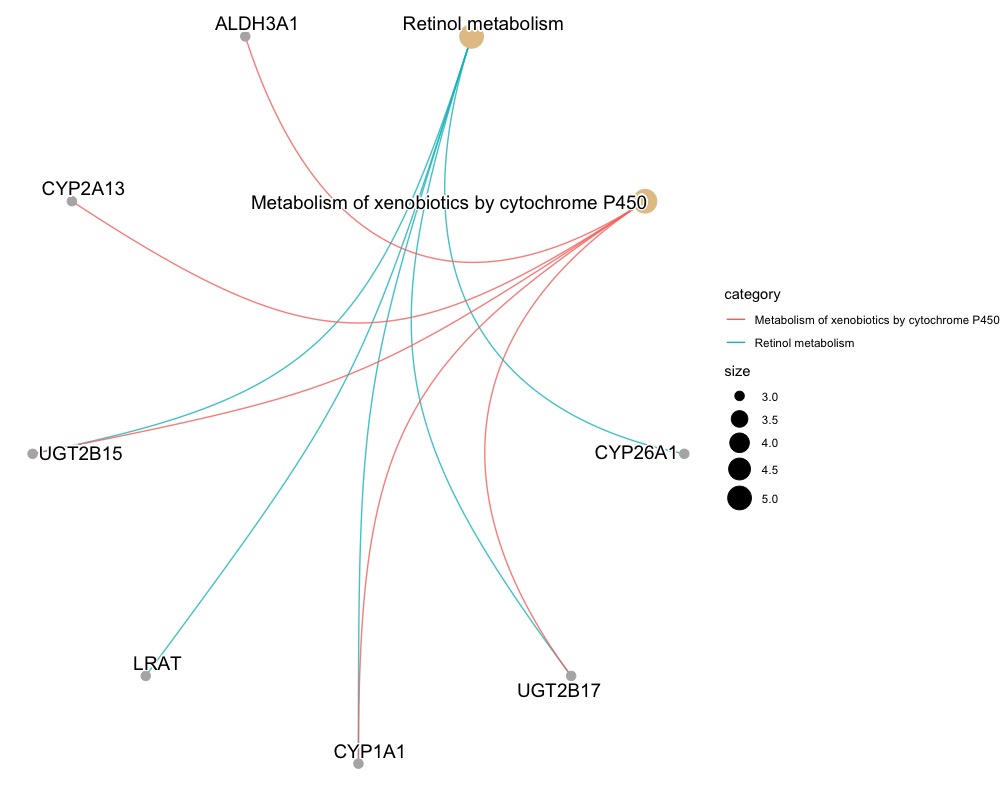}
         \caption{}
         \label{}
     \end{subfigure}
      \caption{Results of the analysis of the TCGA-BRCA data. GO-EA and KEGG-EA for the latency part of the model based on MCM(GMIFS): 
      (a) Barplot of significantly enriched GO terms (b)  Network plot of enriched KEGG pathway terms and related selected genes}
    \label{betap_MCM_gmifs_Scenario1}
\end{figure}

For the incidence part, we detect only one significant enriched GO term, namely the ``GO:0001533 cornified envelope'', and no significant enrichment terms are identified in KEGG-EA.
For the latency part, the significantly enriched GO-EA and KEGG-EA pathway terms are given in Figure \ref{betap_MCM_gmifs_Scenario1}.

\subsubsection*{S-2.1.3 penCox.1se: enrichment analysis results}

When focusing on the selected genes based on the penCox.1se method, no significant enrichment terms are identified in either GO-EA or KEGG-EA. 

\clearpage
 
\subsection*{S-2.2 Results for Scenario $2$}
\addcontentsline{toc}{subsection}{S-2.2 Results for Scenario $2$}

\subsubsection*{S-2.2.1 penMCFM(EM): additional enrichment analysis results}
\addcontentsline{toc}{subsubsection}{S-2.2.1 penMCFM(EM): additional enrichment analysis results}

\begin{figure}[!h]
     \centering    
     \begin{subfigure}[b]{0.48\textwidth}
         \centering
         \includegraphics[width=\textwidth]{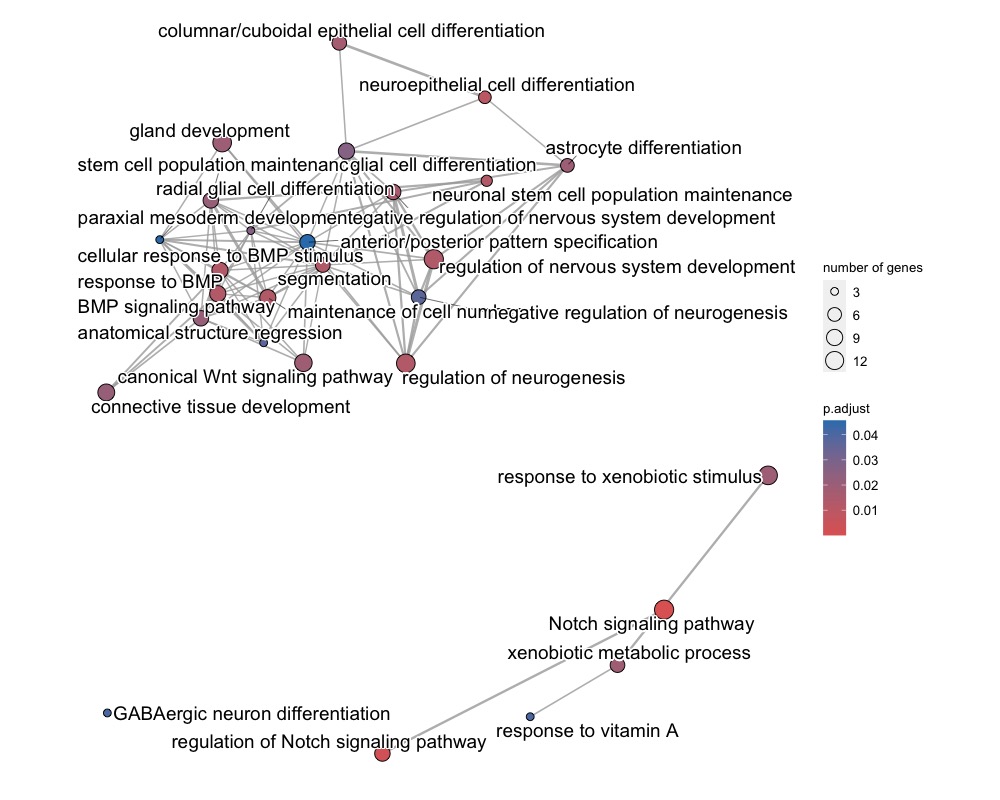}
         \caption{}
         \label{??}
     \end{subfigure}
          \hfill
     \begin{subfigure}[b]{0.51\textwidth}
         \centering
         \includegraphics[width=\textwidth]{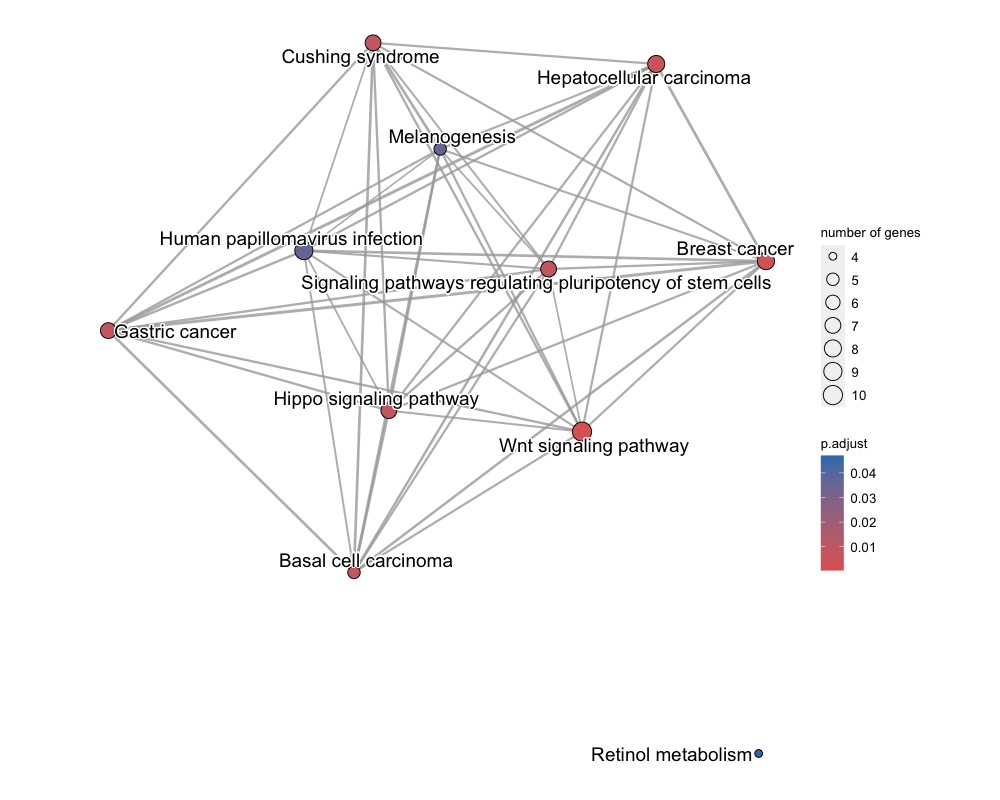}
         \caption{}
         \label{??}
     \end{subfigure}
     \caption{Results of the analysis of the TCGA-BRCA data. (a) Network plot of enriched GO terms (b) Network plot of enriched KEGG pathway terms}
       \label{betap_penMCFM_EM_Scenario2}
 \end{figure}

\begin{figure}[!h]
         \centering
         \includegraphics[width=0.95\textwidth]{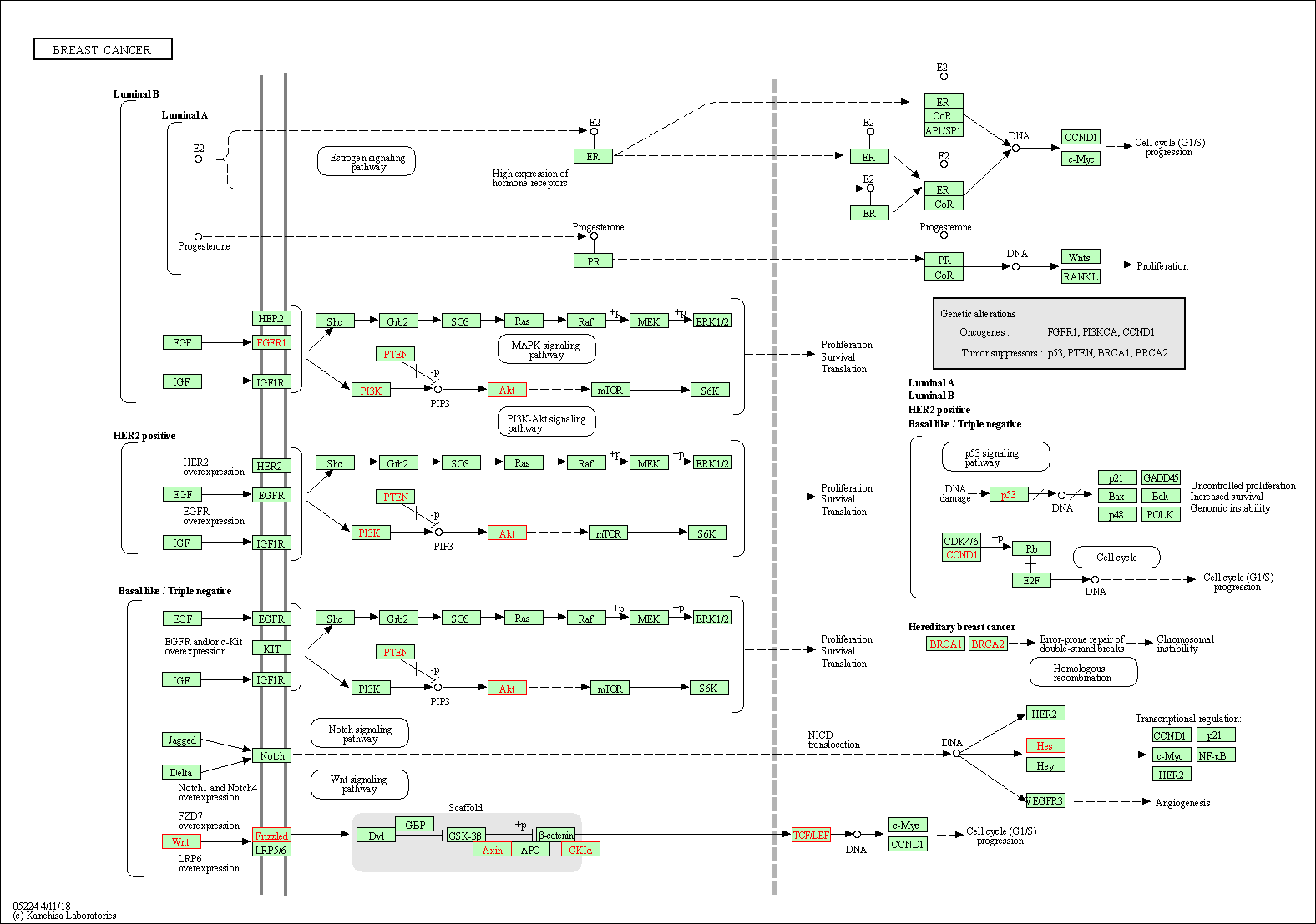}
    \caption{Results of the analysis of the TCGA-BRCA data. The Breast Cancer pathway is one of the top pathways enriched by selected biomarkers obtained when using penMCFM(EM).}
     \label{BRCA_pathway_betap_penMCFM_EM_Scenario2}
\end{figure}

Additional EA results and obtained breast cancer pathway enriched by selected biomarkers for the penMCFM(EM) are presented in Figures \ref{betap_penMCFM_EM_Scenario2}-\ref{BRCA_pathway_betap_penMCFM_EM_Scenario2}. 


\subsubsection*{S-2.2.2 penMCFM(GMIFS): enrichment analysis results}
\addcontentsline{toc}{subsubsection}{S-2.2.2 penMCFM(GMIFS): enrichment analysis results}

\begin{figure}[!h]
     \centering
         \includegraphics[width=0.5\textwidth]{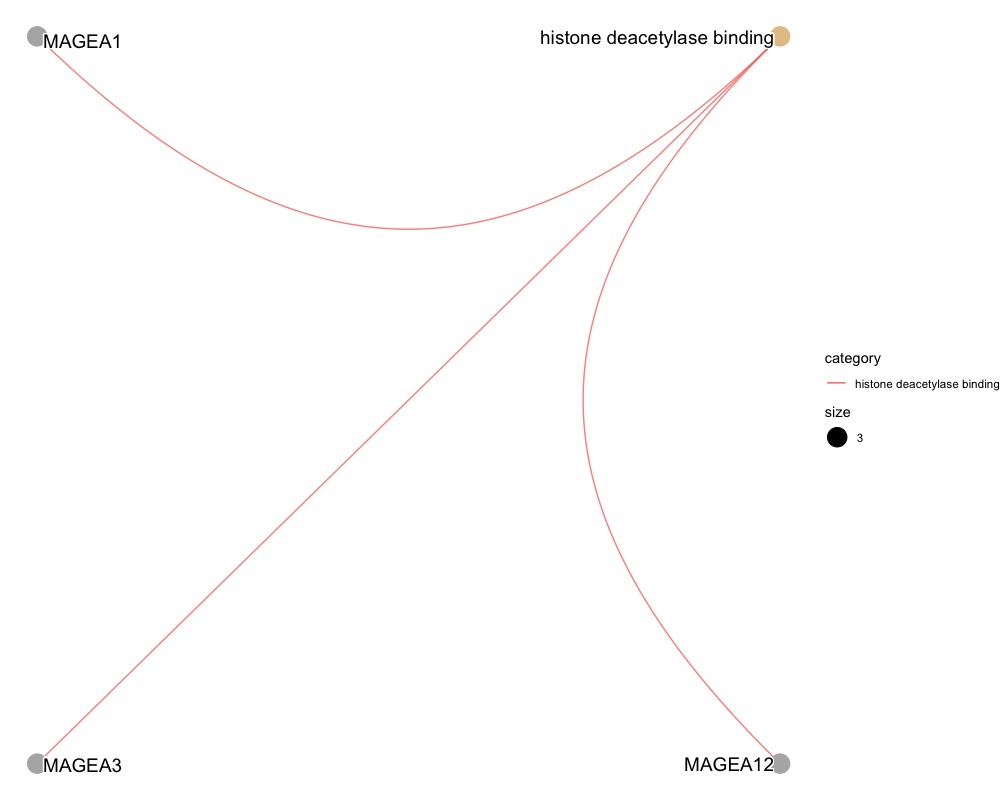}      
      \caption{Results of the analysis of the TCGA-BRCA data. Network plot of enriched GO terms and related selected genes for penMCFM(GMIFS)}
    \label{go_betap_penMCFM_gmifs_Scenario2}
\end{figure}

\begin{figure}[!h]
     \centering
     \begin{subfigure}[b]{0.49\textwidth}
         \centering
         \includegraphics[width=\textwidth]{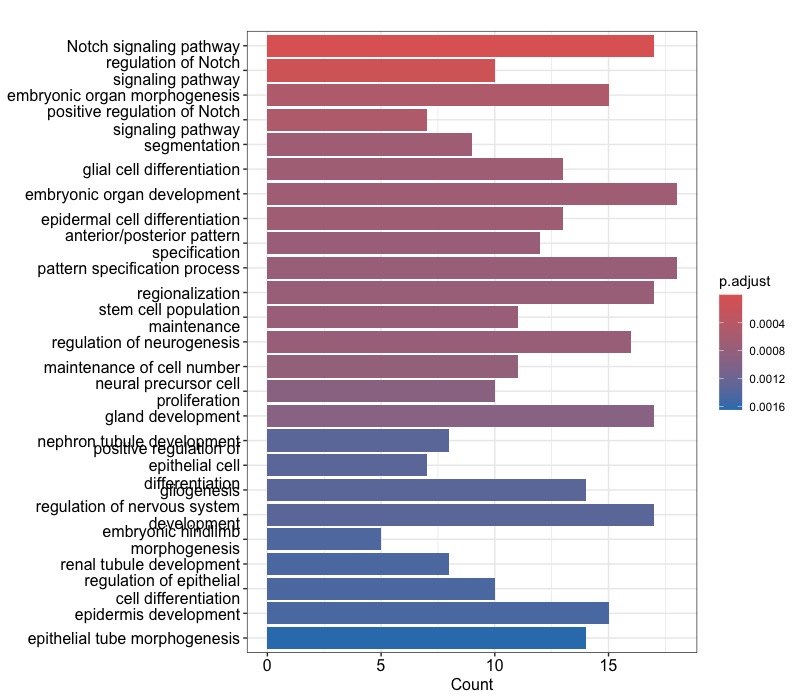}
         \caption{}
         \label{}
     \end{subfigure}
          \begin{subfigure}[b]{0.49\textwidth}
         \centering
         \includegraphics[width=\textwidth]{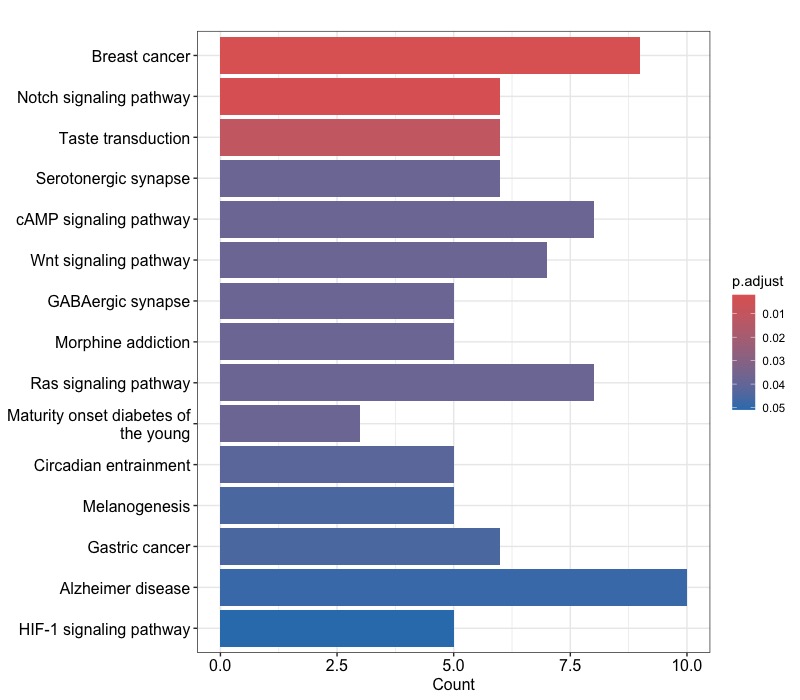}
         \caption{}
         \label{}
     \end{subfigure}
      \caption{Results of the analysis of the TCGA-BRCA data. GO-EA and KEGG-EA for the incidence part of the model based on penMCFM(GMIFS): (a) Barplot of significantly enriched GO terms (b) Barplot of significantly enriched KEGG terms}
    \label{go_kegg_bp_penMCFM_gmifs_Scenario2}
\end{figure}

\begin{figure}[!h]
         \centering
         \includegraphics[width=1\textwidth]{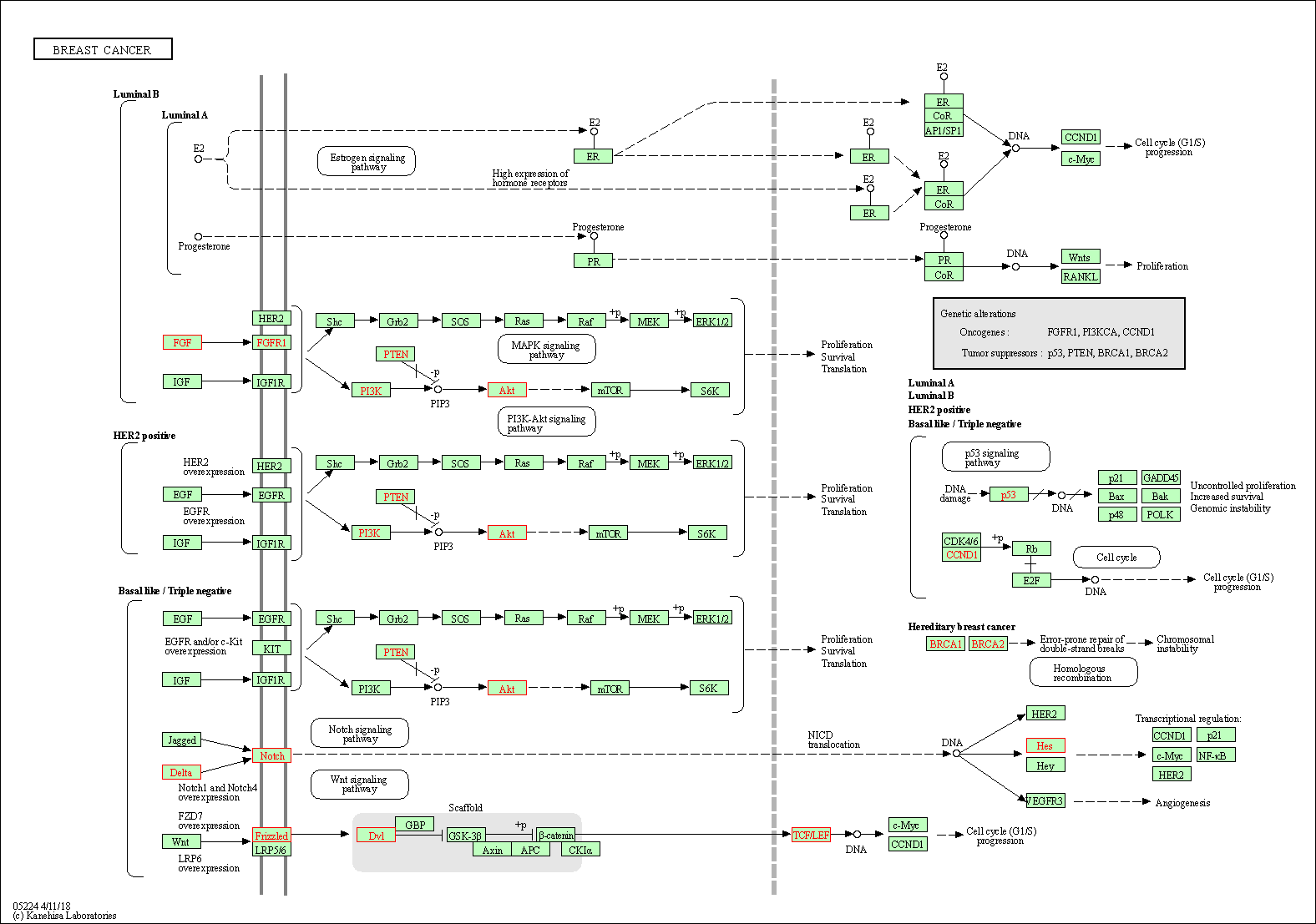}
    \caption{Results of the analysis of the TCGA-BRCA data. The Breast Cancer pathway is the one of top pathways enriched by selected biomarkers obtained from the incidence part of the model when using penMCFM(GMIFS).}
     \label{BRCA_pathway_bp_penMCFM_GMIFS_Scenario2}
\end{figure}

For the latency part, the network plot of enriched GO terms and related selected genes is presented in Figure \ref{go_betap_penMCFM_gmifs_Scenario2}. We have not identified any enriched KEGG-EA pathway terms. 
For the incidence part, GO-EA and KEGG-EA results are given in \ref{go_kegg_bp_penMCFM_gmifs_Scenario2}, and also the breast cancer pathway is presented in Figure \ref{BRCA_pathway_bp_penMCFM_GMIFS_Scenario2}.

\newpage

\subsubsection*{S-2.2.3 MCM(GMIFS): enrichment analysis results}
\addcontentsline{toc}{subsubsection}{S-2.2.3 MCM(GMIFS): enrichment analysis results}

\begin{figure}[!h]
     \centering
     \begin{subfigure}[b]{0.49\textwidth}
         \centering
         \includegraphics[width=\textwidth]{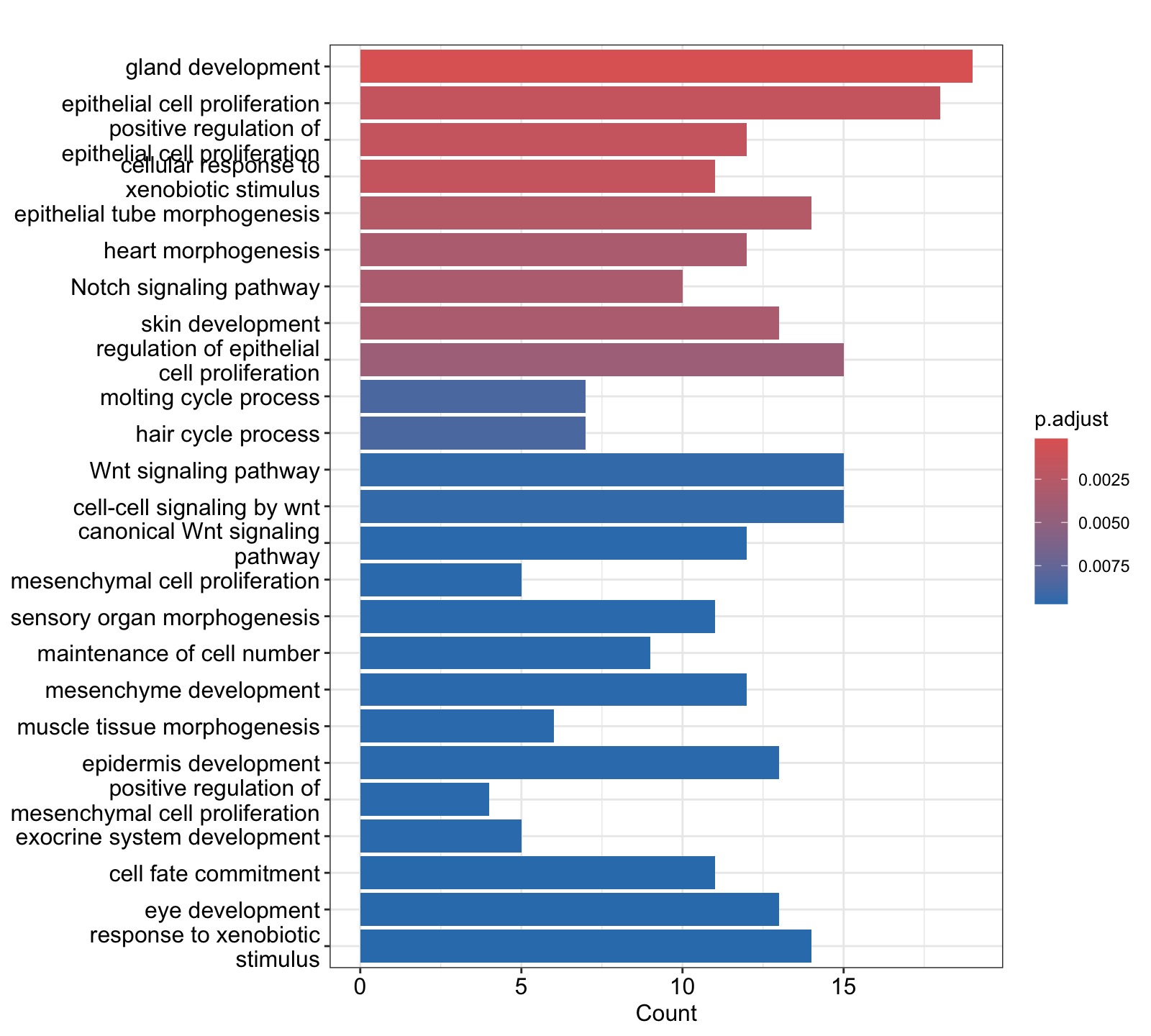}
         \caption{}
         \label{}
     \end{subfigure}
          \begin{subfigure}[b]{0.49\textwidth}
         \centering
         \includegraphics[width=\textwidth]{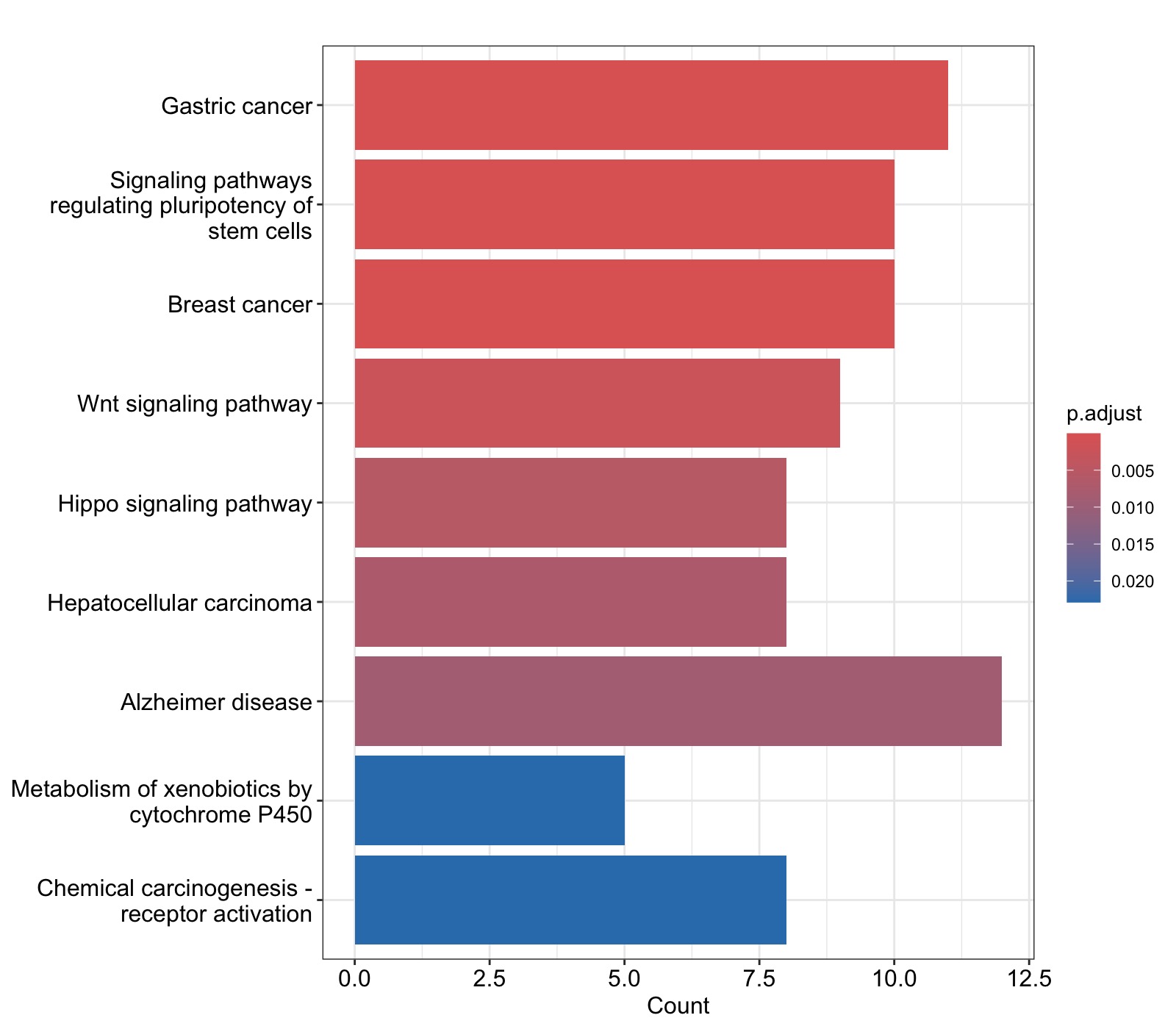}
         \caption{}
         \label{}
     \end{subfigure}
      \caption{Results of the analysis of the TCGA-BRCA data. GO-EA and KEGG-EA for the latency part of the model based on MCM(GMIFS): (a) Barplot of significantly enriched GO terms (b)  Barplot of significantly enriched KEGG terms}
    \label{betap_MCM_gmifs_Scenario2}
\end{figure}

\begin{figure}[!h]
         \centering
         \includegraphics[width=1\textwidth]{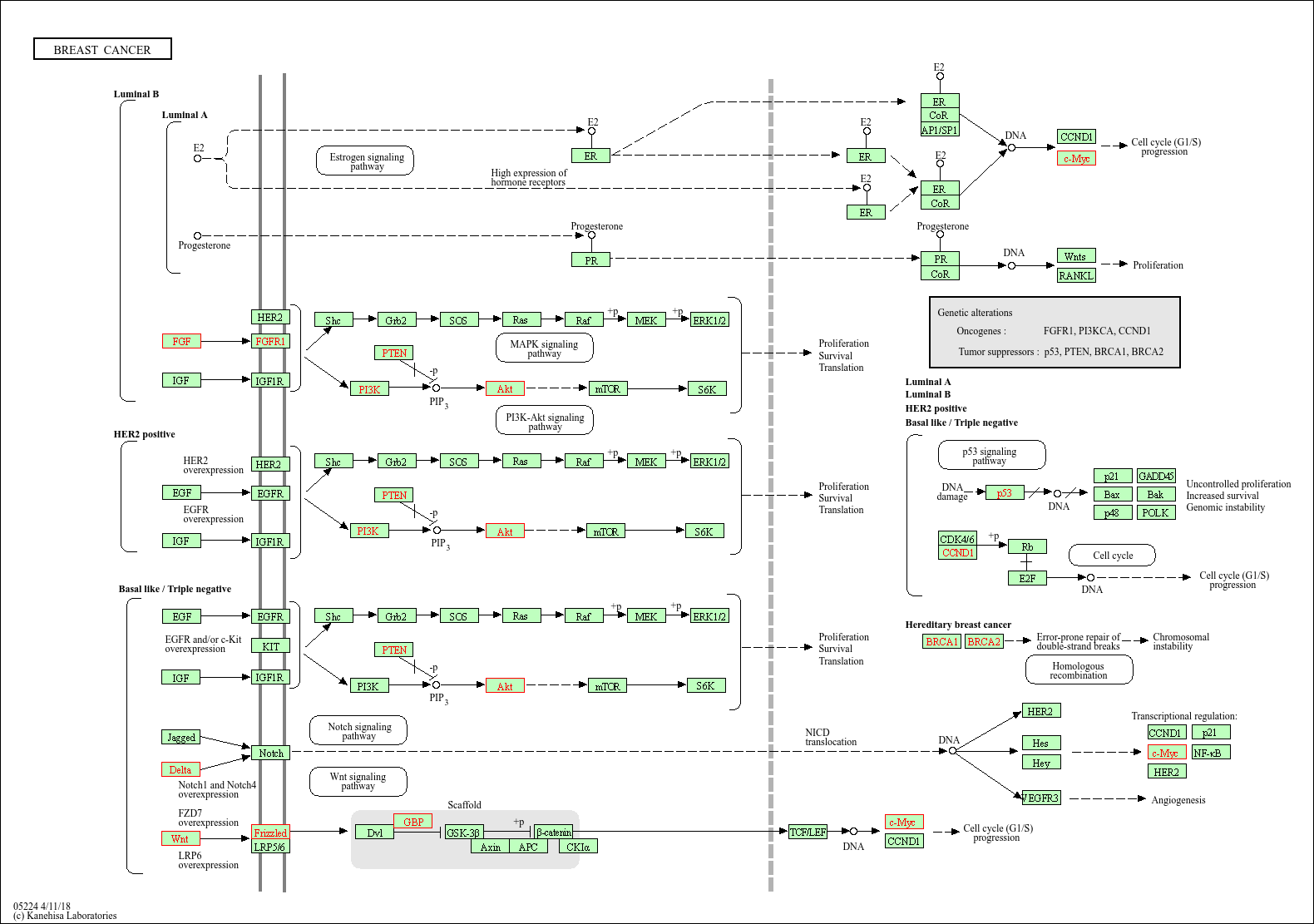}
    \caption{Results of the analysis of the TCGA-BRCA data. The Breast Cancer pathway is the one of top pathways enriched by selected biomarkers obtained from the latency part of the model when using MCM(GMIFS).}
     \label{BRCA_pathway_betap_MCM_GMIFS_Scenario2}
\end{figure}

\begin{figure}[!h]
     \centering
    \includegraphics[width=0.5\textwidth]{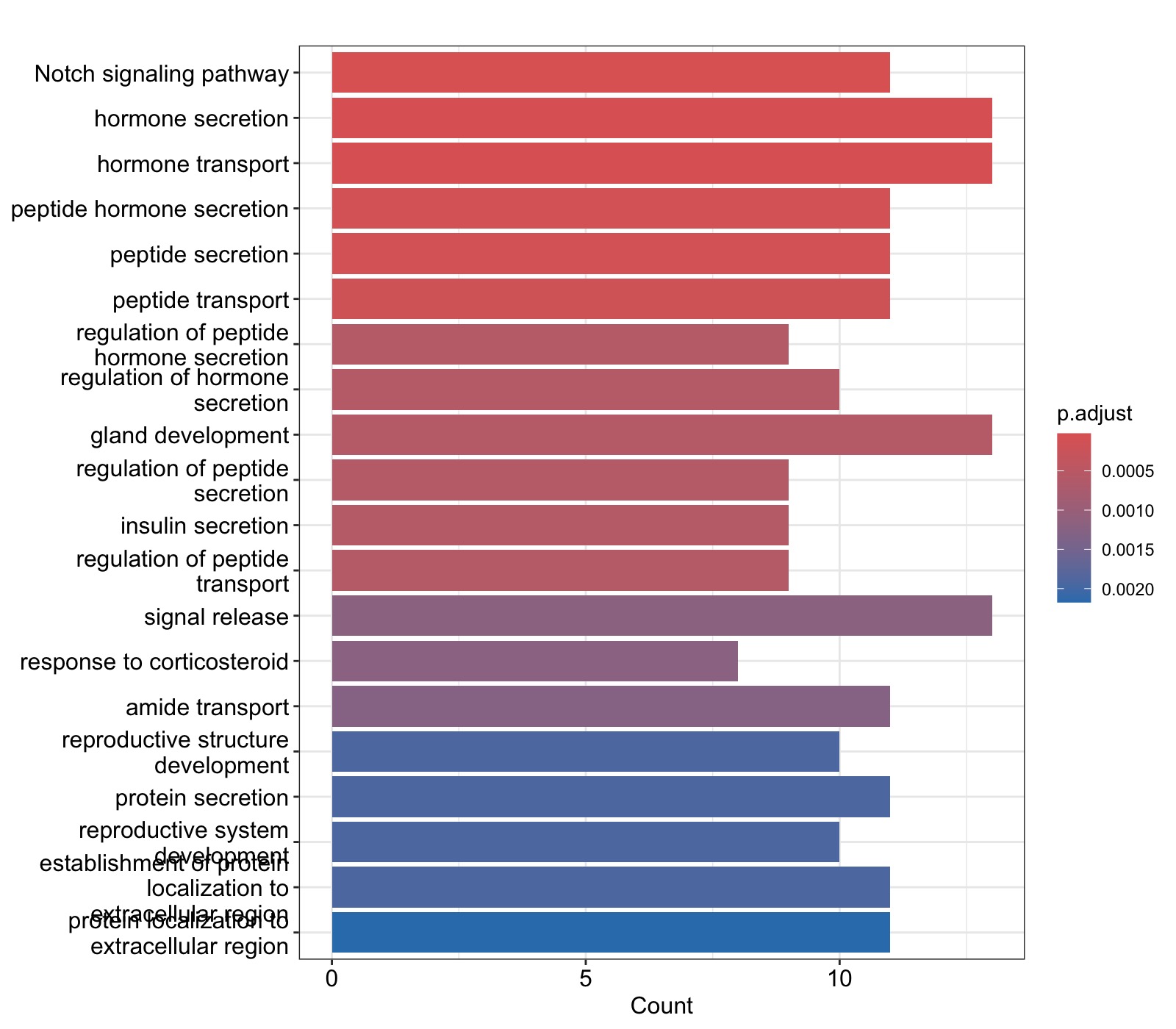}
     \caption{Results of the analysis of the TCGA-BRCA data. GO-EA for the incidence part of the model based on MCM(GMIFS): Barplot of significantly enriched GO terms}
    \label{go_bp_MCM_gmifs_Scenario2}
\end{figure}

For the latency part, GO-EA and KEGG-EA results, and the obtained breast cancer pathway are presented in Figures \ref{betap_MCM_gmifs_Scenario2}-\ref{BRCA_pathway_betap_MCM_GMIFS_Scenario2}.
For the incidence part, the barplot of significantly enriched GO terms is given in Figure \ref{go_bp_MCM_gmifs_Scenario2}. We have not identified any enriched KEGG-EA pathway terms.

\newpage
\hspace*{2cm}

\subsubsection*{S-2.2.4 penCox.1se: enrichment analysis results}
\addcontentsline{toc}{subsubsection}{S-2.2.4 penCox.1se: enrichment analysis results}

\begin{figure}[!h]
     \centering
     \begin{subfigure}[b]{0.49\textwidth}
         \centering
         \includegraphics[width=\textwidth]{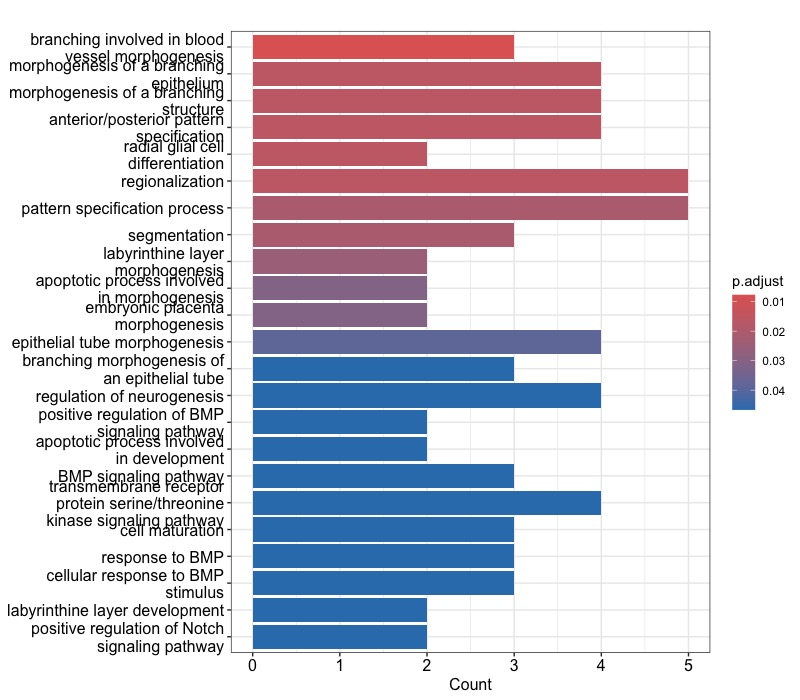}
         \caption{}
         \label{}
     \end{subfigure}
          \begin{subfigure}[b]{0.49\textwidth}
         \centering
         \includegraphics[width=\textwidth]{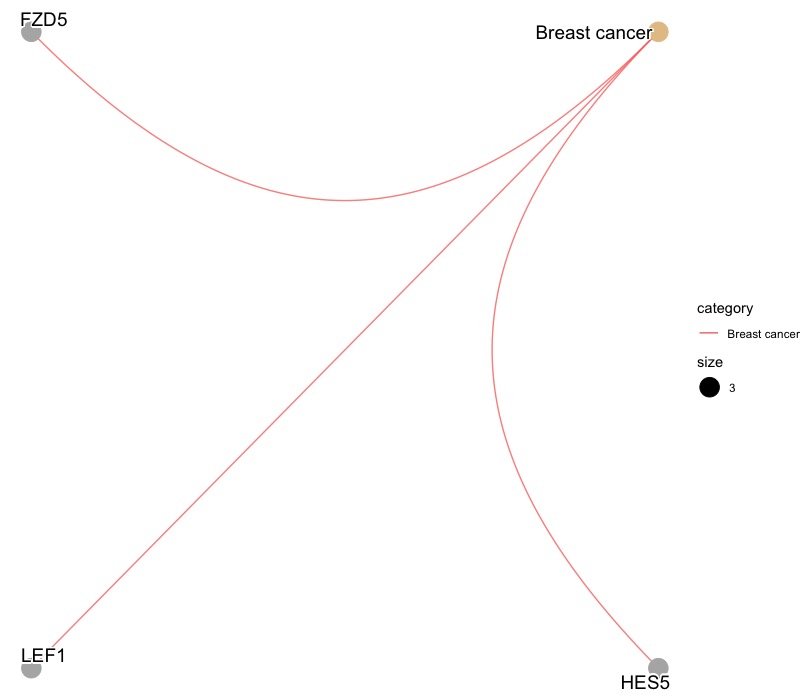}
         \caption{}
         \label{}
     \end{subfigure}
      \caption{Results of the analysis of the TCGA-BRCA data. GO-EA and KEGG-EA for the latency part of the model based on penCox.1se: (a) Barplot of significantly enriched GO terms (b) Network plot of enriched KEGG pathway terms and related selected genes}
    \label{betap_penCox_Scenario2}
\end{figure}

\begin{figure}[!h]
         \centering
         \includegraphics[width=1\textwidth]{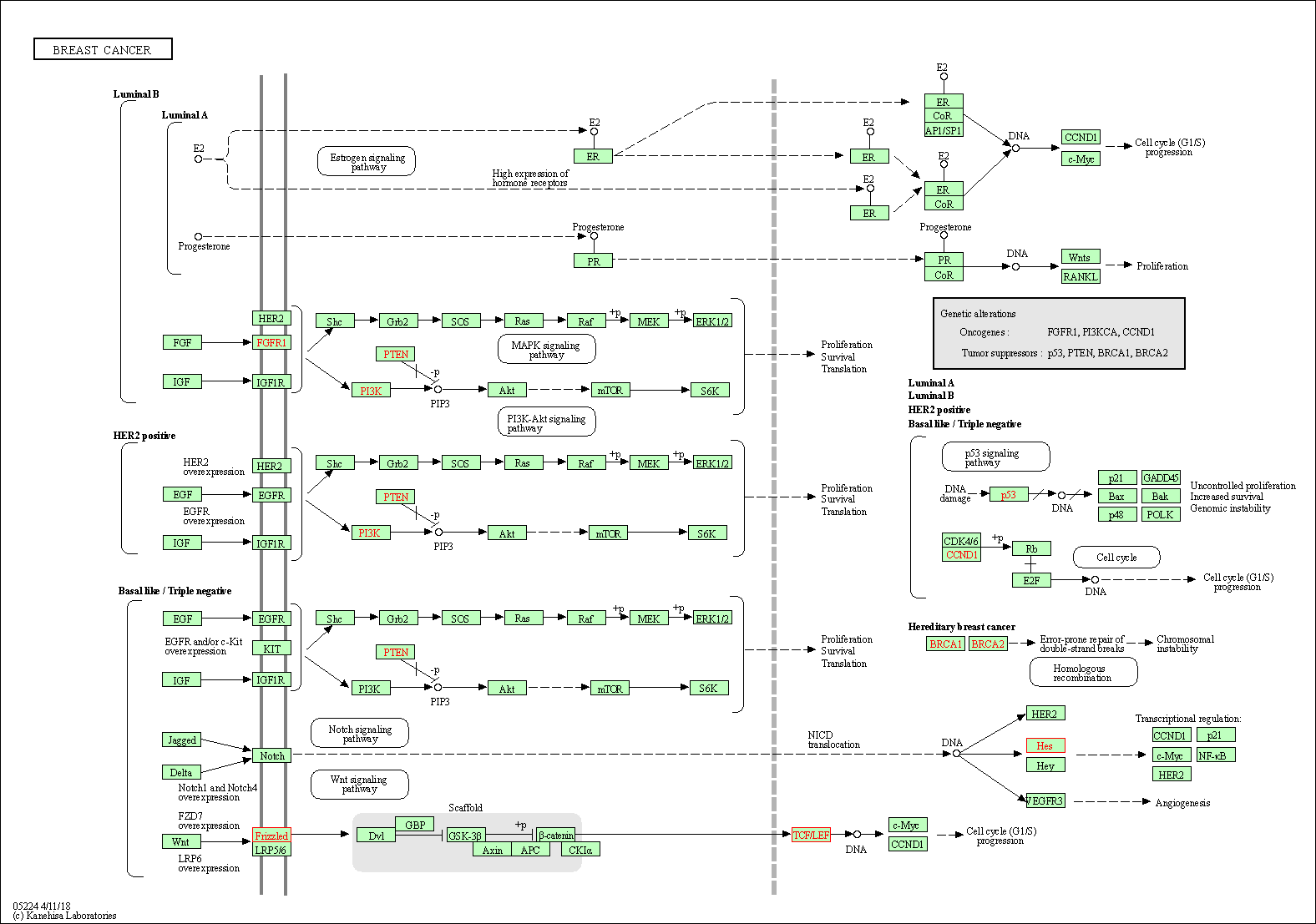}
    \caption{Results of the analysis of the TCGA-BRCA data. The Breast Cancer pathway is the one of top pathways enriched by selected biomarkers obtained when using penCox.1se.}
     \label{BRCA_pathway_betap_penCox_Scenario2}
\end{figure}

In Figures \ref{betap_penCox_Scenario2}-\ref{BRCA_pathway_betap_penCox_Scenario2}, we present GO-EA and KEGG-EA results, and the related breast cancer pathway.

\subsubsection*{S-2.2.5 Additional results for the prognostic risk score analyses}
\addcontentsline{toc}{subsubsection}{S-2.2.5 Additional results for the prognostic risk score analyses}

\begin{figure}[!h]
         \centering
         \begin{subfigure}[b]{0.49\textwidth}
          \centering
         \includegraphics[width=\textwidth]{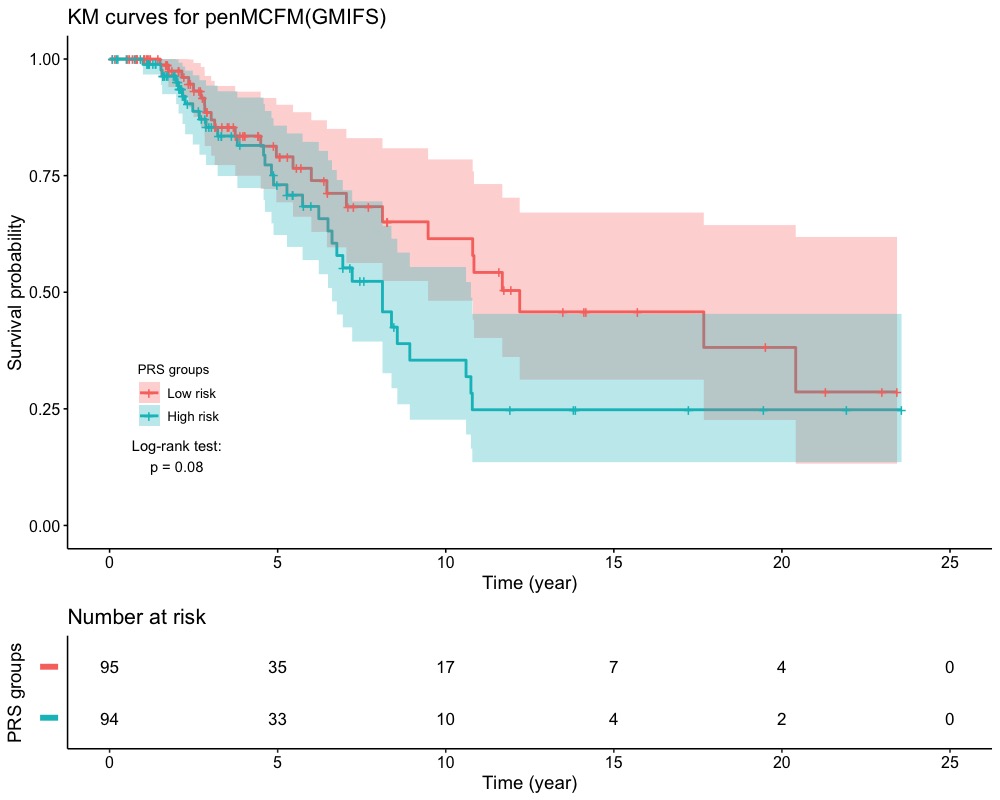}
          \caption{}
         \label{}
     \end{subfigure}
          \begin{subfigure}[b]{0.49\textwidth}
         \centering
         \includegraphics[width=\textwidth]{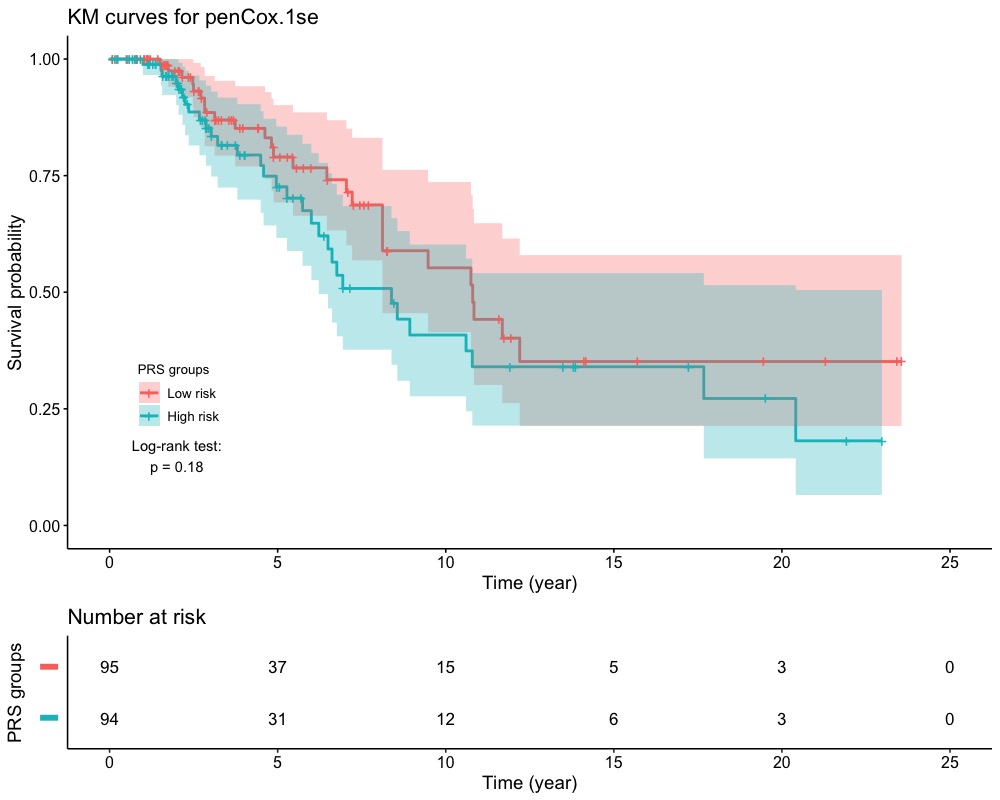}
         \caption{}
         \label{}
     \end{subfigure}
    \caption{Results of the analysis of the TCGA-BRCA data. KM curves for the TCGA-BRCA patients in the validation dataset when dichotomized into two groups by the median PRS  using the average results of $20$ repeats in Scenario $2$: (a) penMCFM(GMIFS) and (b) penCox.1se}
     \label{KM_plots_penMCFM_GMIFS_penCox}
\end{figure}

\begin{figure}[!h]
     \centering    
     \begin{subfigure}[b]{\textwidth}
      \centering   
    \includegraphics[width=0.8\textwidth]{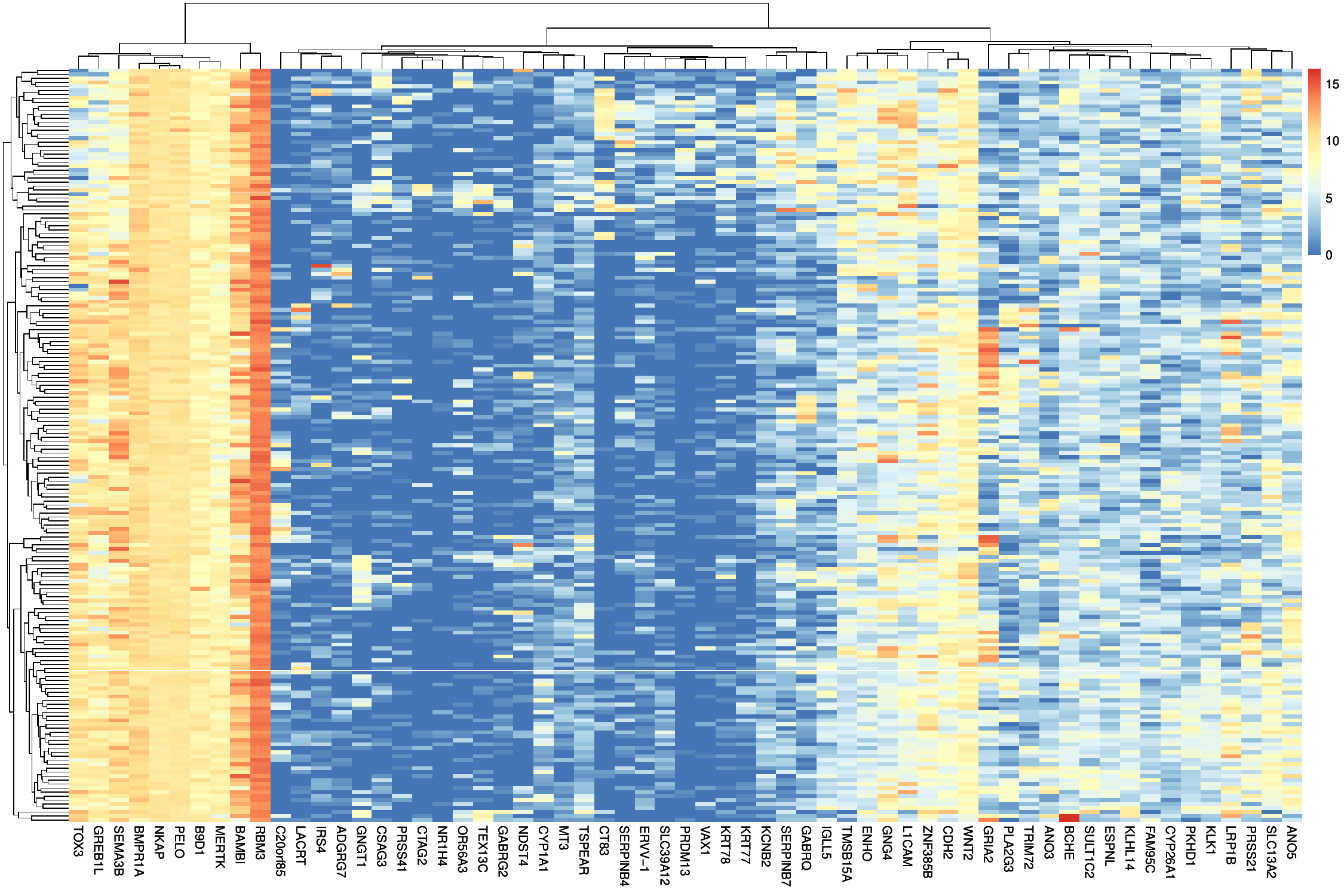}
    \caption{}
        \label{}
    \end{subfigure}
          \hfill
     \begin{subfigure}[b]{0.8\textwidth}
         \centering
          \includegraphics[width=\textwidth]{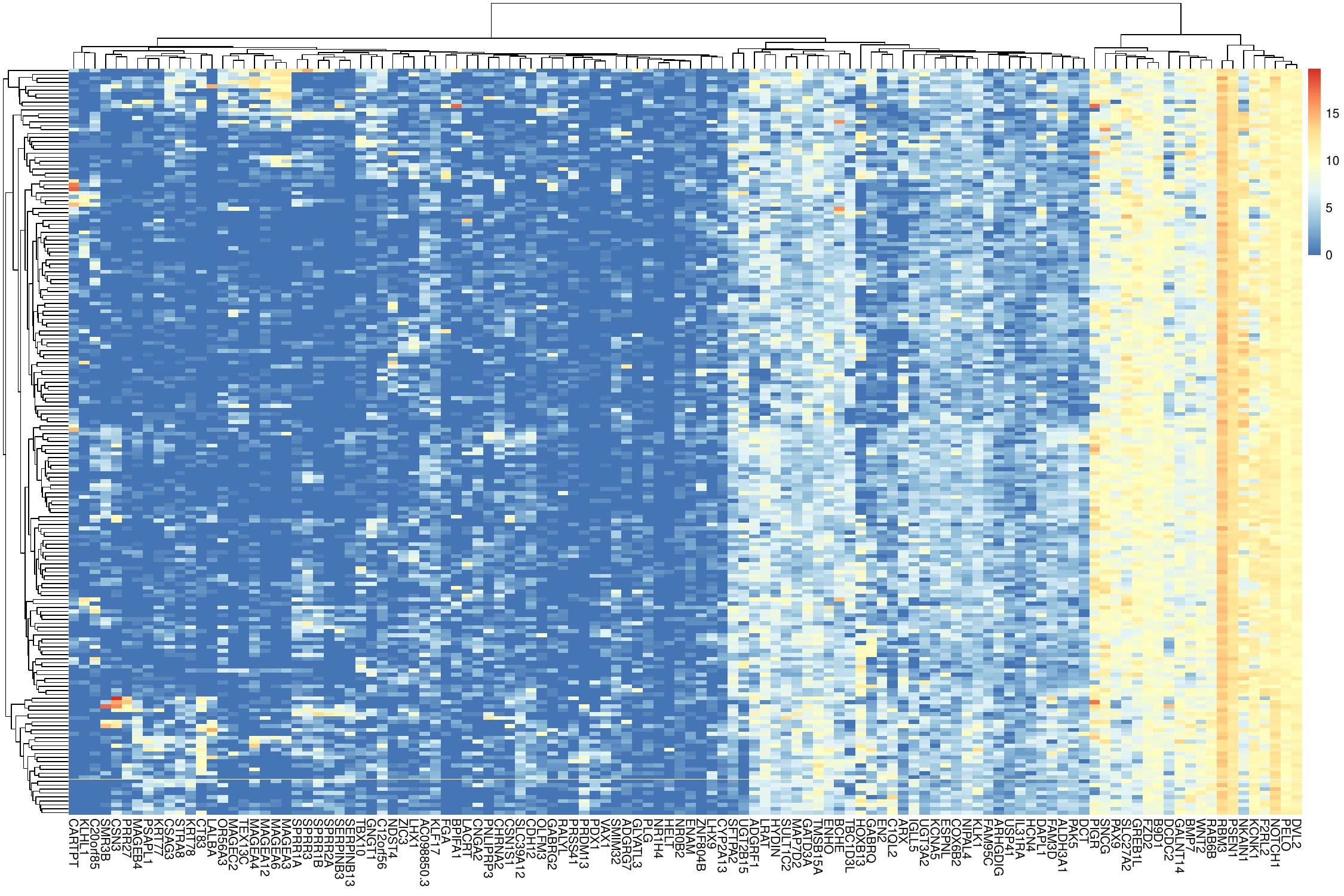}
    \caption{}
        \label{}
    \end{subfigure}
     \caption{Results of the analysis of the TCGA-BRCA data. Heatmap of the expression values of the selected genes that are selected in at least $2$ over $20$ repeats in Scenario $2$ for the validation dataset: (a) penMCFM(EM) and (b) MCM(GMIFS)}
  \label{heatmaps_validation_data}
\end{figure}

The KM curves of low and high risk groups are presented in Figure \ref{KM_plots_penMCFM_GMIFS_penCox} for penMCFM(GMIFS) and penCox.1se. The heatmaps of the expression values of the selected genes via penMCFM(EM) and MCM(GMIFS) based on the validation dataset are presented in Figure \ref{heatmaps_validation_data}.

\clearpage

\begin{table}[!h]
  \centering 
   \caption{Additional information on the genes list selected by penMCFM(EM) when considering the intersection of results
from both scenarios, also in  relation to the existing literature on BRCA.}
\begin{tabular}{C{2cm}|L{11cm}|C{2cm}|C{1.5cm}} \hline
 Biomarker & Description  & Reported literature & KEGG pathway \\  \hline
IGLL5 & The IGLL5 gene can serve not only as a prognostic biomarker for breast cancer patients but also as a potential source of therapeutic strategies for preventing breast cancer recurrence. Its expression is also correlated with tumor‐infiltrating immune cells. & \cite{ascierto2012signature, lee2021novel, liang2015molecular, xia2021igll5} & \\ 
 GABRQ & It has been demonstrated that the mRNA expression of GABRQ could serve as a prognostic marker for clear cell renal cell carcinoma and colon adenocarcinoma. It has been identified as one of the upregulated genes in triple-negative breast cancer (TNBC). &  \cite{lee2019gabrq, yan2020distinct, li2023delta} & \checkmark \\
 L1CAM &  L1CAM is expressed in numerous human cancers and is frequently associated with poor prognosis. It is suggested that L1CAM can be utilized for breast cancer diagnosis, indicating a potential correlation between L1CAM expression and the overall adverse prognosis of TNBC. & \cite{doberstein2014l1cam, altevogt2016l1cam, barron2022gene}  & \checkmark \\
 ADGRG7 & The role of this gene in head and neck squamous cell carcinoma, uterine corpus endometrial carcinoma, and breast cancers are investigated. It is observed that the gene is directly involved in breast tumor metastasis to bone tissues. & \cite{meng2021screening, lei2022correlation, singh2022genome} & \\
 VAX1 & It has been identified as one of the $10$ upregulated genes in TNBC. Its association with lung squamous cell carcinoma has also been observed, suggesting its potential as a prognostic biomarker for evaluating risk assessment. & \cite{gao2019prognostic, li2018combined} & \\
 CSAG3 & CSAG3, also known as TRAG-3 (Taxol Resistance Associated Gene-3), is reported to be upregulated in numerous tumors, including gastric cancer, urothelial carcinoma of the bladder, ovarian carcinoma, and melanoma. & \cite{zhang2019identification} & \\
 ENHO & Adropin is encoded by the energy homeostasis-associated (ENHO) gene. Its association with the development of colorectal cancer has also been investigated. GPR19 is activated by adropin is studied for the breast tumor cells. & \cite{rao2017g, jia2023low} & \\ 
 OR56A3 & The Olfactory Receptor (OR) family has gained attention as a potential biomarker for cancer. The significance of transcript abundance in certain OR genes is examined in an invasive breast carcinoma population. However, a substantial portion of the roles of OR genes in breast cancer remains understudied. &  \cite{weber2018olfactory, masjedi2019olfactory} & \checkmark \\ 
 C20orf85 & It has been implicated in various cancer types, including lung cancer, lower-grade glioma of the brain, ovarian cancer, and breast cancer. Further investigations are needed to explore its implications.  & \cite{hong2007inactivation, shih2018identification, bose2022computing, furrer2022association} & \\  
GREB1L & It has been observed that the GREB1L gene is implicated in the development of breast cancer, and it is proposed as a potential molecular marker for predicting the prognosis of breast cancer. It has been also demonstrated to exhibit a high correlation with both estrogen receptor and androgen receptor expression in breast/prostate cancer cell lines and primary tumors. & \cite{brophy2017gene, dong2023greb1l} & \\
BMPR1A & The Bone Morphogenetic Protein Receptor (BMPR) genes associated with various cancer types is explored. The role of the BMPR1A gene in breast cancer growth and metastasis is also investigated. & \cite{o2016bmp2, pickup2015deletion, hermawan2023bioinformatics} & \checkmark \\
TSPEAR & The TSPEAR gene has not been reported in any cancer study, except for one preprint study related to colorectal cancer. Further investigations are required to delve into its implications.  & \cite{li2022increased}& \\
KRT77 & The differential expression of KRT genes is investigated across various cancers. Its potential role as a biomarker in head and neck squamous cell carcinoma is also explored.  & \cite{dhakal2021divergent, takan2023light} & \\
 \hline
 \end{tabular}
 \label{table:genes}
\end{table}

\captionsetup{justification=raggedright, singlelinecheck=false}
\begin{table}[t]
\vspace{-13cm}
  \centering 
   \caption*{Table S$3$ Continued.}
\begin{tabular}{C{2cm}|L{11cm}|C{2cm}|C{1.5cm}} \hline
 Biomarker & Description  & Reported literature & KEGG pathway \\  \hline
 GNG4 & It has been reported as a potential biomarker in various cancer types, including bladder, colorectal, gastric, and breast cancer. & \cite{zhao2021identifying, mao2021identification, duan2022g, barron2022gene} & \checkmark \\
SERPINB7 & The SERPINB family genes are differentially expressed in the tumor tissues. The SERPINB7 gene has been studied in various cancer types, including breast, cervical, and lung cancers. & \cite{chou2012suppression, wei2023serpinb7, ou2023serpine1} & \\
GABRG2 & GABA receptor genes (GABR) constitute a group of genes associated with developmental and epileptic encephalopathies. They have also been identified to be associated with recurrent breast, colon, and laryngeal cancer samples. & \cite{hu2008cancer, tang2019signature, yan2020distinct, nwosu2022variable} & \checkmark \\ 
ZNF385B & The potential effects of ZNF385B expression in breast cancer have been determined in recent studies, suggesting its utility as a potential diagnostic and prognostic biomarker for breast cancer. It is also observed to be correlated with the overall survival of ovarian cancer patients. & \cite{elgaaen2012znf385b, yan2021downregulated, zhong2022identification}& \\
CT83 & CT83 is highly expressed in gastric, triple-negative breast, lung, and hepatocellular cancers. It is observed to be significantly associated with the overall survival of TNBC patients. & \cite{zhong2020identification, chen2021multiomics, li2022natural} & \\

 \hline
 \end{tabular}
\end{table}

\end{document}